\documentclass[onecolumn,showpacs,superscriptaddress,amsmath,amssymb]{revtex4}
\usepackage{graphicx}% Include figure files
\usepackage{dcolumn}% Align table columns on decimal point
\usepackage{bm}% bold math
\usepackage{color}
\usepackage{mathrsfs}
\usepackage{ulem}
\usepackage[german]{babel}

\def\bd{
\begin{document}} \def\ed{\end{document}}
\def\bmp{\begin{minipage}} \def\emp{\end{minipage}}
\def\bcc{\begin{center}} \def\ecc{\end{center}}     \def\npg{\newpage}
\def\beq{\begin{equation}} \def\eeq{\end{equation}} \def\hph{\hphantom}
\def\be{\begin{equation}} \def\ee{\end{equation}} \def\r#1{$^{[#1]}$}
\def\n{\noindent} \def\ni{\noindent} \def\pa{\parindent}
\def\hs{\hskip} \def\vs{\vskip} \def\hf{\hfill} \def\ej{\vfill\eject}
\def\cl{\centerline} \def\ob{\obeylines}  \def\ls{\leftskip}
\def\underbar#1{$\setbox0=\hbox{#1} \dp0=1.5pt \mathsurround=0pt
   \underline{\box0}$}   \def\ub{\underbar}    \def\ul{\underline}
\def\f{\left} \def\g{\right} \def\e{{\rm e}} \def\o{\over} \def\d{{\rm d}}
\def\vf{\varphi} \def\pl{\partial} \def\cov{{\rm cov}} \def\ch{{\rm ch}}
\def\la{\langle} \def\ra{\rangle} \def\EE{e$^+$e$^-$} \def\pt{p_{\rm t}}
\def\pti{p_{{\rm t},i}} \def\vti{v_{{\rm t},i}}
\def\ptj{p_{{\rm t},j}}\def\Pt{P_{\rm t}} \def\vt{v_{\rm t}}

\def\bitz{\begin{itemize}} \def\eitz{\end{itemize}}
\def\btbl{\begin{tabular}} \def\etbl{\end{tabular}}
\def\btbb{\begin{tabbing}} \def\etbb{\end{tabbing}}
\def\beqar{\begin{eqnarray}} \def\eeqar{\end{eqnarray}}
\def\\{\hfill\break} \def\dit{\item{-}} \def\i{\item}
\def\bbb{\begin{thebibliography}{9} \itemsep=-1mm}
\def\ebb{\end{thebibliography}} \def\bb{\bibitem}
\def\bpic{\begin{picture}(260,240)} \def\epic{\end{picture}}
\def\akgt{\cl{\bf ACKNOWLEDGMENTS}}
\def\fgn{\noindent{\bf\large\bf figure captions}}
%%%%%%%%%%%%%%%%%%%%%%%%%%%%%%%%%%%%%%%%%%%%%%%%%%%%%%%%%%%%%%%%%%%%%%
\def\m1{\langle N_p\rangle} \def\u2{\langle N_{\bar p}\rangle} \def\Nap{N_{\bar
p}}
%%%%%%%%%%%%%%%%%%%%%%%%%%%%%%%%%%%%%%%%%%%%%%%%%%%%%%%%%%%%%%%%%%%%%%%%%%%%%%
\def\lan{\langle}
\def\ran{\rangle}
\def\p{\pi}
\def\ifmath#1{\relax\ifmmode #1\else $#1$\fi}%
\def\rc{\ifmath{{\mathrm{c}}}}
\def\cut{\ifmath{{\mathrm{cut}}}}
\def\rF{\ifmath{{\mathrm{F}}}}
\def\rK{\ifmath{{\mathrm{K}}}}
\def\rp{\ifmath{{\mathrm{p}}}}
\def\rt{\ifmath{{\mathrm{t}}}}
\def\LAB{\ifmath{{\mathrm{LAB}}}}
\def\cut{\ifmath{{\mathrm{cut}}}}
\def\beq{\begin{equation}}
\def\eeq{\end{equation}}

\newcommand{\cinst}[2]{$^{\mathrm{#1}}$~#2\par}
\newcommand{\crefi}[1]{$^{\mathrm{#1}}$}
\newcommand{\crefii}[2]{$^{\mathrm{#1,#2}}$}
\newcommand{\crefiii}[3]{$^{\mathrm{#1,#2,#3}}$}
\newcommand{\HRule}{\rule{0.5\linewidth}{0.5mm}}

\bd
\title{Universal critical behavior of generaliezd susceptibilities of net-baryon number at small quark mass}

\author{Xue Pan}\email{panxue1624@163.com}
\affiliation{School of Electronic Information and Electrical Engineering, Chengdu University, Chengdu 610106, China}
\author{Peng Yang}
\affiliation{School of Electronic Engineering, Chengdu Technological University, Chengdu 611730, China}
\author{Lingling Cao}
\affiliation{School of Electronic Engineering, Chengdu Technological University, Chengdu 611730, China}

\begin{abstract}
%\sout{}
In the limit of small quark masses, the angle between the temperature axis and the applied magnetic field direction in the three-dimensional Ising model vanishes as $m_q^{2/5}$ when mapped onto the QCD $T-\mu_B$ phase plane. By selecting two distinct small angles and projecting the Ising model results onto QCD, we have investigated the universal critical behavior of the sixth-, eighth-, and tenth-order susceptibilities of the net-baryon number. When considering only the leading critical contribution, the negative dip in the $\mu_B$ dependence of the generalized susceptibilities is not universal, in contrast to the observation in the case where the angle is $90^{\circ}$. Its existence depends on the mapping parameters and the distance to the phase transition line. After incorporating the sub-leading critical contribution, the negative dip is enhanced to some extent but remains a non-robust feature. In contrast, the positive peak structure persists in all cases and represents a robust characteristic of generalized susceptibilities of the net-baryon number near the critical point.
\end{abstract}


\maketitle

\section{Introduction}

One of the main goals of current relativistic heavy-ion collision experiments is to reveal the phase diagram of quantum chromo-dynamics (QCD)~\cite{maingoal}. Where the location of the critical point, which is a unique character of the QCD phase diagram, is the most important.

High-order cumulants of multiplicity distributions of conserved charges, such as net-baryon, net-charge and net-strangeness, are suggested to search for the QCD critical point~\cite{stephanov-prl91, koch, Stephanov-prl102, Karsch-EPJC71}. They are more sensitive to the correlation length, and can be measured in experiments and also calculated in theories.

Recently, lattice QCD predict that, if the QCD critical point exists, the critical temperature ($T_C$) should be lower than $135$ MeV~\cite{chiraltransitionT1,chiraltransitionT2}. On the other side, the functional renormalization group and Dyson-Schwinger equations approaches show that the transition from hadronic matter to quark gluon plasma (QGP) is a crossover with increasing $\mu_B$ for $\mu_B/T \lesssim 4$~\cite{Fuweijie-PRD101,Fischer,Isserstedt,Gaofei1,Gaofei2}, where $\mu_B$ and $T$ are the net-baryon chemical potential and the QCD temperature, respectively. A QCD critical point is found at larger $\mu_B$. For example, the QCD critical point is at $(T_C,\mu_{BC})=(117, 488))$ MeV in Ref.~\cite{Fischer}, $(T_C,\mu_{BC})=(107, 635))$ MeV in Ref.~\cite{Fuweijie-PRD101}, $(T_C,\mu_{BC})=(109, 610))$ MeV in Ref.~\cite{Gaofei2}, where $\mu_{BC}$ are the net-baryon chemical potential at the QCD critical point.

However, it is still challenging to get quantitative reliable results of the high-order cumulants at large $\mu_B$ for $\mu_B/T > 4$ from lattice QCD, the functional renormalization group and Dyson-Schwinger equations approaches~\cite{Gaofei2,LQCD1,LQCD2,LQCD3,Fuweijie-PRD104}. An alternative method is based on the universality of the critical behavior~\cite{Stephanov-prl107,stephanov-prc103}.

The QCD critical point is in the same universality class with that of the three-dimensional Ising model. The Ising variables, reduced temperature ($t$) and magnetic field ($h$), can be mapped onto the QCD $T-\mu_B$ phase plane to investigate the critical features of QCD~\cite{Asakawa-PRC71,Stephanov-PRD100,stephanov-prc101}. The $t$ axis is tangential to the QCD first-order phase transition line at the critical point. Generally, the $h$ axis will deform when mapped onto the QCD $T-\mu_B$ phase plane. But it is not
clear how this occurs. The mapping parameters are not universal. The common assumption in existing literature is that the $h$ axis is orthogonal to the $t$ axis~\cite{Asakawa-PRC71, PRC92.034912}.

Through mapping the Ising results to that of QCD, the static universal critical behavior of susceptibilities of the net-baryon number can be predicted. It is found that the fourth-, sixth-, eighth- and tenth-order susceptibilities all has a negative dip when the critical point is approached from the crossover side when considering only the leading critical contribution~\cite{Stephanov-prl107,cpc2016,cpc2025}. If incorporating the sub-leading critical contribution, the negative dip in the fourth-order susceptibilities disappears~\cite{stephanov-prc103}. While for the sixth-, eighth, and tenth-order susceptibilities, the existence of the negative dip depends on the distance to the phase transition line and scaling parameters of the mapping from the Ising model to QCD~\cite{cpc2025}.

Recently, it is pointed out in Ref.~\cite{Stephanov-PRD100} that in the limit of small quark masses, the mapping parameters show universal dependence on the quark masses when the critical point is close to the tricritical point. The angel between the $t$ axis and $h$ axis vanishes in the QCD $T-\mu_B$ phase plane as $m_q^{2/5}$, where $m_q\equiv m_u \approx m_d$ is the light quark mass. In the random matrix model, it shows the angle follow the relation $0<\alpha_2<\alpha_1$~\cite{Stephanov-PRD100}. $\alpha_1$ and $\alpha_2$ represent the angles between the horizontal axis (where $T$ is a constant) and $t$ axis and $h$ axis in the Ising model when they are mapped onto the QCD $T-\mu_B$ phase plane, respectively. A sketch of such mapping is shown in Fig. 1. The solid black line represents the QCD first-order phase transition line. The red point is the QCD critical point.

\begin{figure}[hbt]
	\centering
	\includegraphics[width=0.65\textwidth]{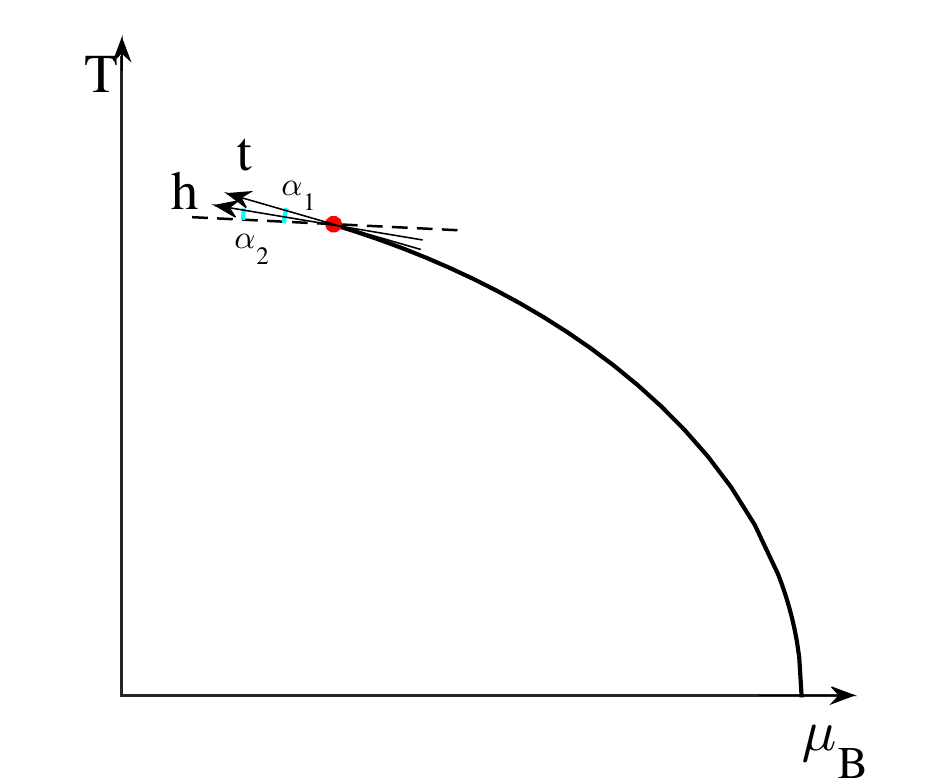}
	\caption{\label{fig1}(Color online) A sketch of mapping the Ising temperature $t$ and magnetic field $h$ onto the QCD $T-\mu_B$ phase plane in the limit of small quark masses. The red point represents the QCD critical point at the end of the first-order phase transition line.}
\end{figure}

With this kind of mapping, as shown in Ref.~\cite{stephanov-prc103}, when the sub-leading critical contribution is included, the existence of negative dip in the fourth-order susceptibility of net-baryon number on the freeze-out line may exist. But it depends on the distance to the phase transition line and scaling parameters. Given the significant deviation of the mapping angles with that in the common assumption, it would be worthwhile to investigate how sub-leading critical contribution influences other high-order susceptibilities—both when they are incorporated and when they are neglected.

In this paper, on the assumption of the equilibrium of the QCD system and  $0<\alpha_2<\alpha_1$, using parametric representation of the three-dimensional Ising model, we have studied the critical contribution on the behavior of sixth-, eighth- and tenth-order susceptibilities of the net-baryon number. Both the leading as well as sub-leading critical contribution is discussed.

The paper is organized as follows. In section 2, the parametric representation of the three-dimensional Ising model is introduced. Furthermore, the linear mapping from the Ising model to QCD is presented. The expression of generalized susceptibilities of net-baryon number is deduced. In section 3, with $\alpha_2=1.8^{\circ}$, the critical behavior of net-baryon number susceptibilities is systematically examined both with and without the sub-leading critical contribution. The density plots on the QCD $T-\mu_B$ phase plane as well as the $\mu_B$ dependence of the susceptibilities along the freeze-out curves are presented and analyzed in detail. In section 4, by setting $\alpha_2=7.8^{\circ}$, the critical behavior of net-baryon number susceptibilities is further examined. Finally, conclusions and summary are given in section 5.

\section{The linear mapping from Ising model to QCD}

Magnetization ($M$) and reduced temperature ($t$) can be parameterized by two variables $R$ and $\theta$ in the parametric representation of the three-dimensional Ising model~\cite{linearpara, linearpara3},
\begin{equation}\label{parametric}
	M=m_0R^{\beta}\theta,~~~~~~t=R(1-\theta^2).
\end{equation}
The equation of state can be expressed by $R$ and $\theta$ as follows,
\begin{equation}\label{equation state}
	h=h_0R^{\beta\delta}\widetilde h(\theta).
\end{equation}
Where $m_0$ in Eq.~\eqref{parametric} and $h_0$ in Eq.~\eqref{equation state} are normalization constants. By imposing the normalization conditions $M(t=-1,h=+0)=1$ and $M(t=0,h=1)=1$, their values can be fixed at $0.605$ and $0.364$, respectively. $\beta\simeq0.326$ and $\delta\simeq4.8$ are critical exponents of the three-dimensional Ising universality class~\cite{Isingexponents}. $\widetilde h(\theta)=\theta(1-0.76201\theta^{2}+0.00804\theta^{4})$. The parameters are within the range $R\geqslant0$ and $|\theta|\leqslant1.154$.

The free energy density can be defined as~\cite{stephanov-prc101},
\begin{equation}\label{free energy density}
	F(M,t)=h_0m_0R^{2-\alpha}g(\theta),
\end{equation}
where $\alpha\simeq0.11$ is another critical exponent of the three-dimensional Ising universality class. The relation $2-\alpha=\beta(\delta-1)$ holds, and
\begin{equation}\label{gtheta}
	g(\theta)=c_0+c_1(1-\theta^2)+c_2(1-\theta^2)^2+c_3(1-\theta^2)^3,
\end{equation}
with
\begin{equation}
	\begin{split}
		&c_0=\frac{\beta}{2-\alpha}(1+a+b),\\
		&c_1=-\frac{1}{2(\alpha-1)}[(1-2\beta)(1+a+b)-2\beta(a+2b)],\\
		&c_2=-\frac{1}{2\alpha}[2\beta b-(1-2\beta)(a+2b)],\\
		&c_3=-\frac{1}{2(\alpha+1)}b(1-2\beta).
		\nonumber
	\end{split}
\end{equation}
Then the Gibbs free energy density is
\begin{equation}\label{Gibbs free energy density}
	G(h,t)=F(M,t)-Mh.
\end{equation}

The pressure is equivalent to the Gibbs free energy density up to a minus sign: $P = -G$. Consequently, the expression for the pressure in the Ising model can be formulated as follows,
\begin{equation}\label{Ising pressure}
	P^{Ising}(R,\theta)=h_0m_0R^{2-\alpha}[\theta\widetilde h(\theta)-g(\theta)].
\end{equation}

The $n_{th}$-order susceptibility of the magnetization represented by $R$ and $\theta$ can be got from the derivatives of the pressure with respect to $h$,
\begin{equation}\label{cumulant of magetization}
	\chi_{n}^{M}=\left(\frac{\partial^{n}P^{Ising}}{\partial h^{n}}\right)_t.
\end{equation}
For example,
\begin{equation}\label{M1}
	\chi_1^{M} = \left(\frac{\partial P^{Ising}}{\partial h} \right)_t 
	=\frac{\partial P^{Ising}}{\partial R}\left(\frac{\partial R}{\partial h} \right)_t+\frac{\partial P^{Ising}}{\partial \theta}\left(\frac{\partial \theta}{\partial h}\right)_t.
\end{equation}
Where $\partial R/\partial h$ and $\partial \theta /\partial h$, can be got from Eqs.~\eqref{parametric} and \eqref{equation state}.
The $n_{th}$-order susceptibility of the energy represented by $R$ and $\theta$ can be got from the derivatives of the pressure with respect to $t$,
\begin{equation}\label{cumulant of energy}
	\chi_{n}^{E}=\left(\frac{\partial^{n}P^{Ising}}{\partial t^{n}} \right)_h.
\end{equation}
For example,
\begin{equation}\label{E1}
	\chi_1^{E} = \left(\frac{\partial P^{Ising}}{\partial t} \right)_h =\frac{\partial P^{Ising}}{\partial R}\left(\frac{\partial R}{\partial t} \right)_h+\frac{\partial P^{Ising}}{\partial \theta}\left(\frac{\partial \theta}{\partial t}\right)_h.
\end{equation}
Where $\partial R /\partial t$ and $\partial \theta /\partial t$ can be got from Eqs.~\eqref{parametric} and \eqref{equation state}.

The $n_{th}$-order off-diagonal susceptibility of magnetization and energy represented by $R$ and $\theta$ can be got from the derivatives of the pressure with respect to both $h$ and $t$,
\begin{equation}\label{cumulant of Ising}
	\chi_{n_1,n_2}^{M,E}=\frac{\partial^{n_1+n_2}P^{Ising}}{\partial h^{n_1}\partial t^{n_2}}, n=n_1+n_2, n_1=1,2,3...,n_2=1,2,3...
\end{equation}
For example,
\begin{equation}\label{cumulant of ME}
	\chi_{1,1}^{M,E} =\frac{\partial^{2} P^{Ising}}{\partial h\partial t}  =\frac{\partial \chi_1^{M}}{\partial R}\left(\frac{\partial R}{\partial t} \right)_h+\frac{\partial \chi_1^{M}}{\partial \theta}\left(\frac{\partial \theta}{\partial t}\right)_h.
\end{equation}

In order to map the results of the Ising model to that of QCD, a linear relationship~\cite{linearmap1, stephanov-prc101, stephanov-prc103} including six mapping parameters can be written as follows:
\begin{equation}\label{linear map}
	\frac{T-T_C}{T_C}=w(\rho t\sin\alpha_1+h\sin\alpha_2),
\end{equation}
\begin{equation}\label{linear map1}
	\frac{\mu_B-\mu_{BC}}{T_C}=w(-\rho t\cos\alpha_1-h\cos\alpha_2),
\end{equation}
where $T_C$ and $\mu_{BC}$ are the temperature and net-baryon chemical potential at the QCD critical point, respectively. $w$ and $\rho$ are two scaling parameters of the mapping from Ising model to QCD. $\alpha_1$ and $\alpha_2$ are two angles which have been introduced in Section 1.

As in Ref.~\cite{cpc2025}, the number of mapping parameters can be reduced by supposing that the critical point is located on the QCD phase transition line,
\begin{equation}\label{QCD transition line}
	T= T_0[1-\kappa(\frac{\mu_B}{T_0})^2-\lambda (\frac{\mu_B}{T_0})^4].
\end{equation}
To keep consistent with our previous work, the same input are used as in Ref.~\cite{cpc2025}, which is based on the recent results from lattice QCD~\cite{fodor-plb751,Bazavov2019,Borsanyi2020}, the functional renormalization group and Dyson-Schwinger equations approaches~\cite{Fuweijie-PRD101,Fischer,Isserstedt,Gaofei1,Gaofei2}. That is $T_0=156.5$ MeV, $\kappa=0.015$, $\lambda=0.000256$, and $(T_C,\mu_{BC})=(107, 635)$ MeV.

Because the $t$ axis in the Ising model is tangential to the first-order phase transition line at the QCD critical point, the value of $\alpha_1$ can be fixed at $10.8^{\circ}$. There are two different considerations of value for $\alpha_2$. One is that in the random matrix model, the angle $\alpha_2$ is very small and close to zero~\cite{Stephanov-PRD100}. So $\alpha_2$ is set as $1.8^{\circ}$. In the other case where $m_q \rightarrow 0$, the $h$ axis is nearly parallel to the $t$ axis, $\alpha_2$ is set as $7.8^{\circ}$. At last, there are two unknown parameters $w$ and $\rho$.

Generalized susceptibilities of net-baryon number ($\chi_n^{B}$) can be obtained from the $n_{th}$-order derivatives of the pressure with respect to $\mu_B$ at fixed $T$:
\begin{equation}\label{net-baryon number susceptibility}
	\chi_n^{B}(T,\mu_B)=\left(\frac{\partial^{n} P/T^4}{\partial (\mu_B/T)^n}\right)_T.
\end{equation}

The full QCD pressure can be reconstructed as Ref.~\cite{stephanov-prc101},
\begin{equation}\label{pressure}
	P(T,\mu_B)=T^4 \sum_{n}c_n^{Non-Ising}(T)\left(\frac{\mu_B}{T}\right)^n +P_C^{QCD}(T,\mu_B),
\end{equation}
where the first term on the left side is the Taylor expansion of 
the pressure from the "Non-Ising" contribution. $c_n^{Non-Ising}(T)$ is the corresponding Taylor expansion coefficients. While $P_C^{QCD}(T,\mu_B)$ represents the critical pressure mapped from the three-dimensional Ising model onto QCD. The details can be got from Refs.~\cite{stephanov-prc101, stephanov-prc103}.

In this paper, we consider only the critical point contribution to the behavior of the sixth-, eighth- and tenth-order susceptibilities of net-baryon number, the pressure in Eq.~\eqref{net-baryon number susceptibility} can be written as follows~\cite{stephanov-prc101},
\begin{equation}\label{QCD critical pressure}
	P(T,\mu_B)=T_C^4P^{Ising}(R(T,\mu_B),\theta(T,\mu_B)).
\end{equation}

The $2n$th-order susceptibility of net-baryon number can be written as
\begin{equation}\label{susceptibilities}
	\begin{split}
		&\chi_{2n}^B = T_C^4T^{2n-4}\times \\
		&\left[\left(\frac{\partial h}{\partial \mu_B}\right)^{2n}\chi_{2n}^{M}+ \sum_{k=1}^{2n-1} C(2n,k) \left(\frac{\partial h}{\partial \mu_B}\right)^{2n-k} \left(\frac{\partial t}{\partial \mu_B}\right)^{k}\chi_{2n-k,k}^{M,E} + \left(\frac{\partial t}{\partial \mu_B}\right)^{2n}\chi_{2n}^{E}\right]	
	\end{split}
\end{equation}
where $C(2n,k)=(2n)!/k!/(2n-k)!$. $\partial h /\partial \mu_B$ and $\partial t /\partial \mu_B$ can be got from Eq.~\eqref{linear map1},
\begin{equation}\label{partial h}
	\partial h /\partial \mu_B=-\sin(\alpha_1)/(T_C w \sin(\alpha_1 - \alpha_2)),
\end{equation}
\begin{equation}\label{partial t}
	\partial t /\partial \mu_B=\sin(\alpha_2)/(T_C w \rho \sin(\alpha_1 - \alpha_2)).
\end{equation}
If considering only the leading singular contribution, the corresponding $2n_{th}$-order susceptibility of net-baryon number is as follows,
\begin{equation}\label{six}
	\chi_{2n}^{B,L}=T_C^4T^{2n-4}\left(\frac{\partial h}{\partial \mu_B}\right)^{2n}\chi_{2n}^{M}.
\end{equation}

\begin{figure*}[hbt]
	\centering
	\includegraphics[width=0.32\textwidth]{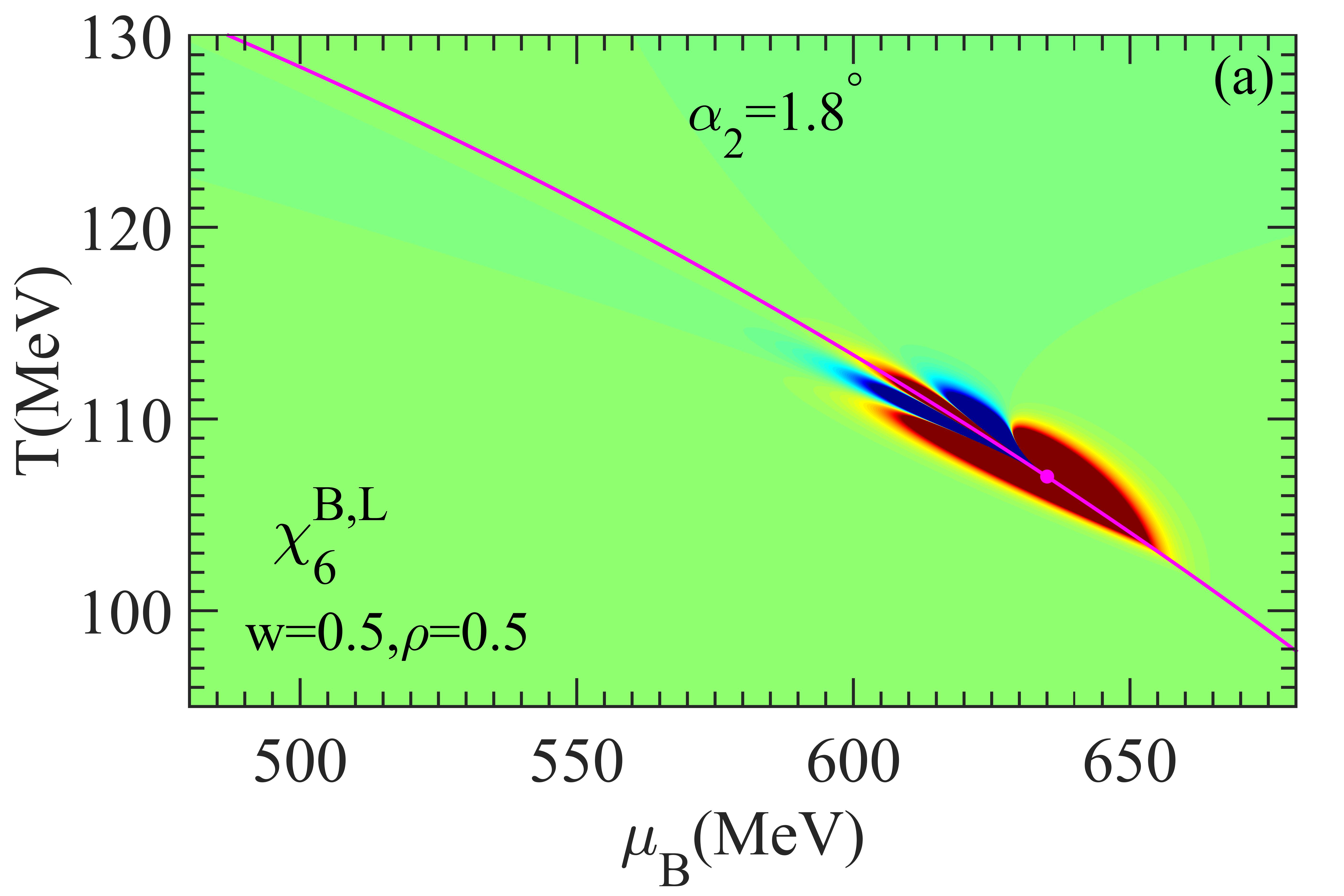}
	\includegraphics[width=0.32\textwidth]{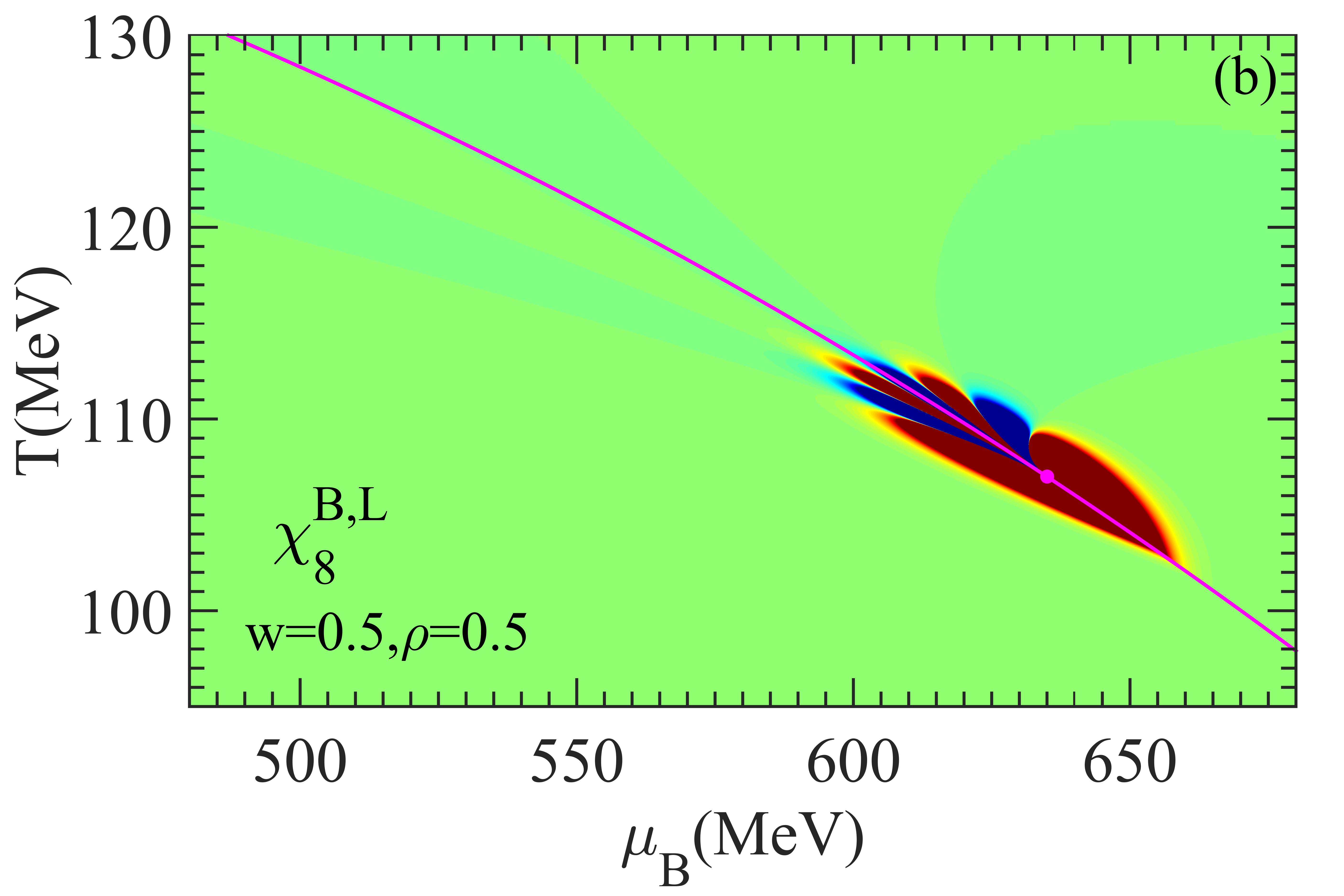}
	\includegraphics[width=0.32\textwidth]{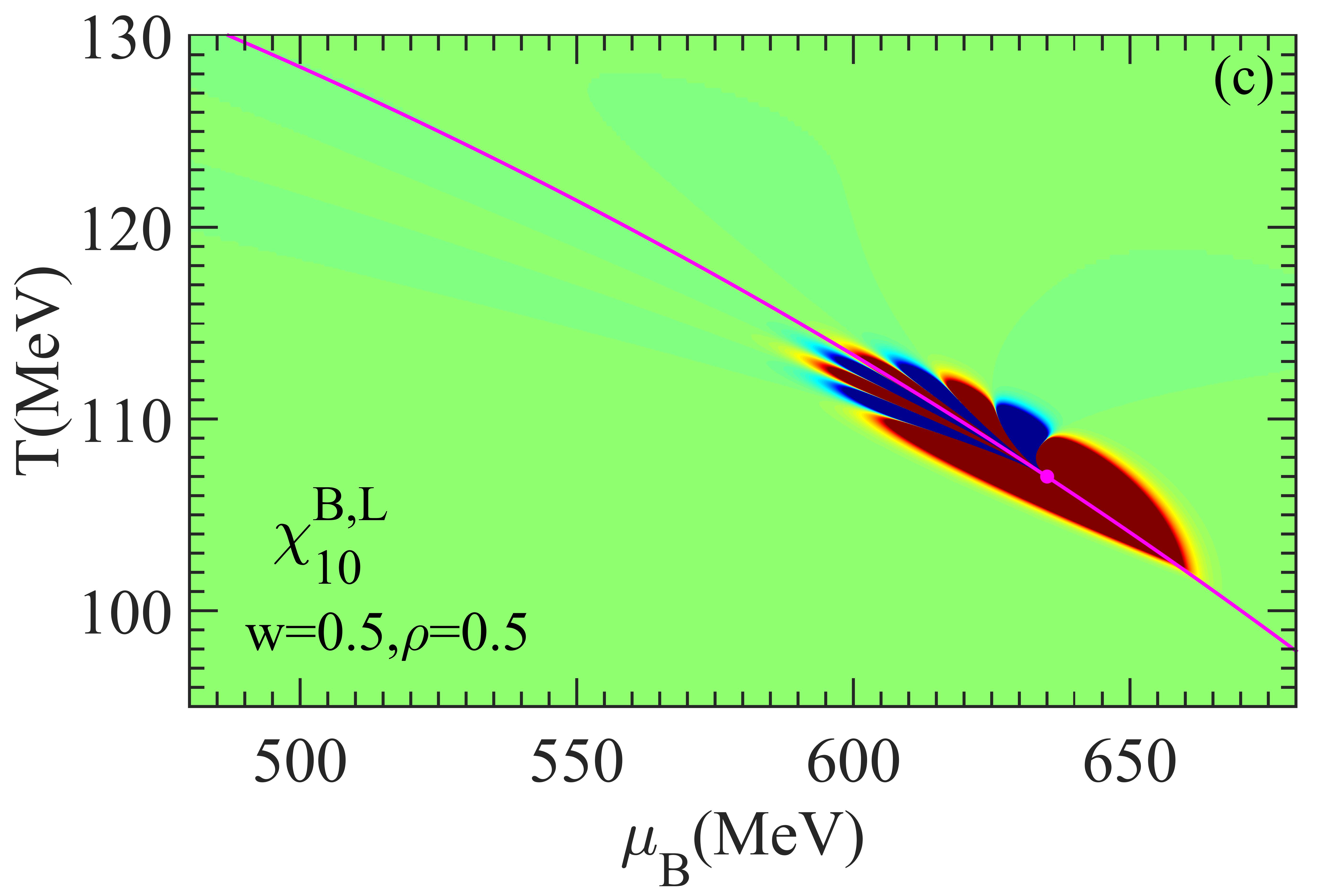}
	\includegraphics[width=0.32\textwidth]{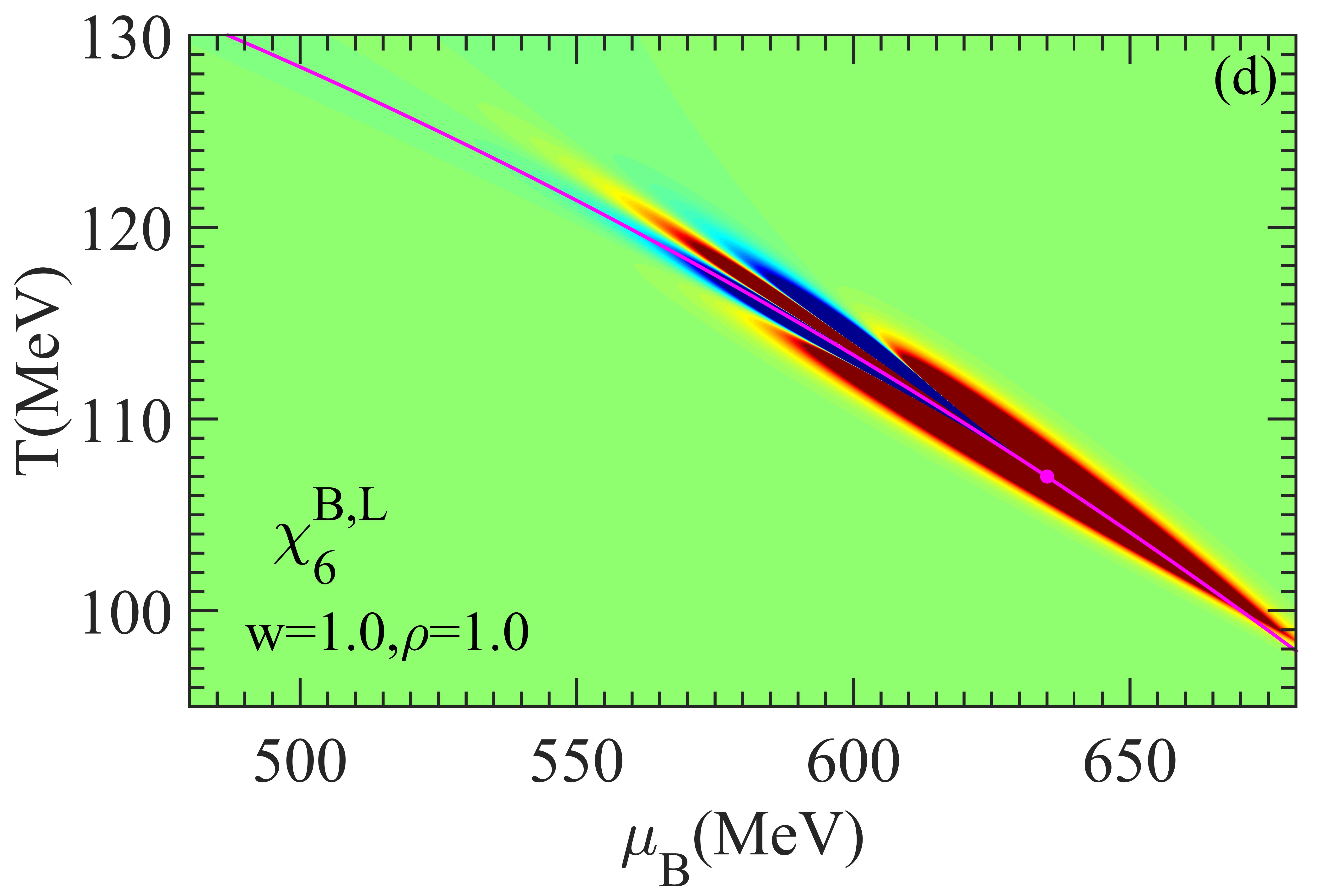}
	\includegraphics[width=0.32\textwidth]{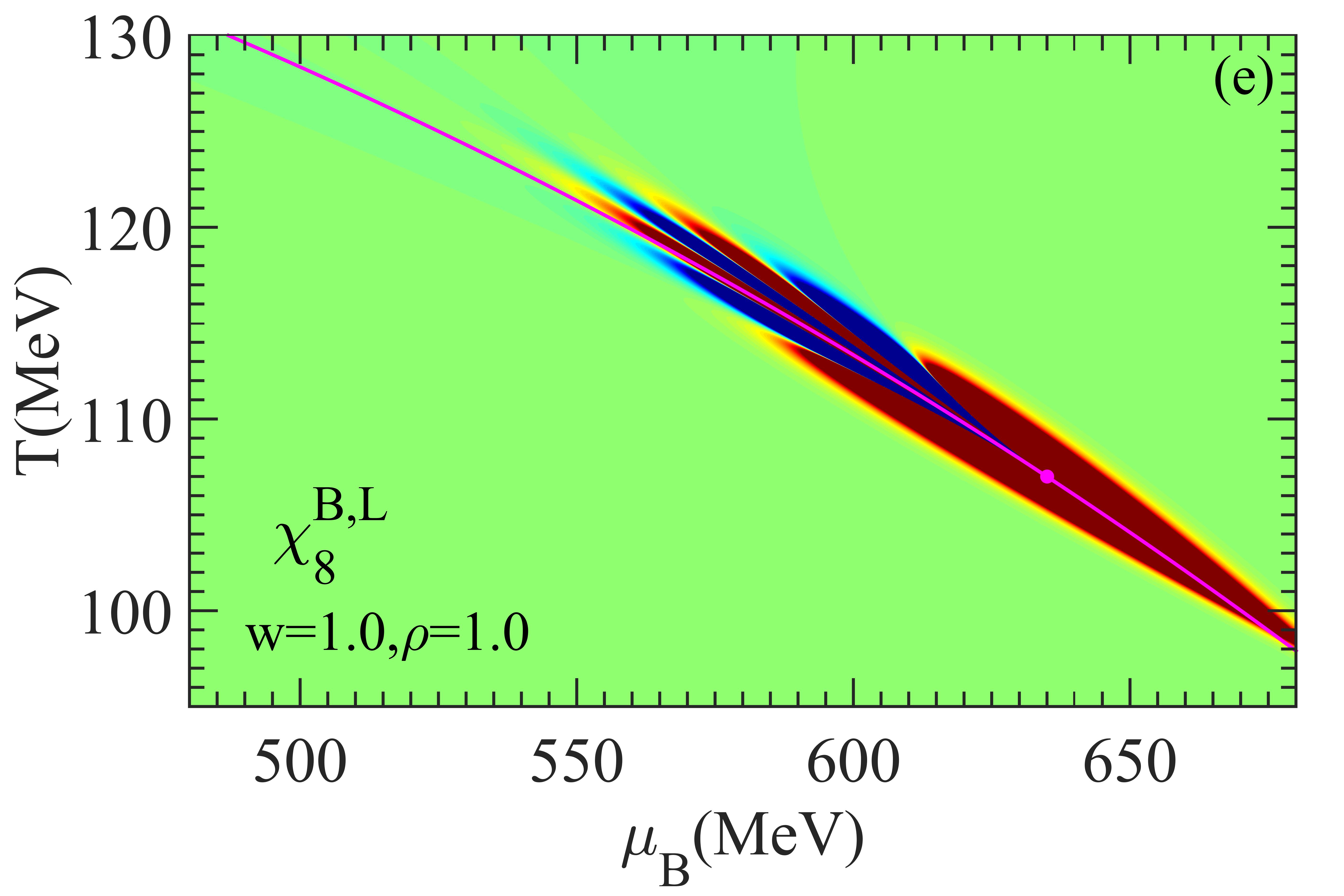}
	\includegraphics[width=0.32\textwidth]{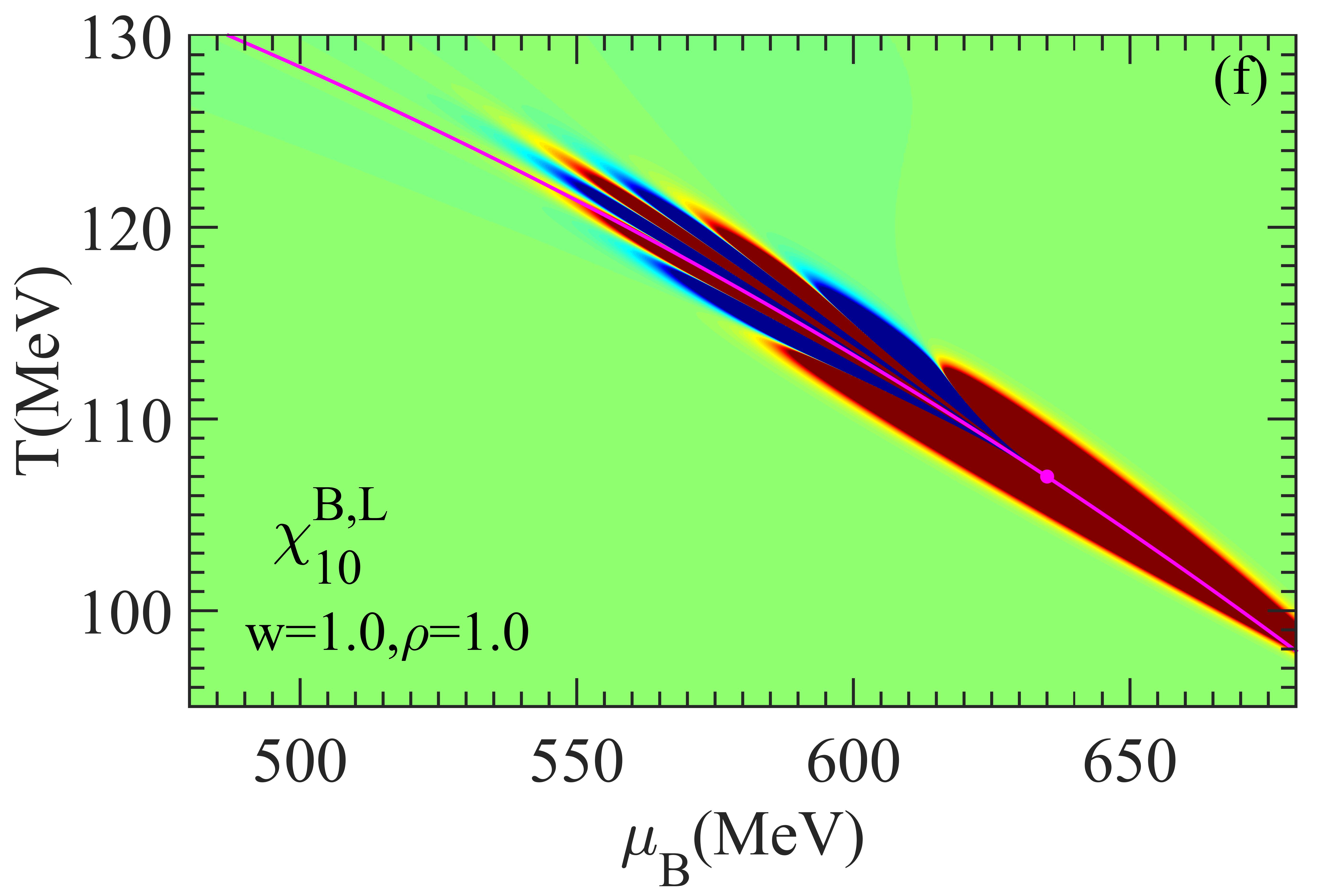}
	\includegraphics[width=0.32\textwidth]{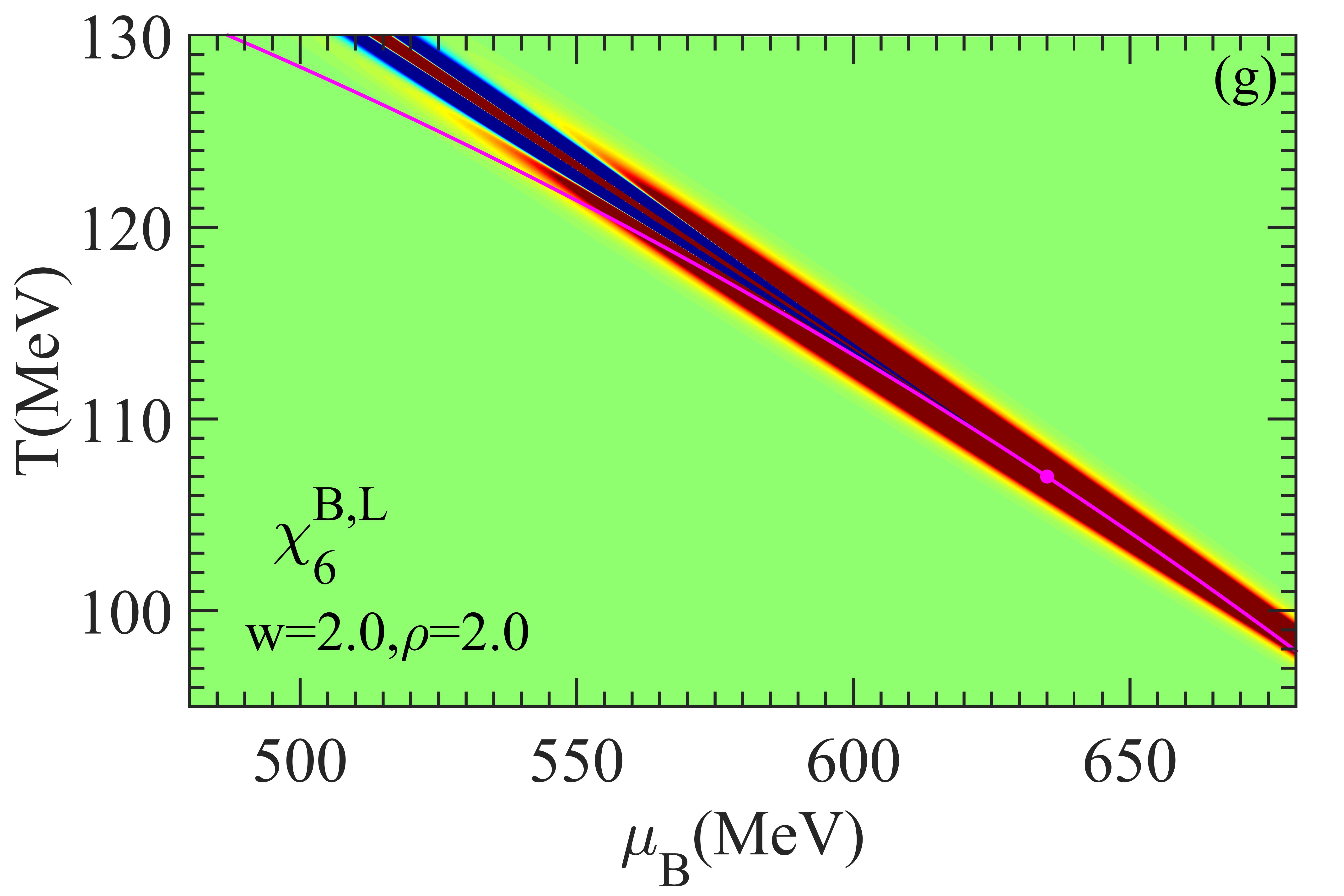}
	\includegraphics[width=0.32\textwidth]{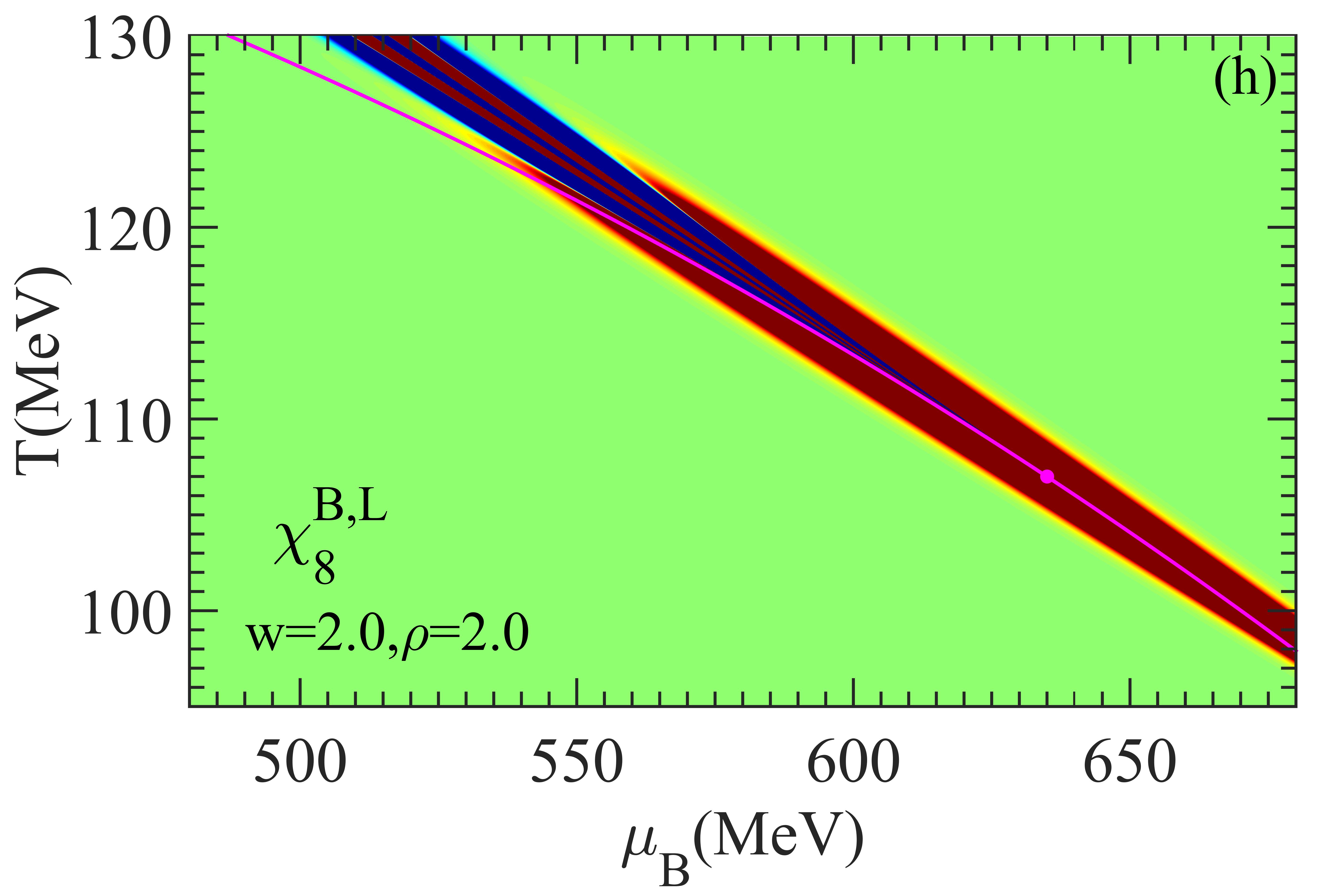}
	\includegraphics[width=0.32\textwidth]{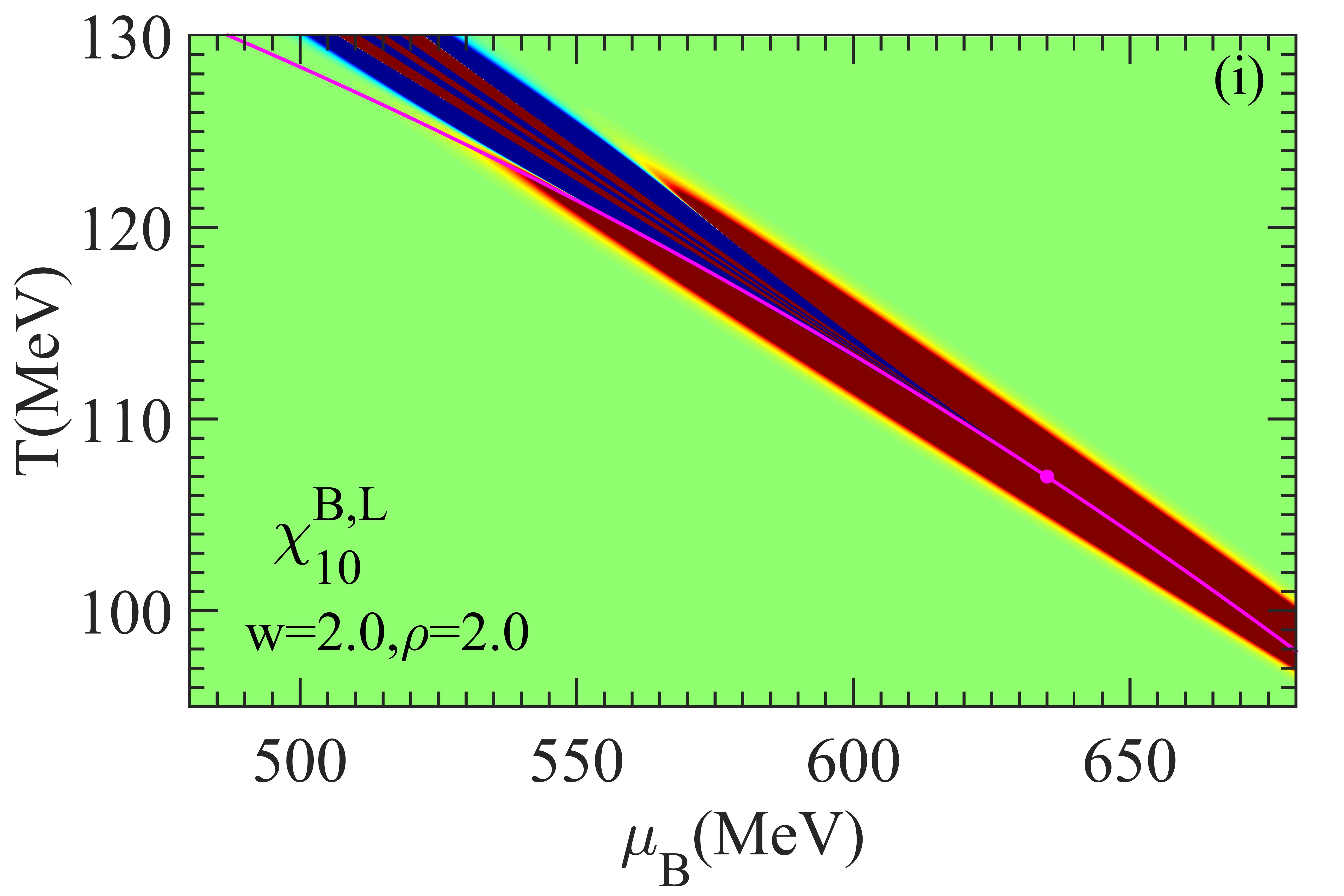}
	\caption{\label{Fig. 2}(Color online). Density plots of critical contribution to $\chi_{6}^{B,L}$, $\chi_{8}^{B,L}$, and $\chi_{10}^{B,L}$ in the QCD $T-\mu_B$ phase plane with $w=0.5, \rho=0.5$ (top row), $w=1.0, \rho=1.0$ (middle row) and $w=2.0, \rho=2.0$ (bottom row) at $\alpha_2=1.8^{\circ}$. The critical point is indicated by a purple dot, while the chiral phase transition line is represented by the solid purple line. The green, yellow and red areas correspond to positive values (the regions where it is the largest and smallest are indicated in red and green, respectively) of the susceptibilities, while the blue ones correspond to negative values (the darker the blue, the larger in magnitude of the susceptibilities).}
\end{figure*}

\begin{figure*}[hbt]
	\centering
	\includegraphics[width=0.32\textwidth]{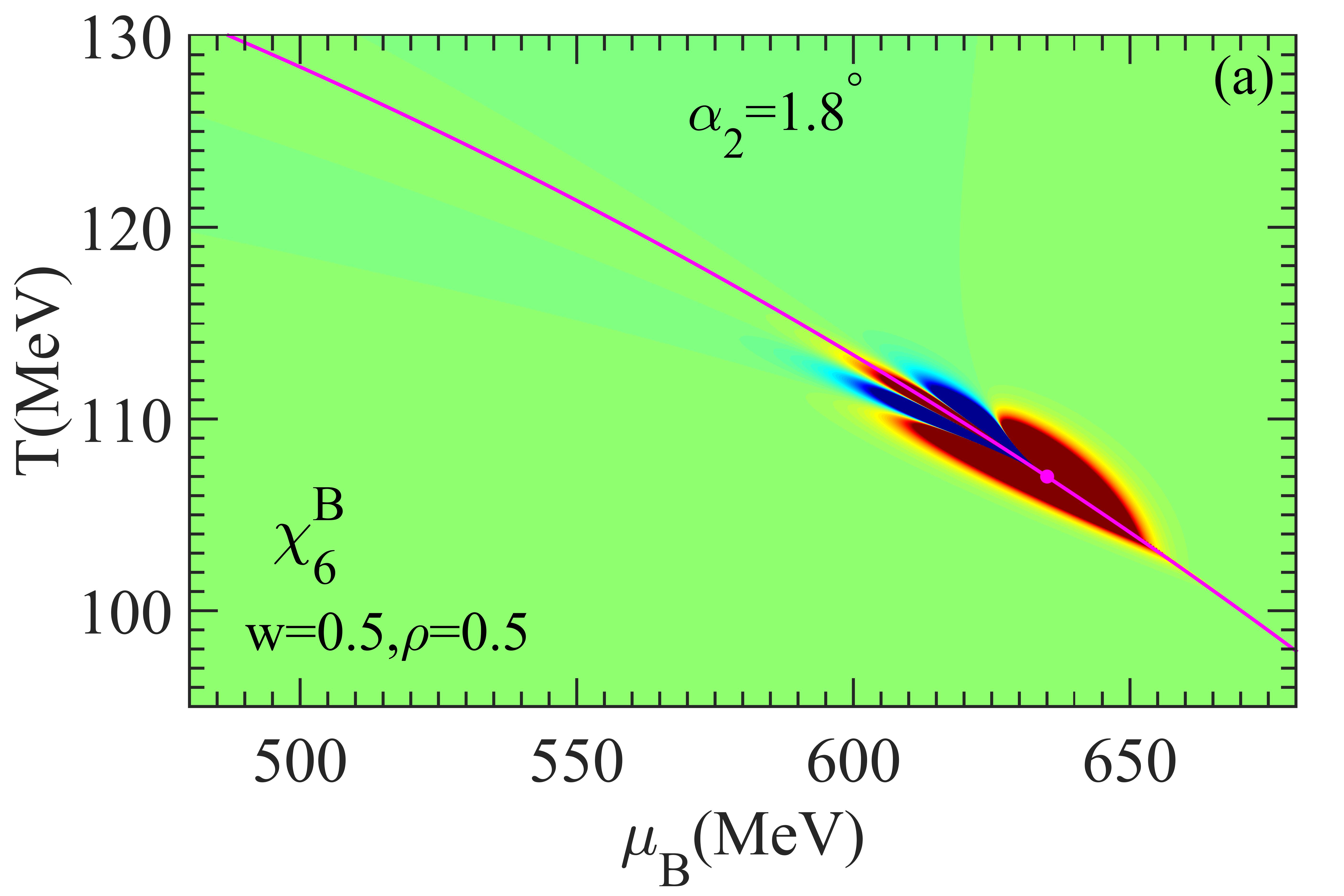}
	\includegraphics[width=0.32\textwidth]{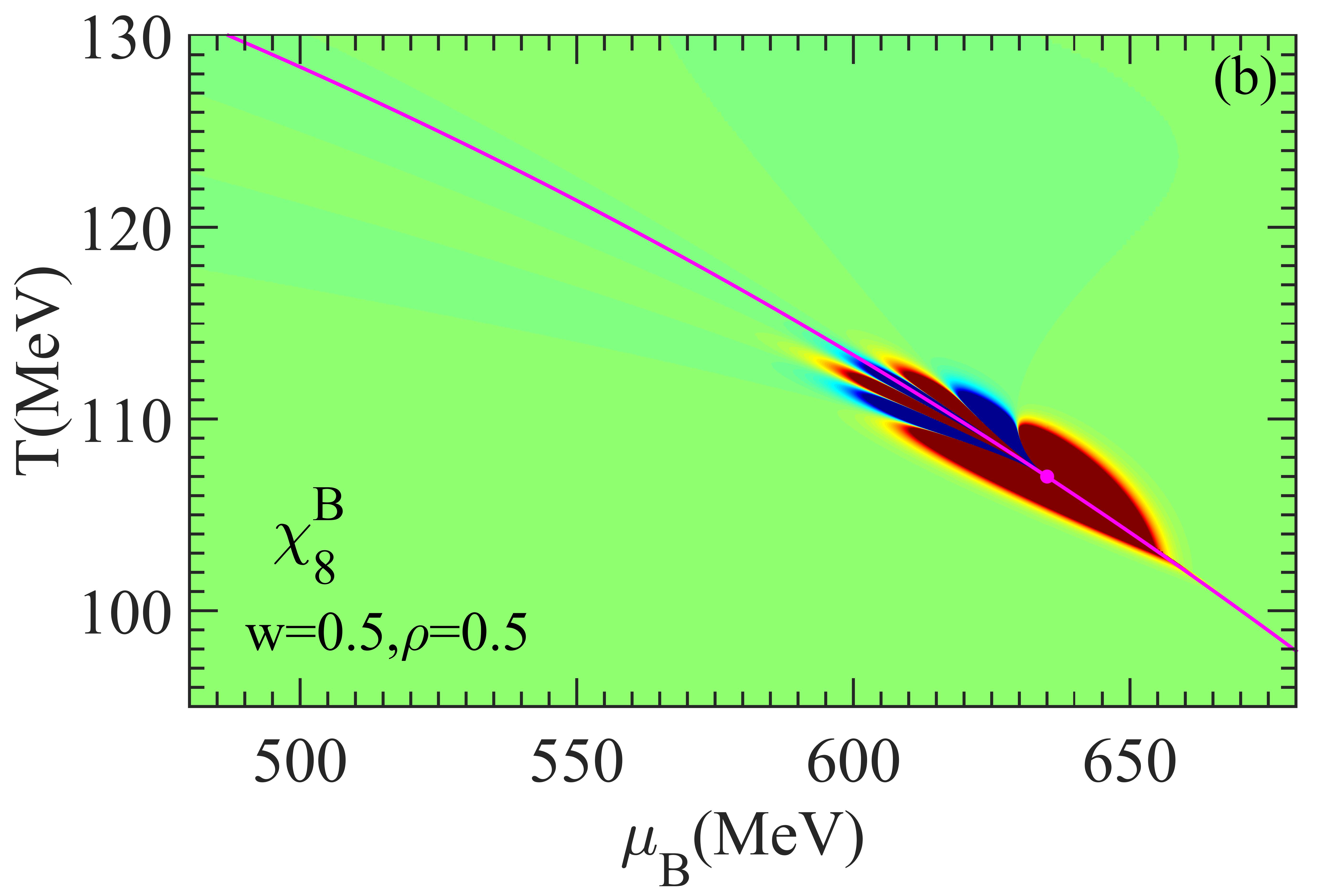}
	\includegraphics[width=0.32\textwidth]{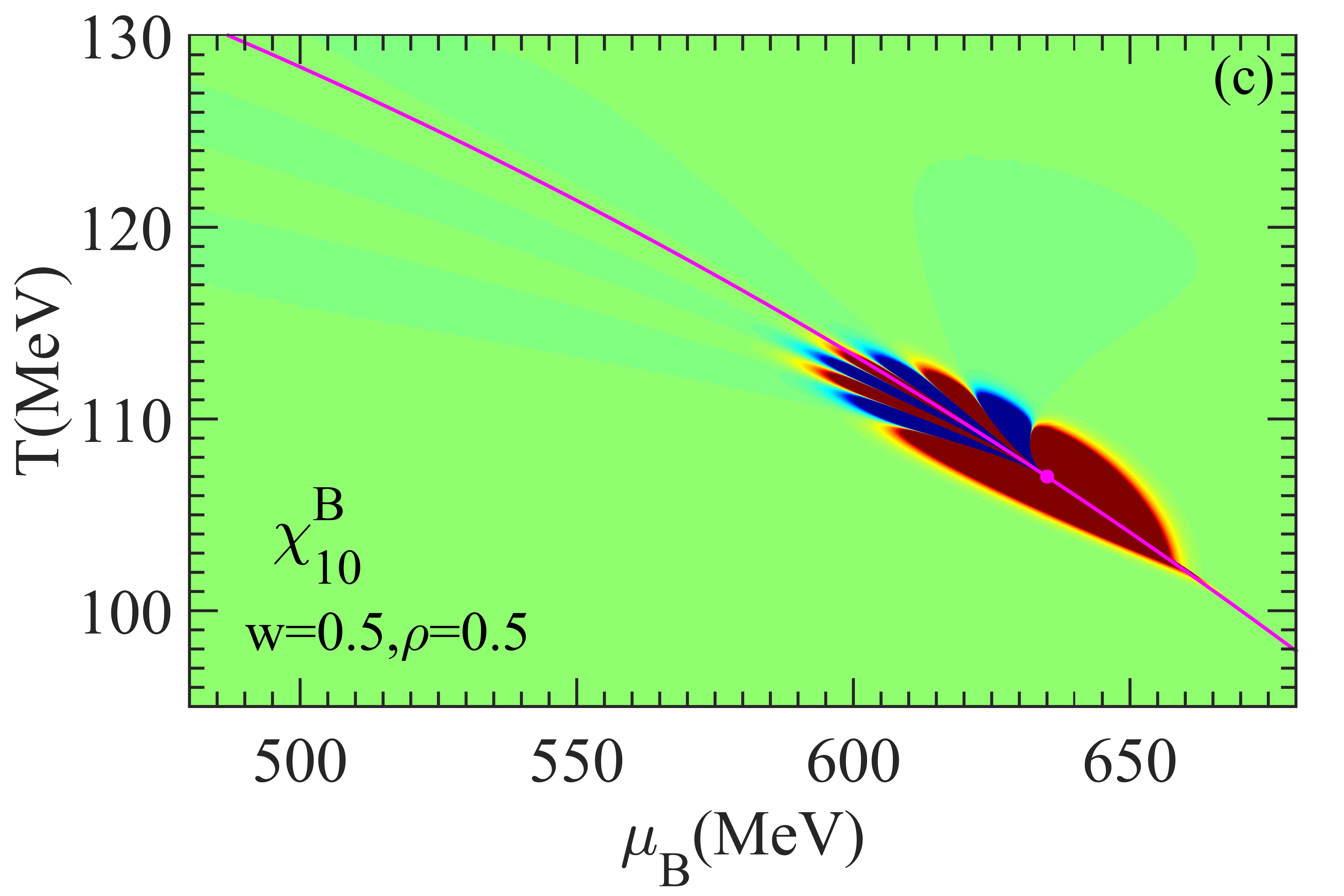}
	\includegraphics[width=0.32\textwidth]{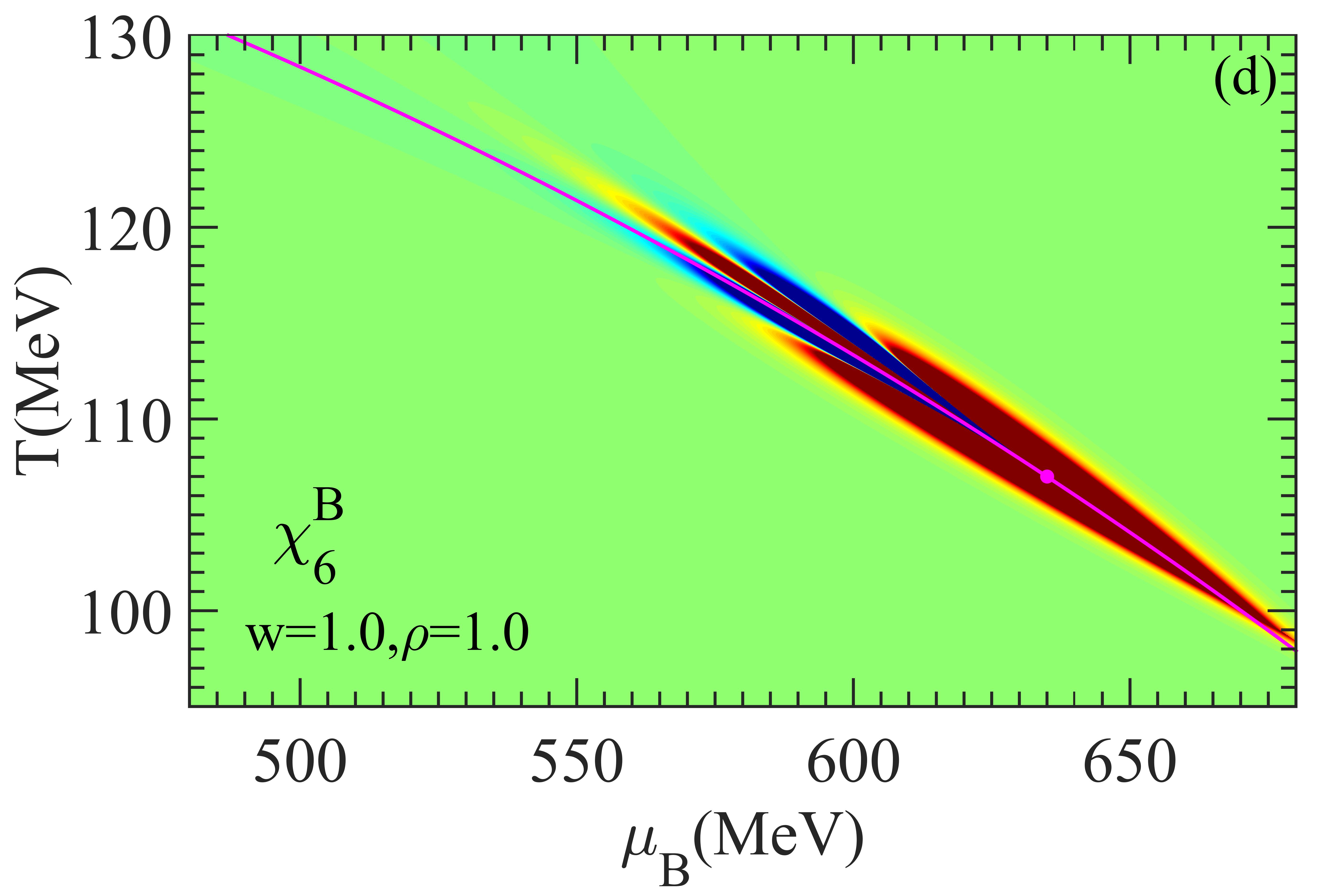}
	\includegraphics[width=0.32\textwidth]{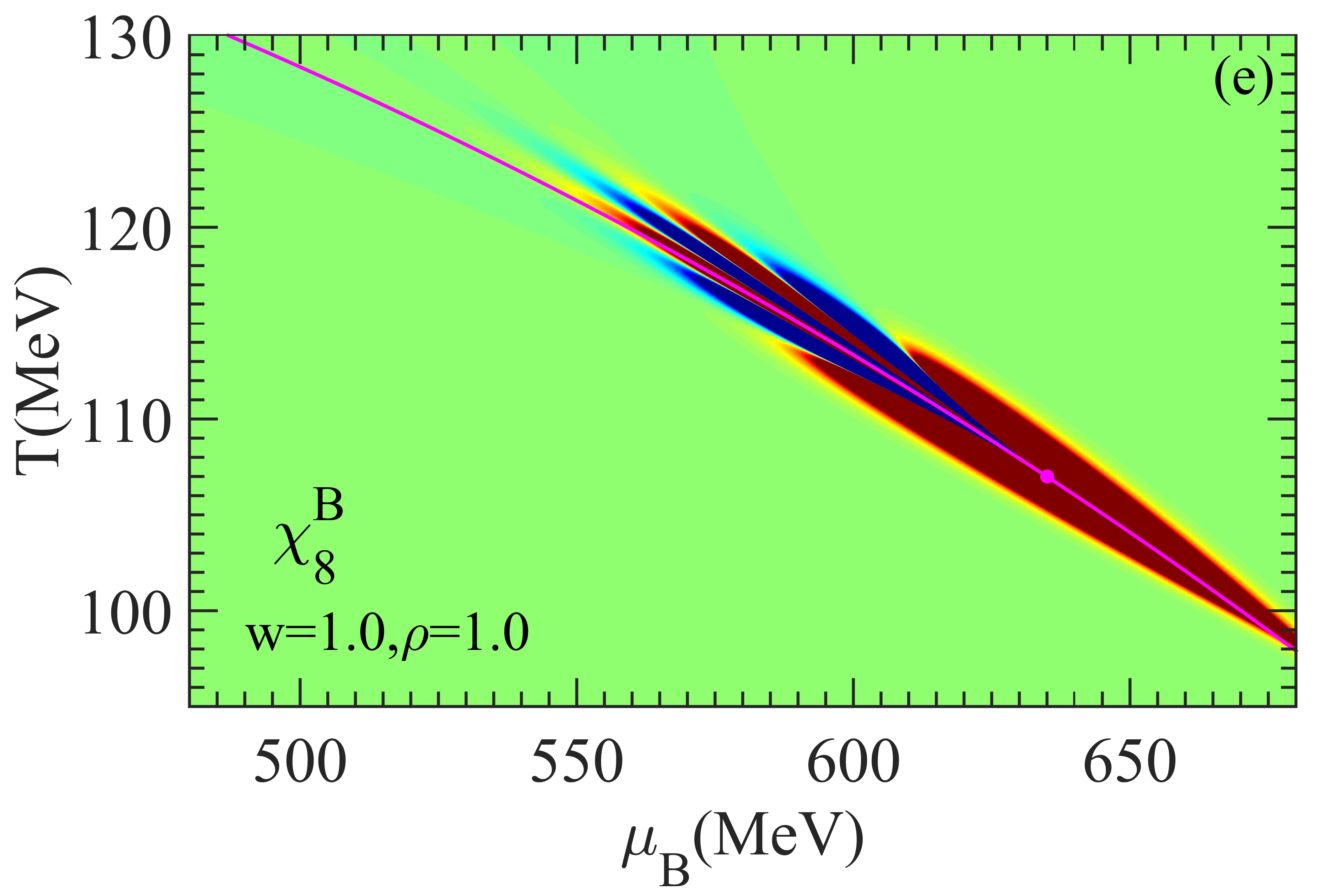}
	\includegraphics[width=0.32\textwidth]{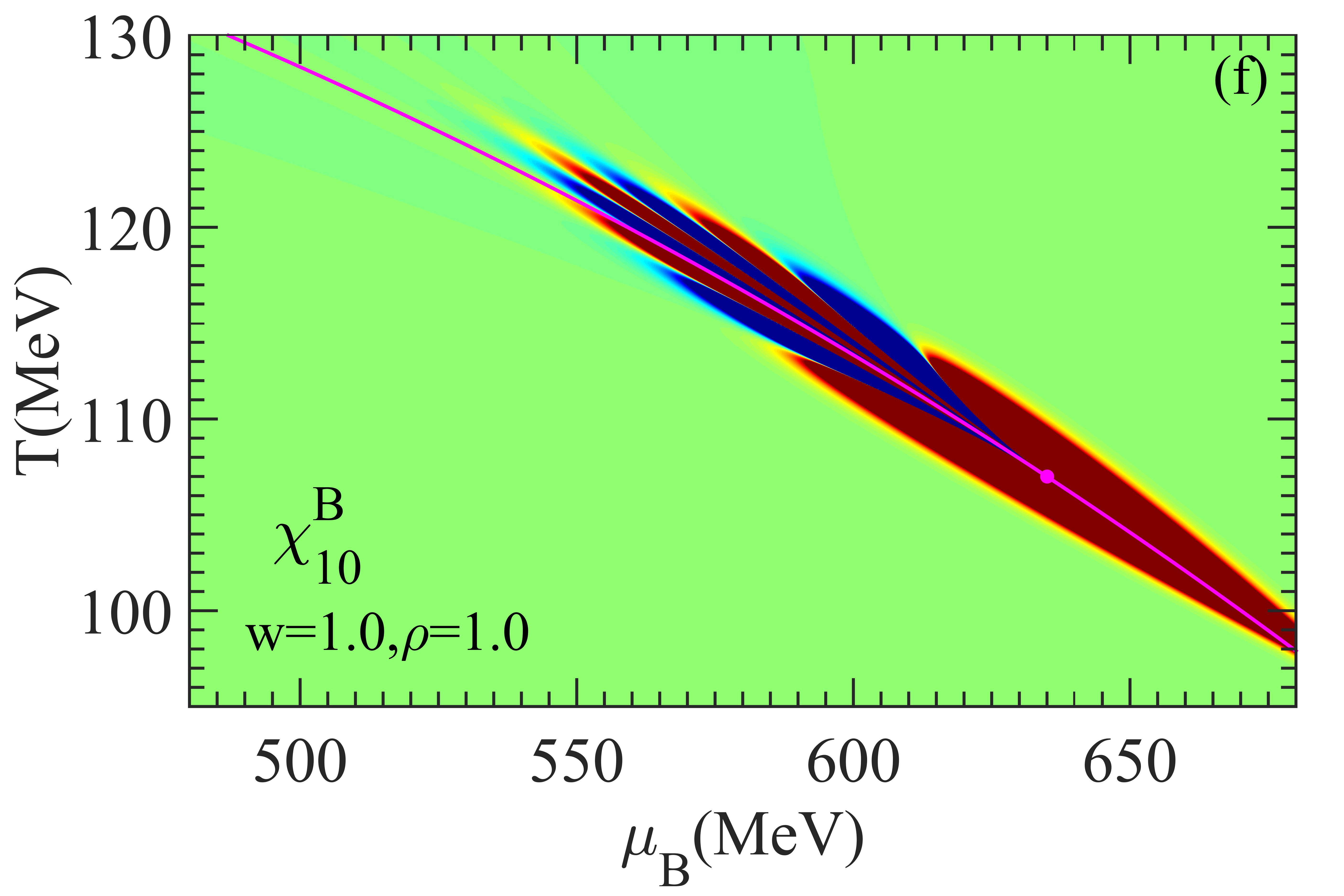}
	\includegraphics[width=0.32\textwidth]{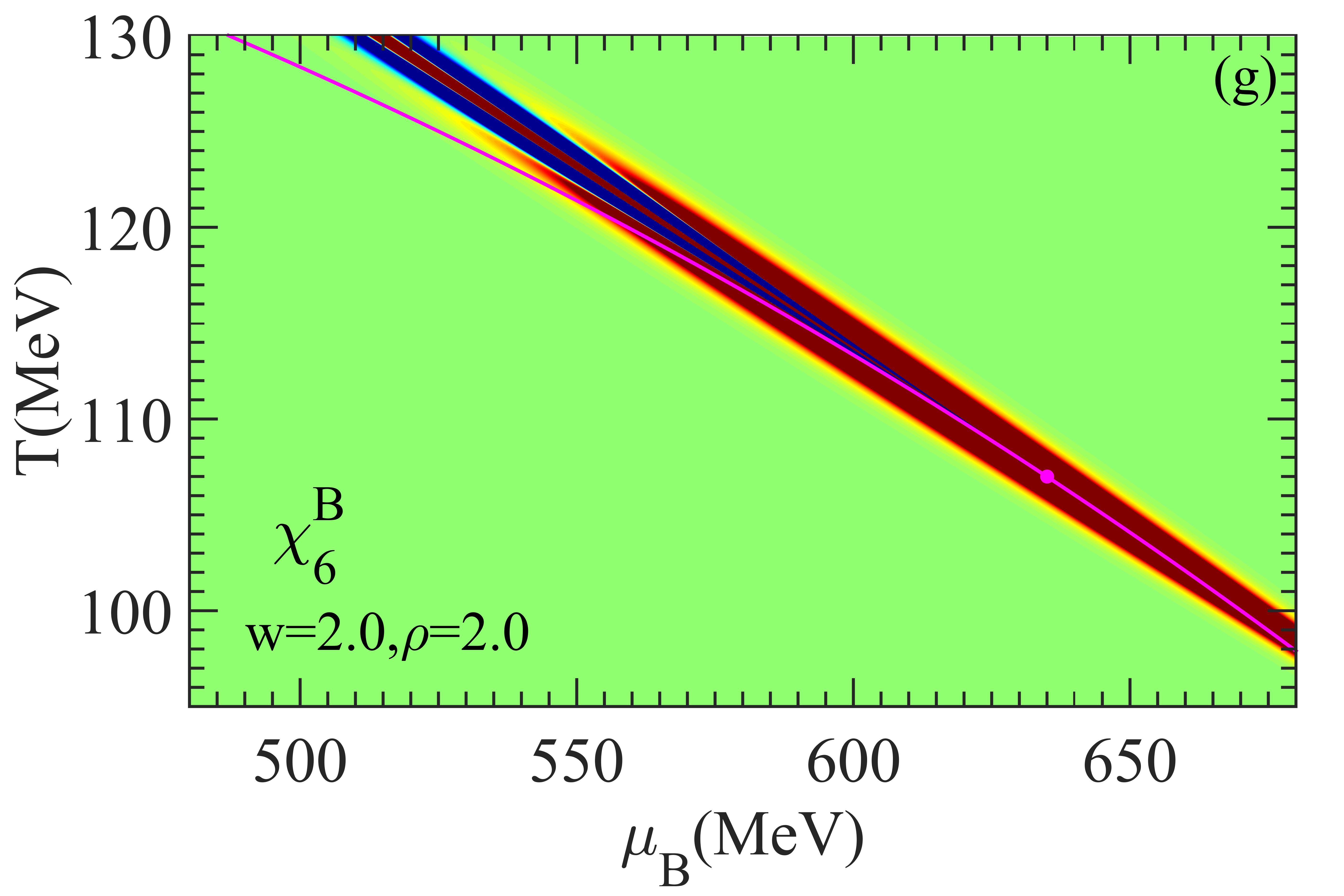}
	\includegraphics[width=0.32\textwidth]{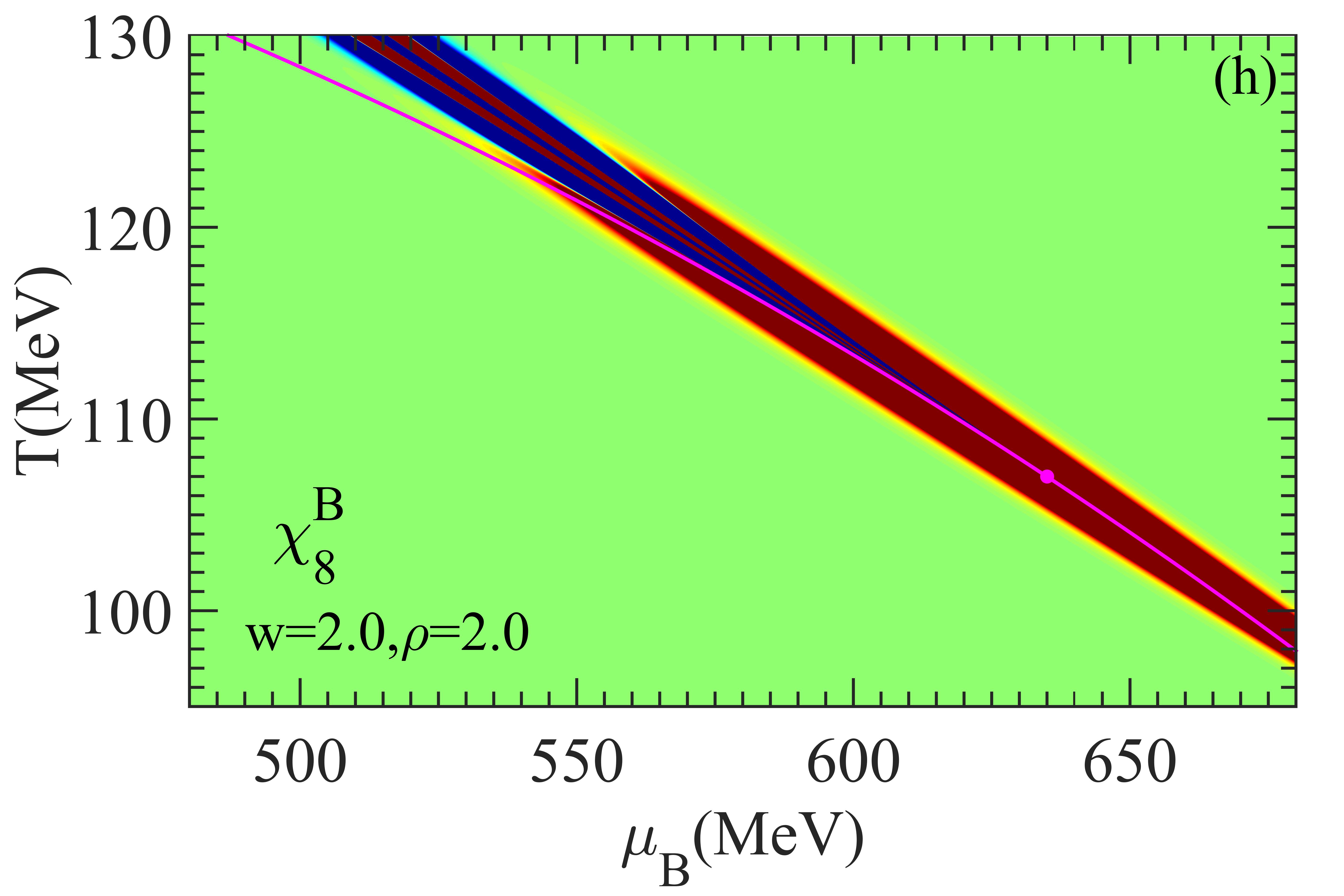}
	\includegraphics[width=0.32\textwidth]{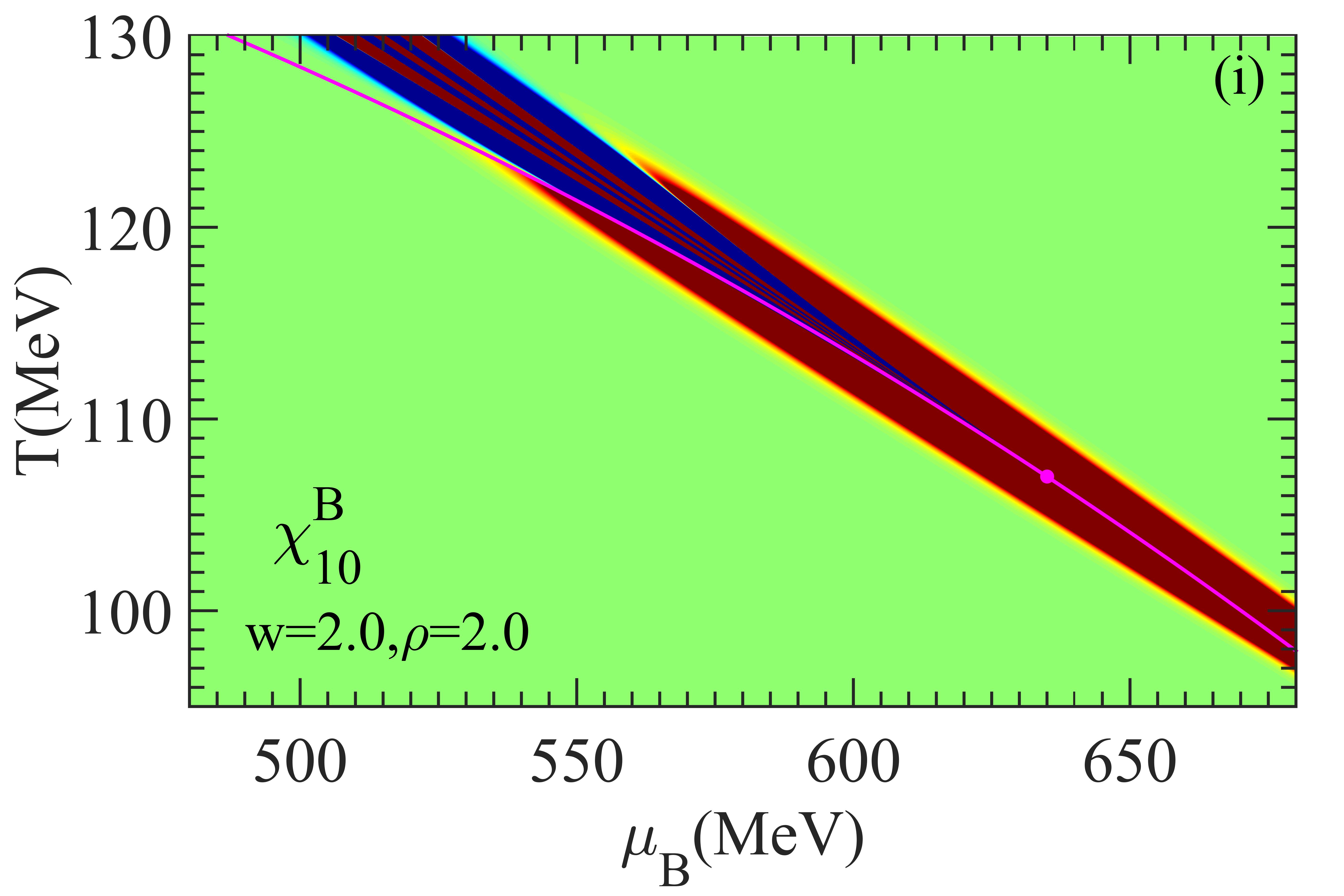}
	\caption{\label{Fig. 3}(Color online). Density plots of critical contribution to $\chi_{6}^{B}$, $\chi_{8}^{B}$, and $\chi_{10}^{B}$ in the QCD $T-\mu_B$ phase plane with $w=0.5, \rho=0.5$ (top row), $w=1.0, \rho=1.0$ (middle row) and $w=2.0, \rho=2.0$ (bottom row) at $\alpha_2=1.8^{\circ}$. The critical point is indicated by a purple dot, while the chiral phase transition line is represented by the solid purple line.}
\end{figure*}

\begin{figure*}[hbt]
	\centering
	\includegraphics[width=0.32\textwidth]{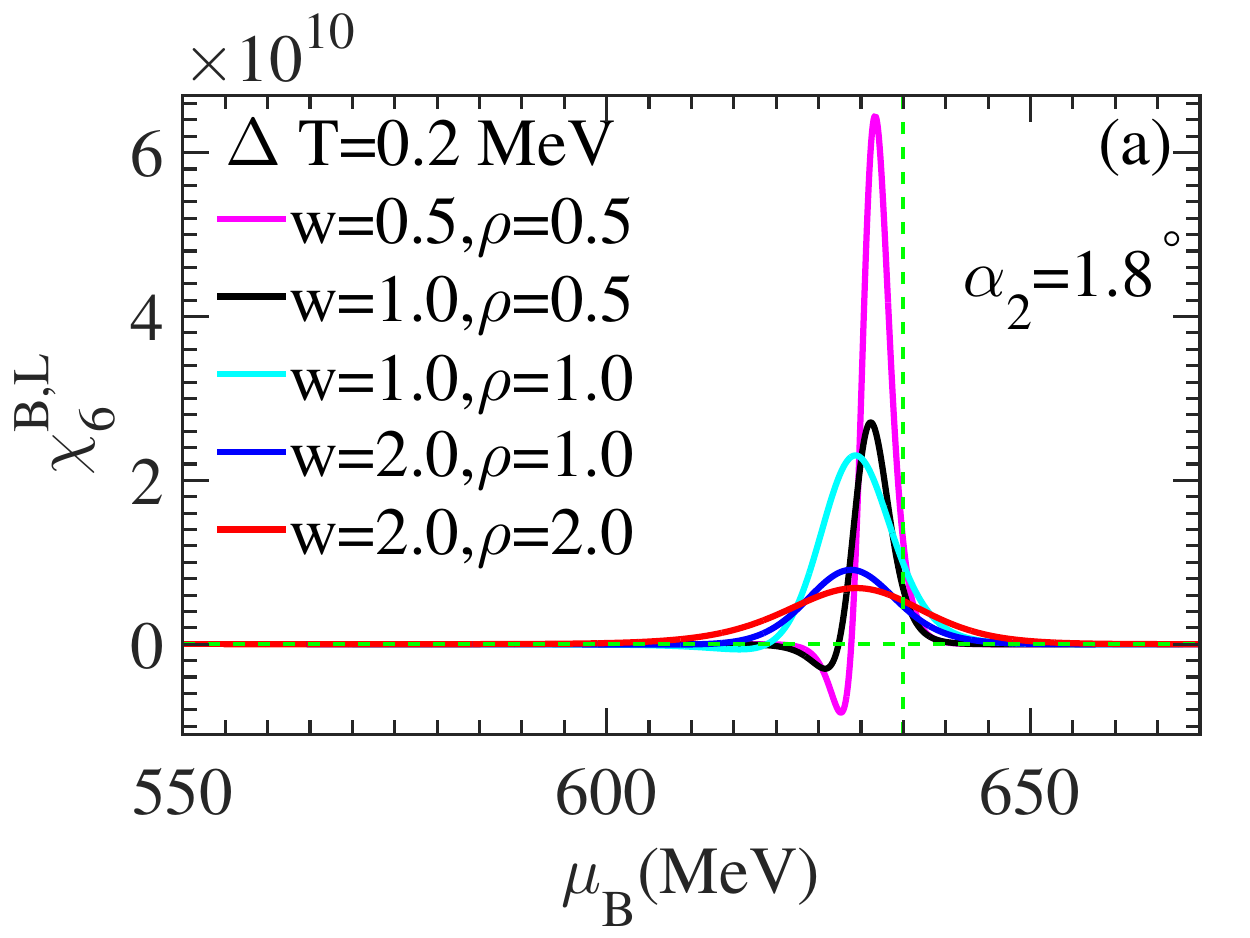}
	\includegraphics[width=0.32\textwidth]{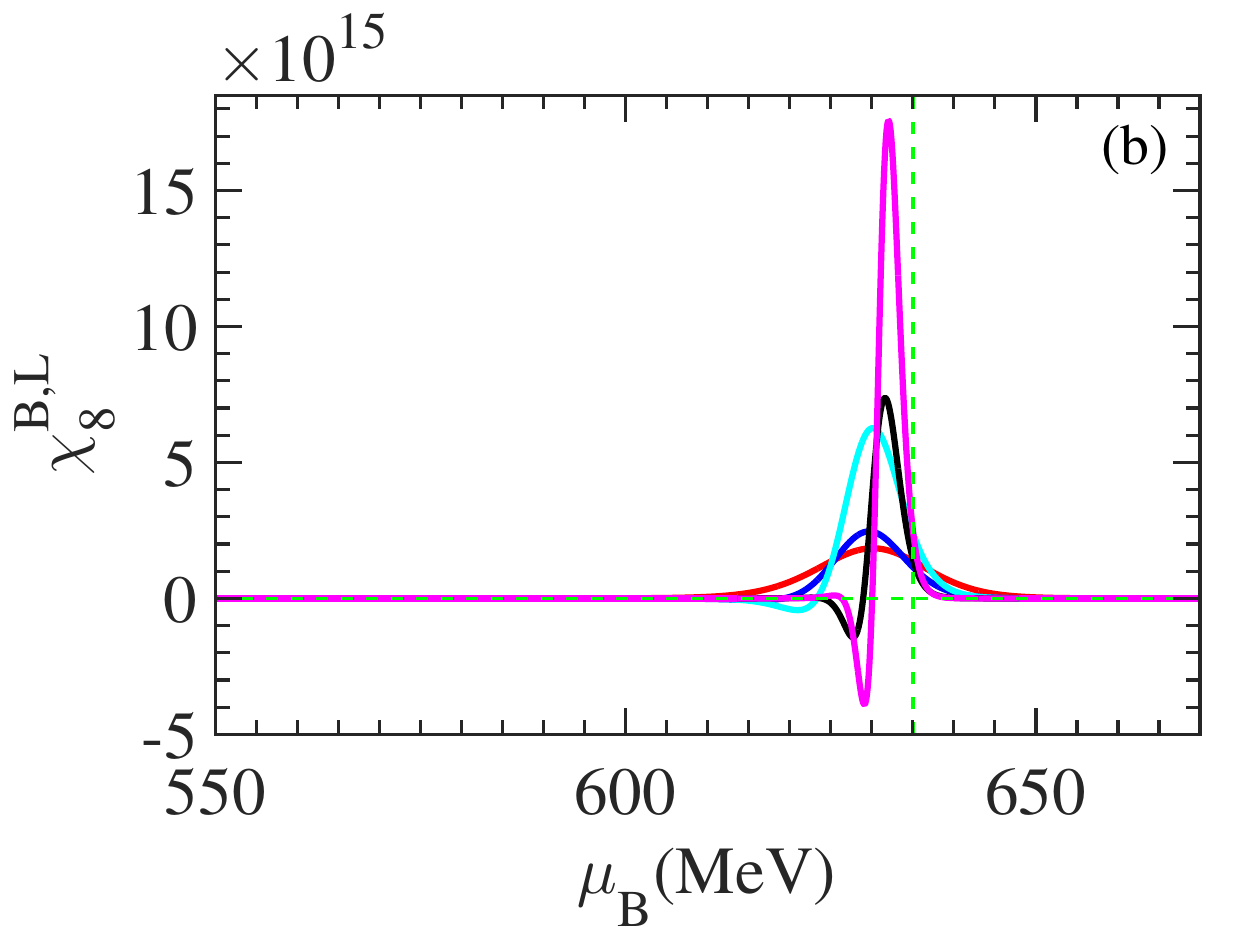}
	\includegraphics[width=0.32\textwidth]{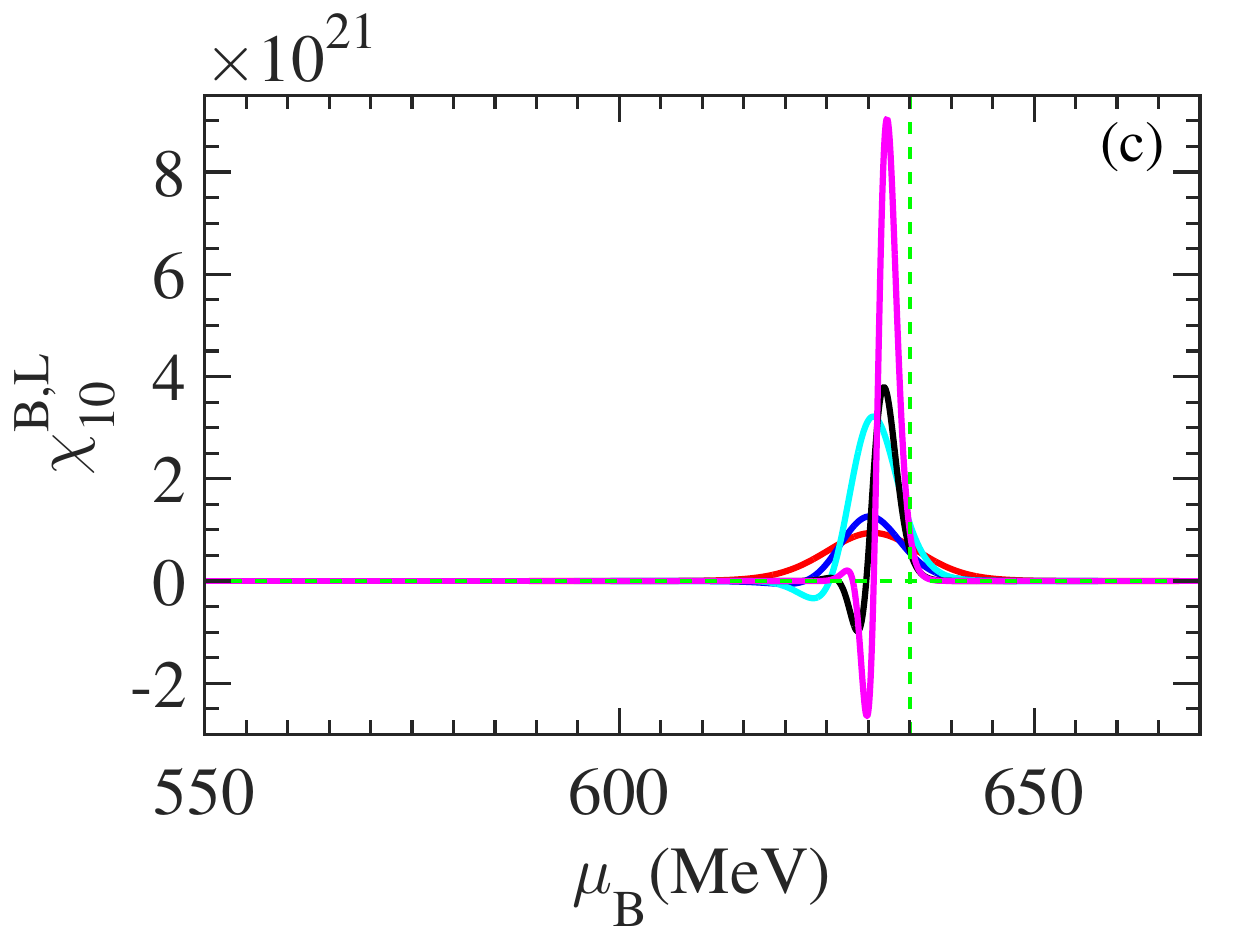}
	\includegraphics[width=0.32\textwidth]{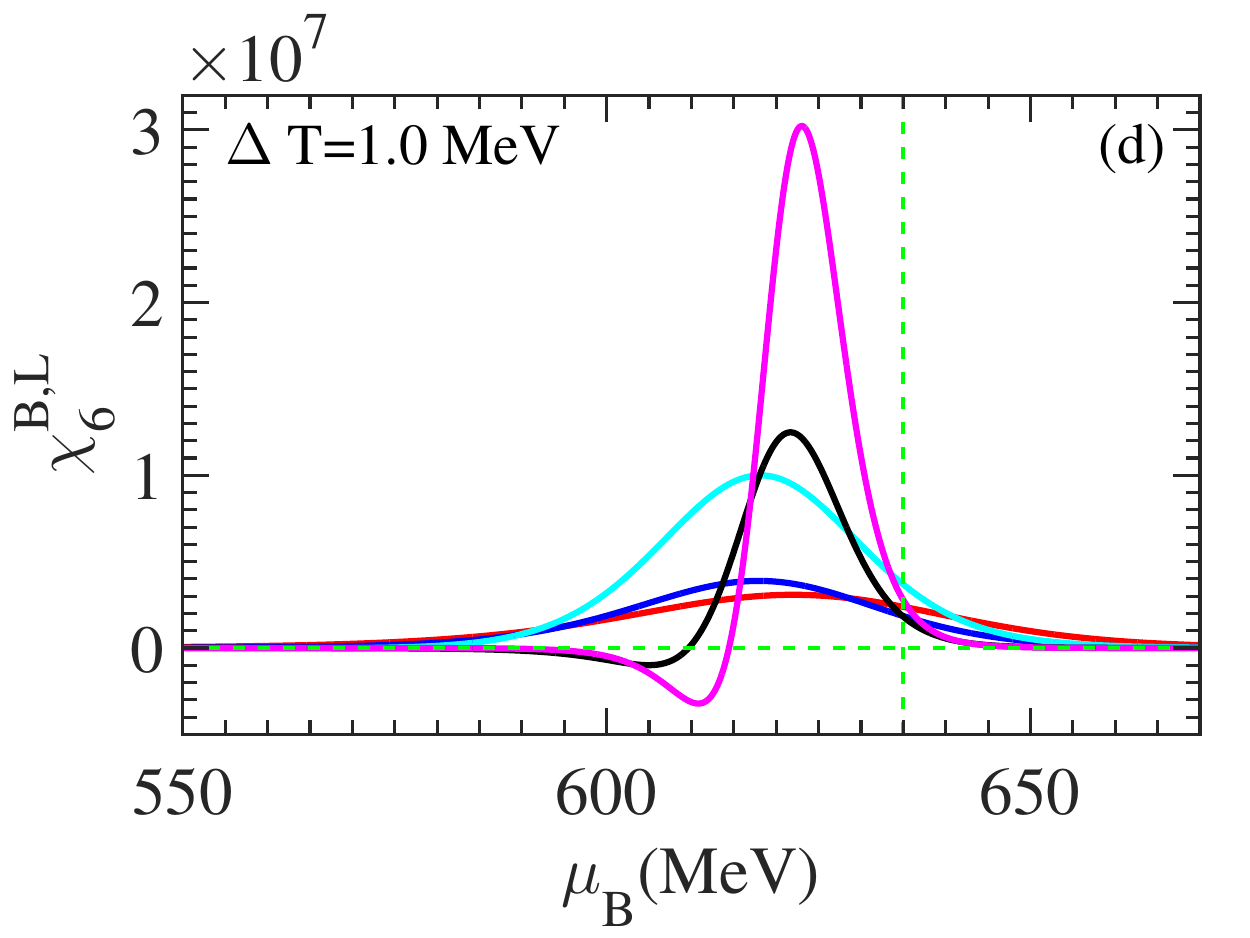}
	\includegraphics[width=0.32\textwidth]{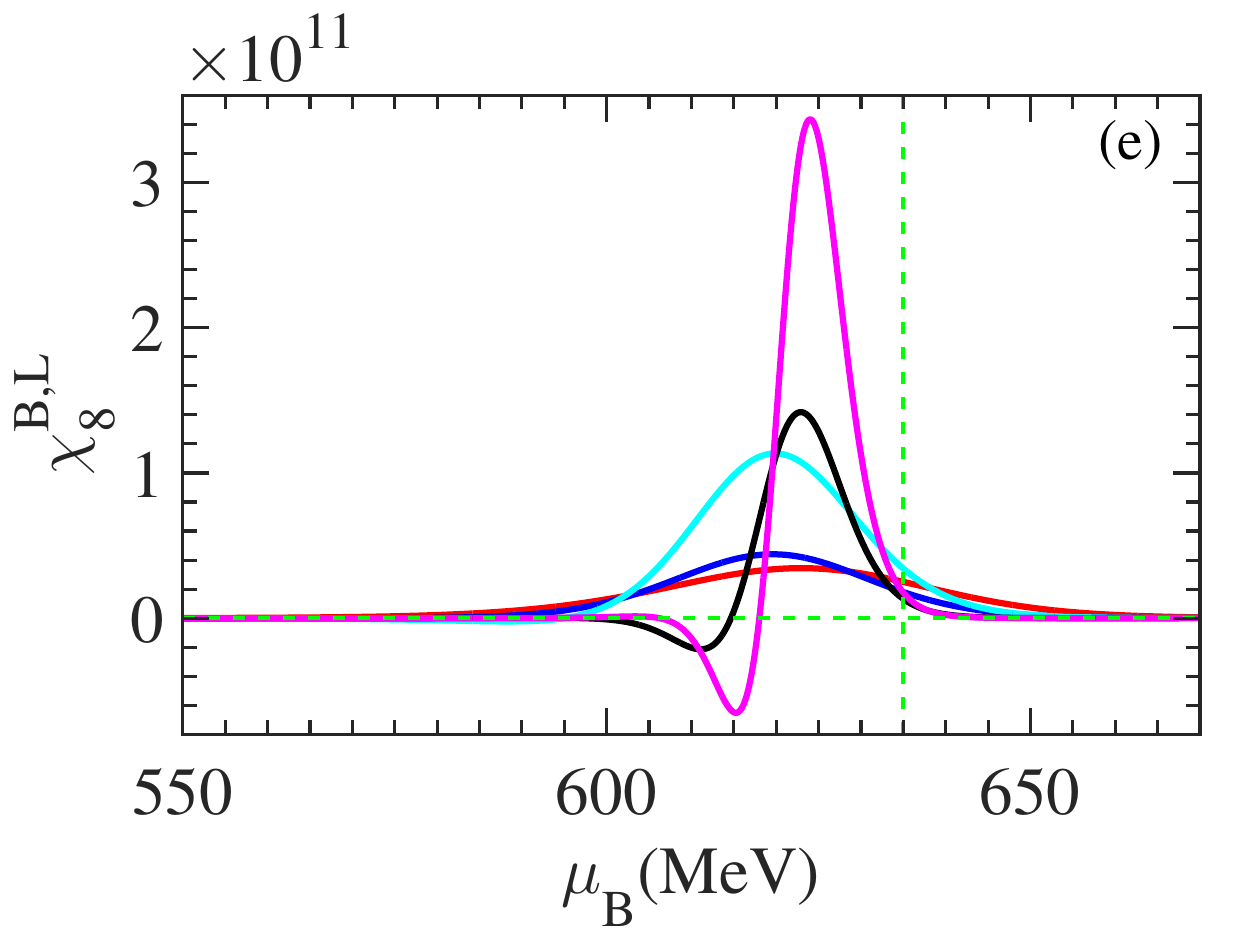}
	\includegraphics[width=0.32\textwidth]{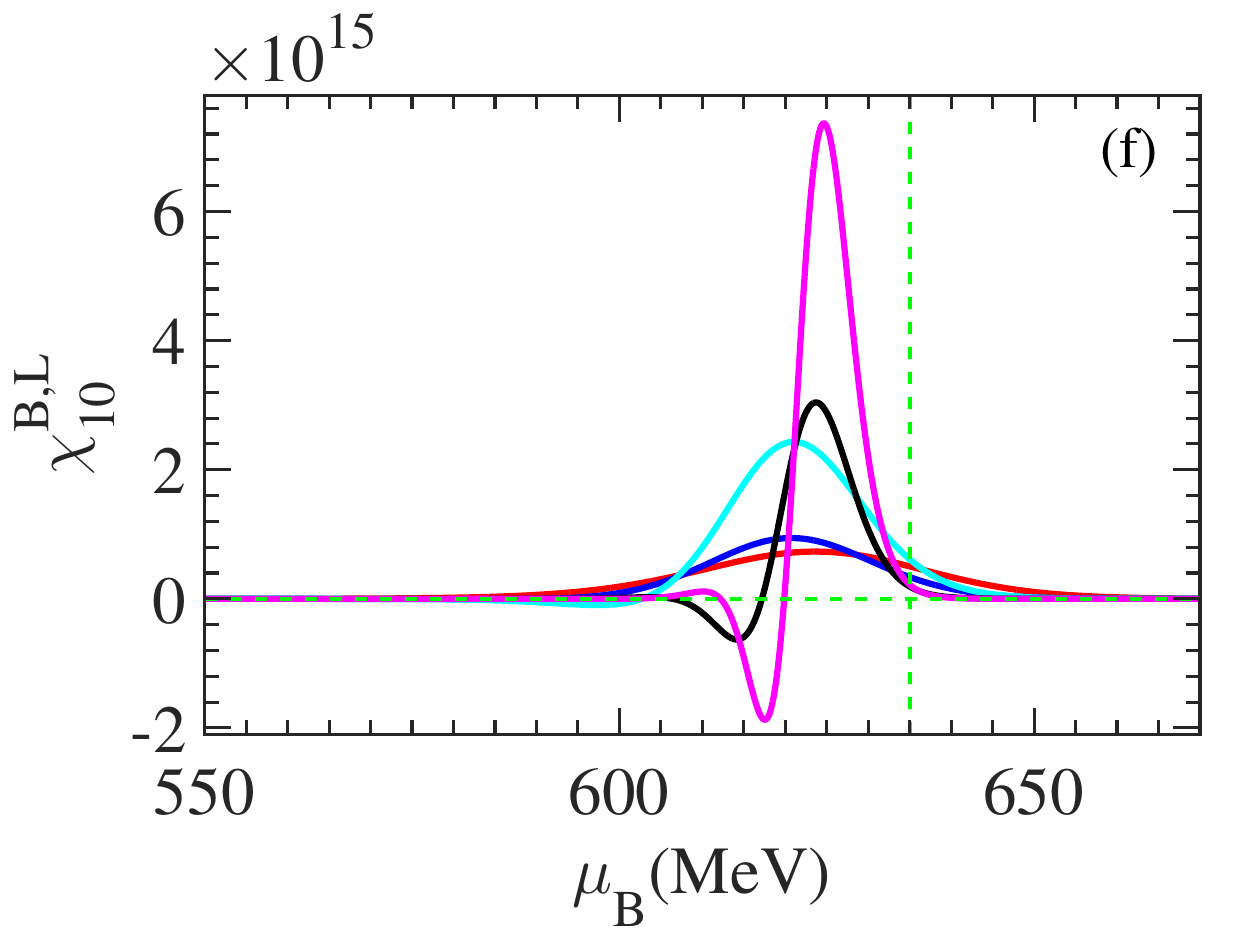}
	\includegraphics[width=0.32\textwidth]{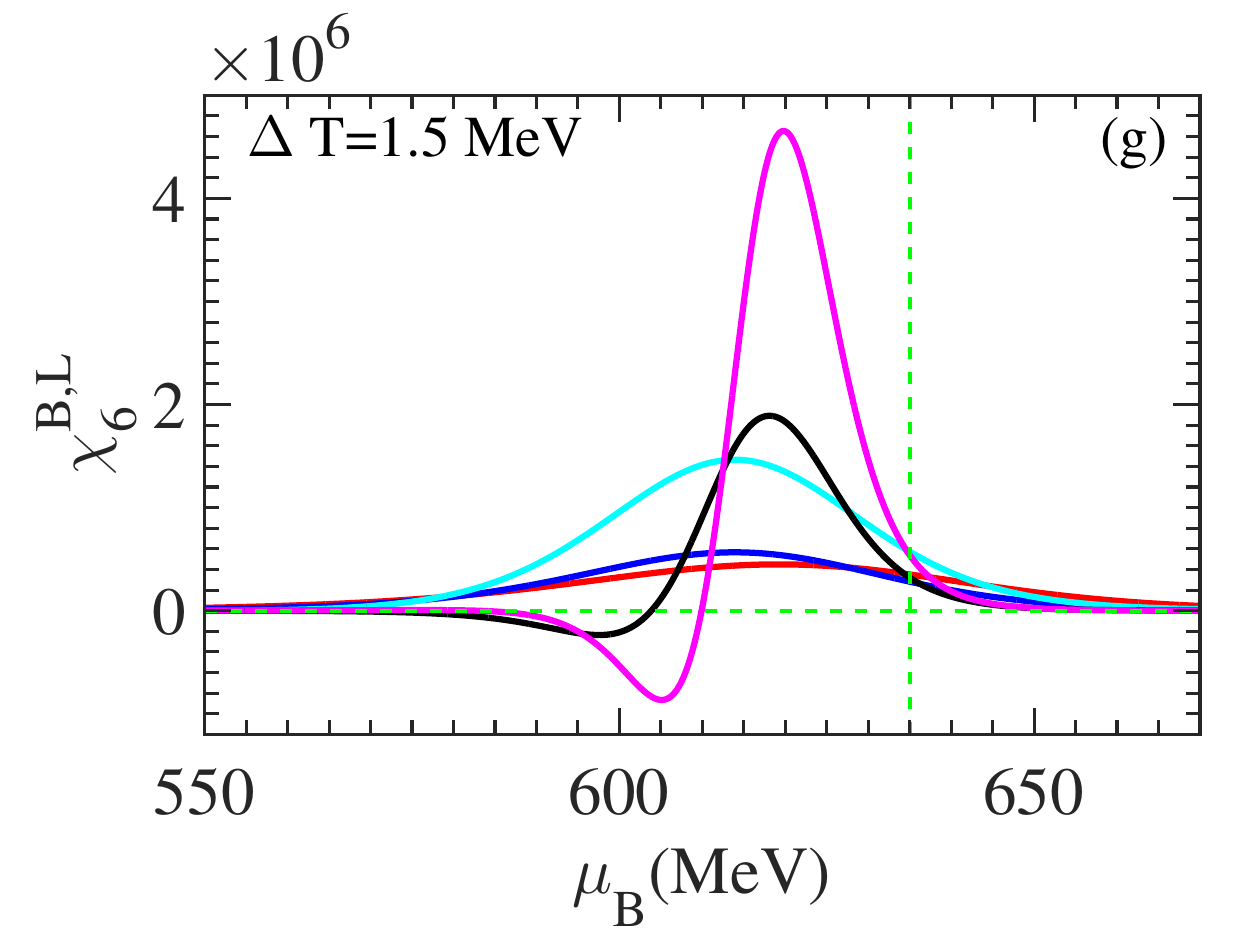}
	\includegraphics[width=0.32\textwidth]{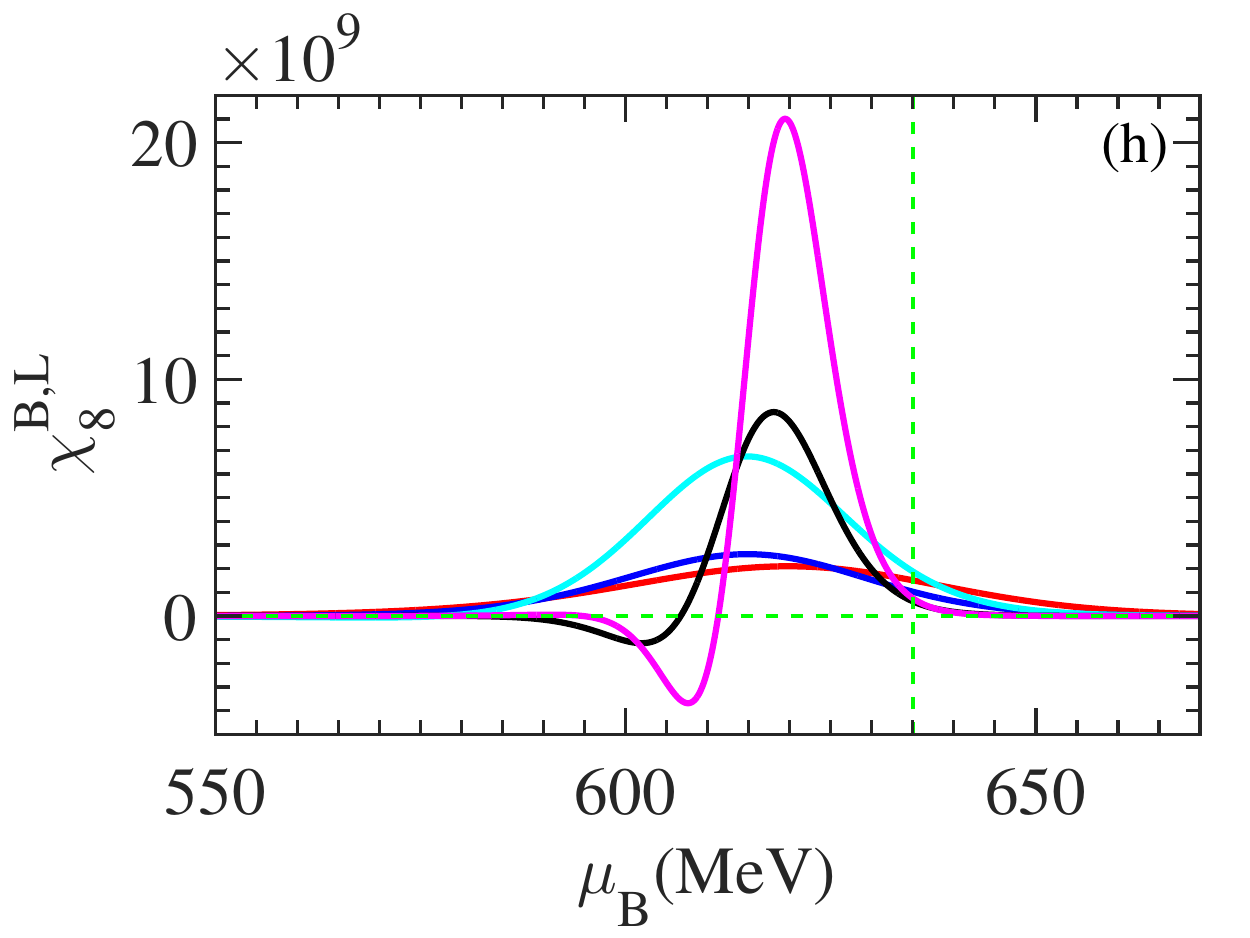}
	\includegraphics[width=0.32\textwidth]{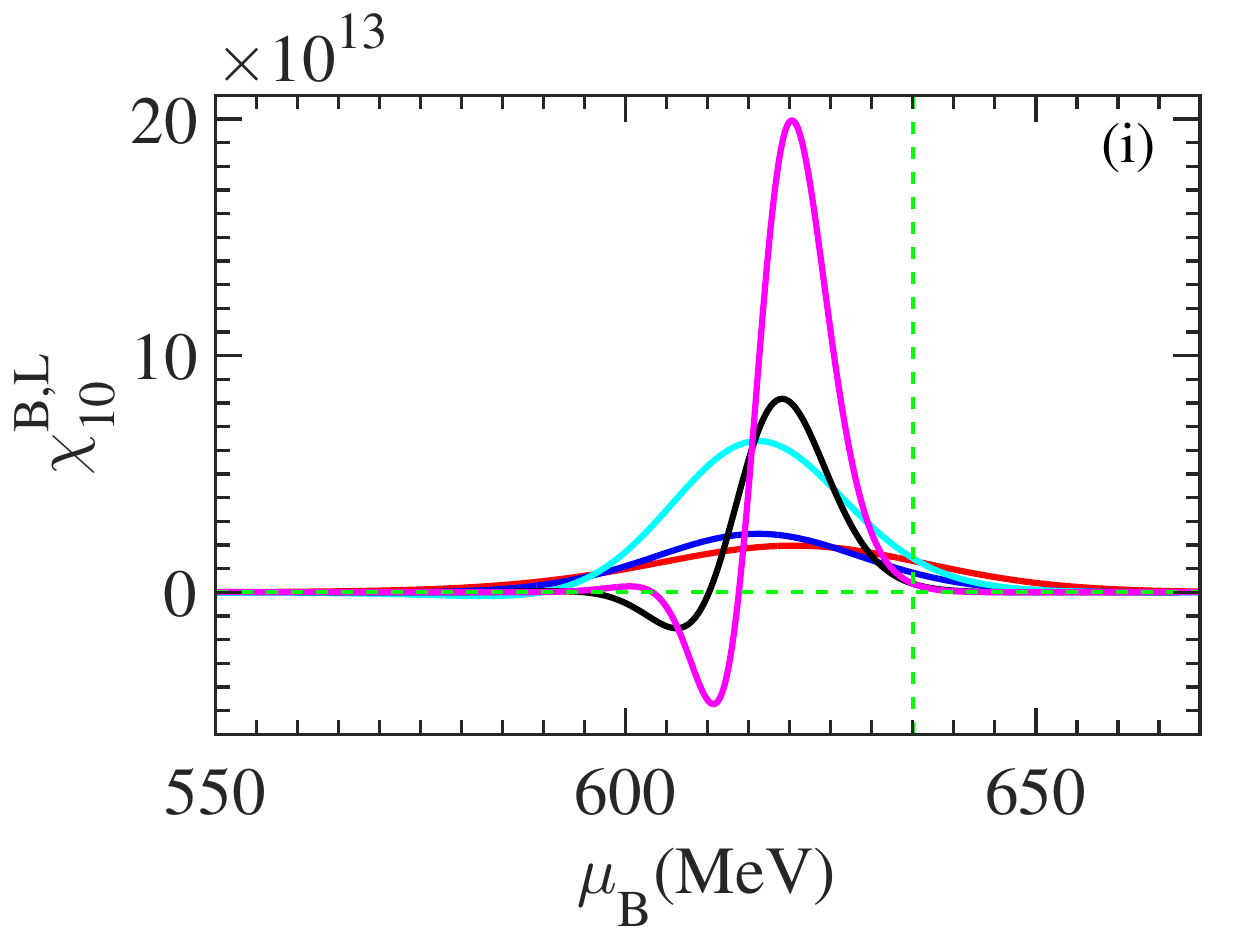}
	\caption{\label{Fig. 4}(Color online). $\mu_B$ dependence of $\chi_{6}^{B,L}$, $\chi_{8}^{B,L}$, and $\chi_{10}^{B,L}$ along the freeze-out curves at $\Delta T=0.2$ MeV (top row), $\Delta T=1.0$ MeV (middle row) and $\Delta T= 1.5$ MeV (bottom row) with different values of $w$ and $\rho$ where $\alpha_2=1.8^{\circ}$.}
\end{figure*}

\begin{figure*}[hbt]
	\centering
	\includegraphics[width=0.32\textwidth]{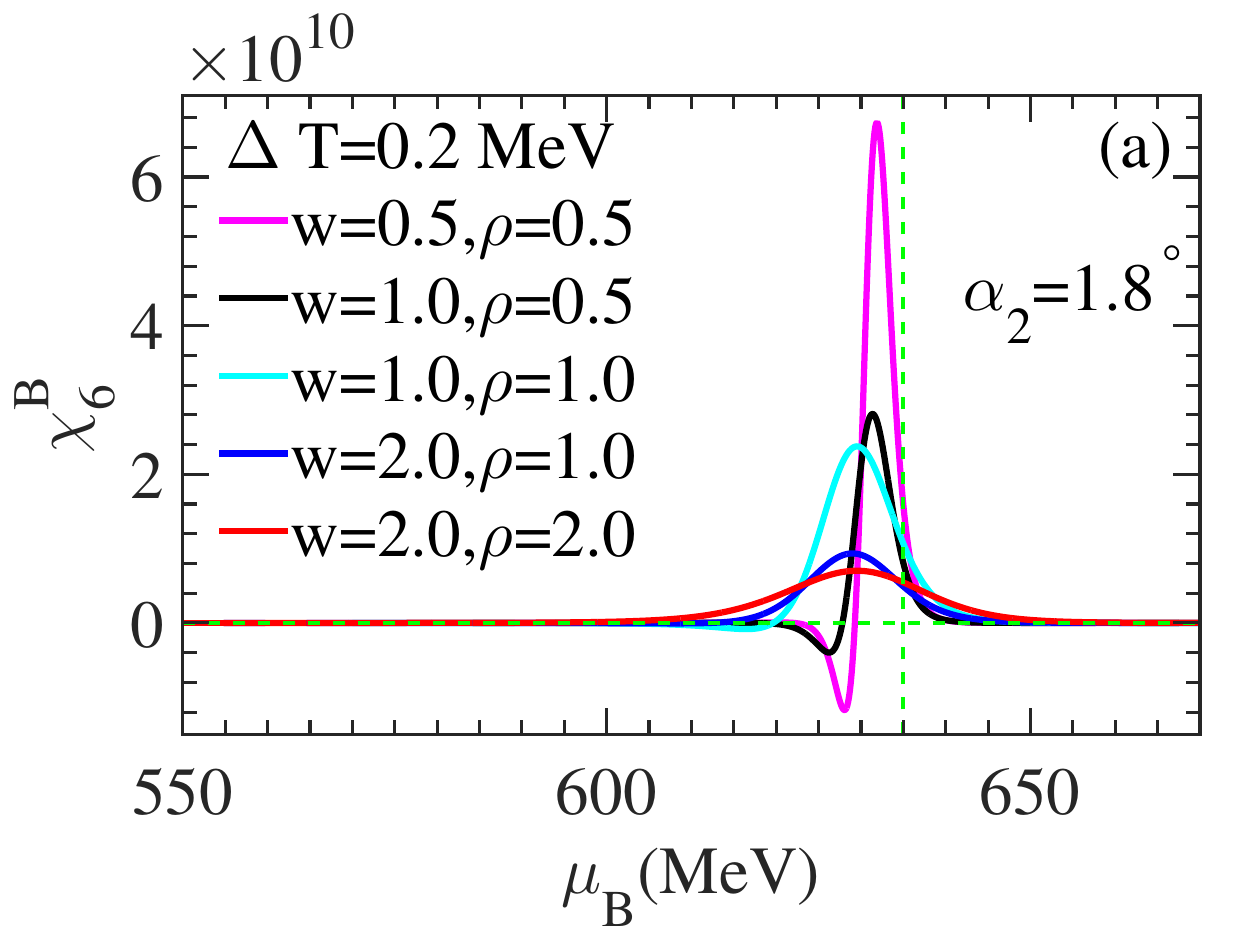}
	\includegraphics[width=0.32\textwidth]{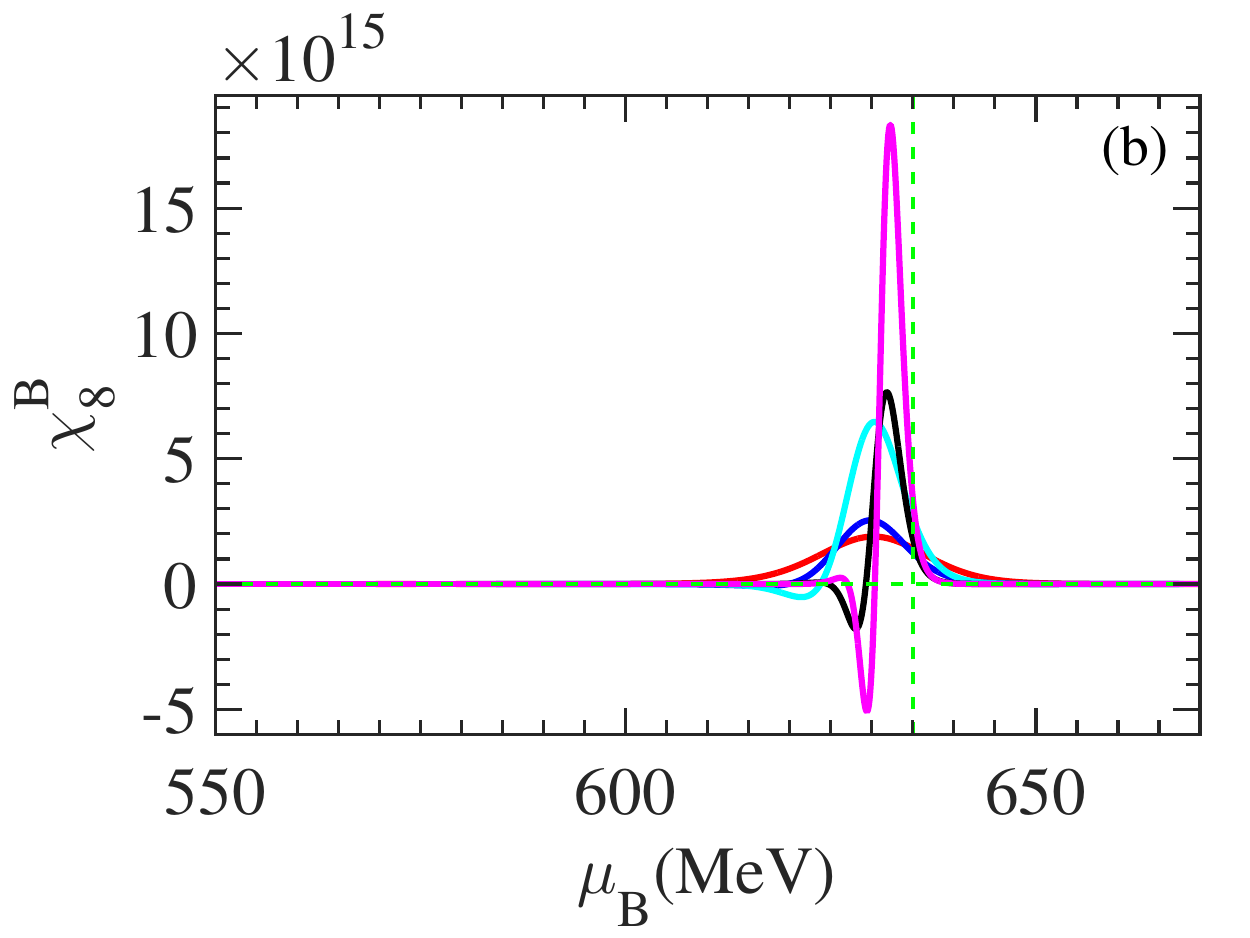}
	\includegraphics[width=0.32\textwidth]{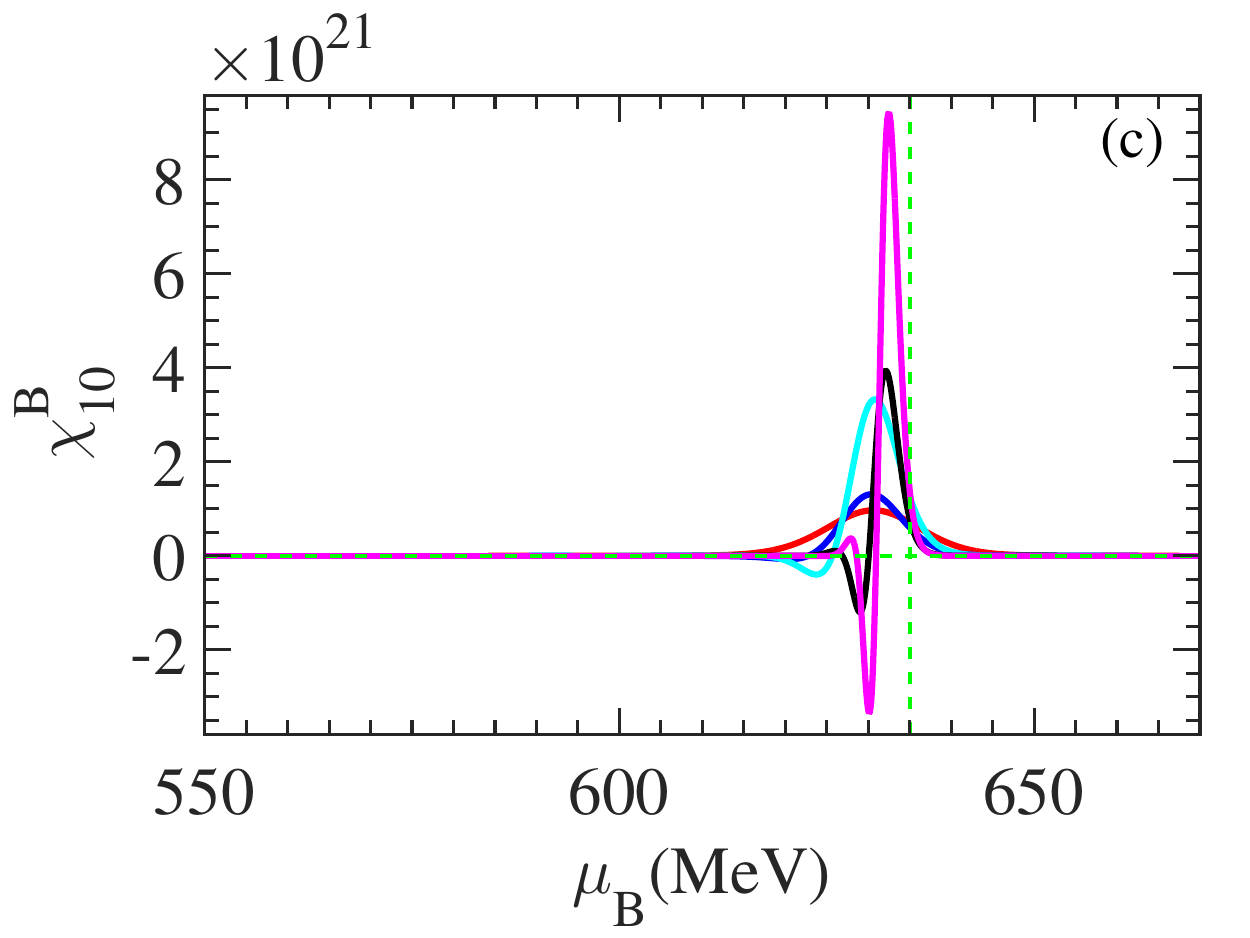}
	\includegraphics[width=0.32\textwidth]{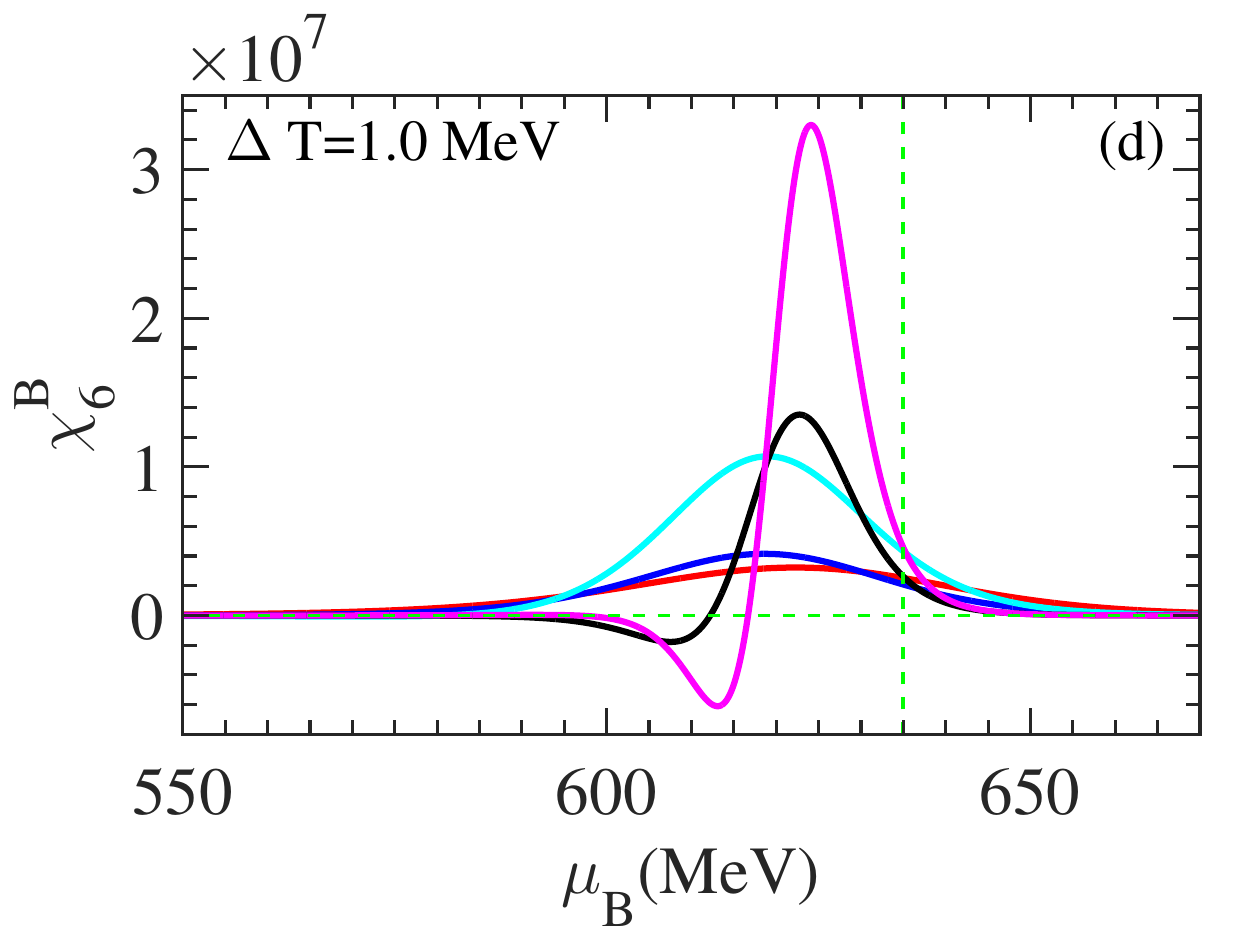}
	\includegraphics[width=0.32\textwidth]{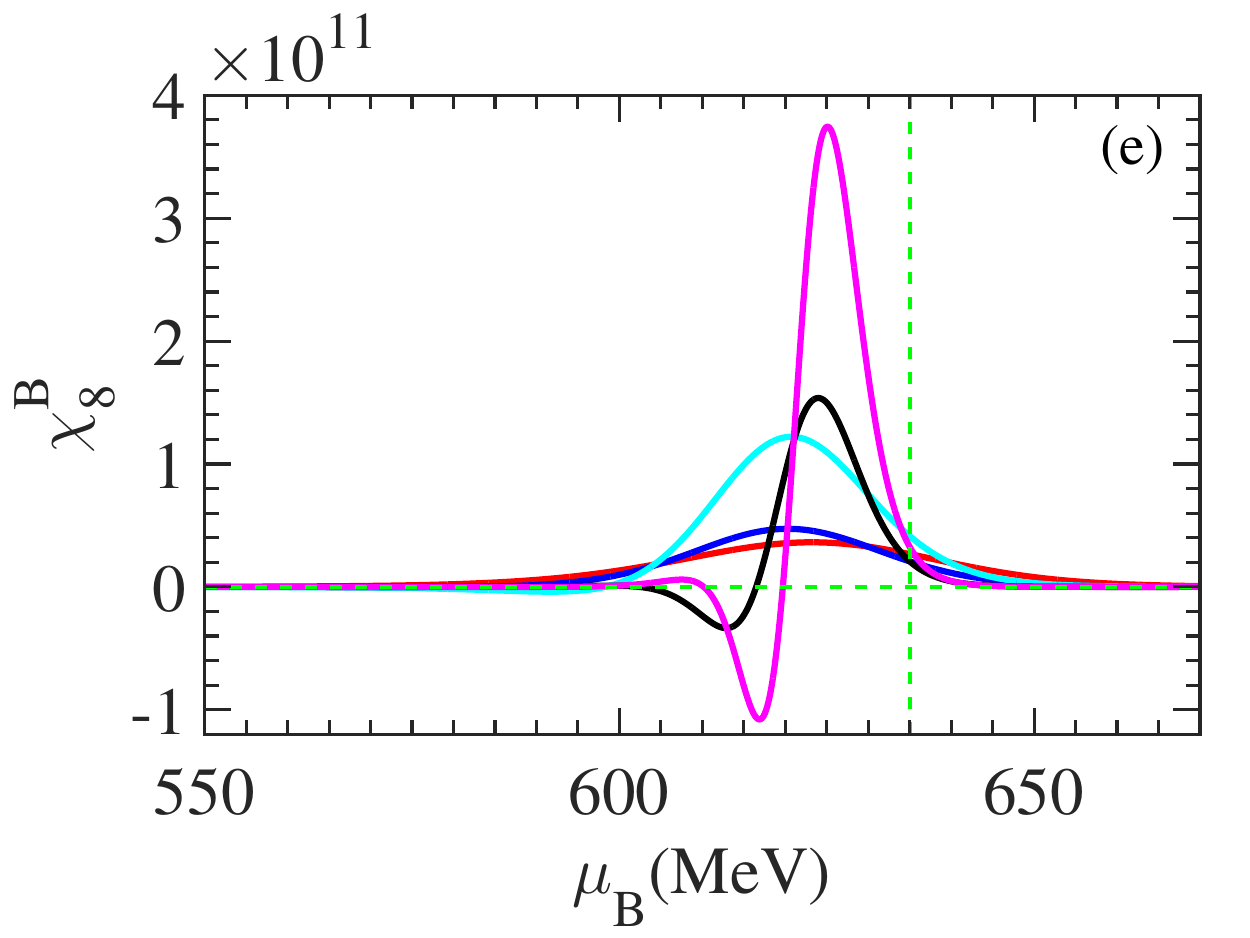}
	\includegraphics[width=0.32\textwidth]{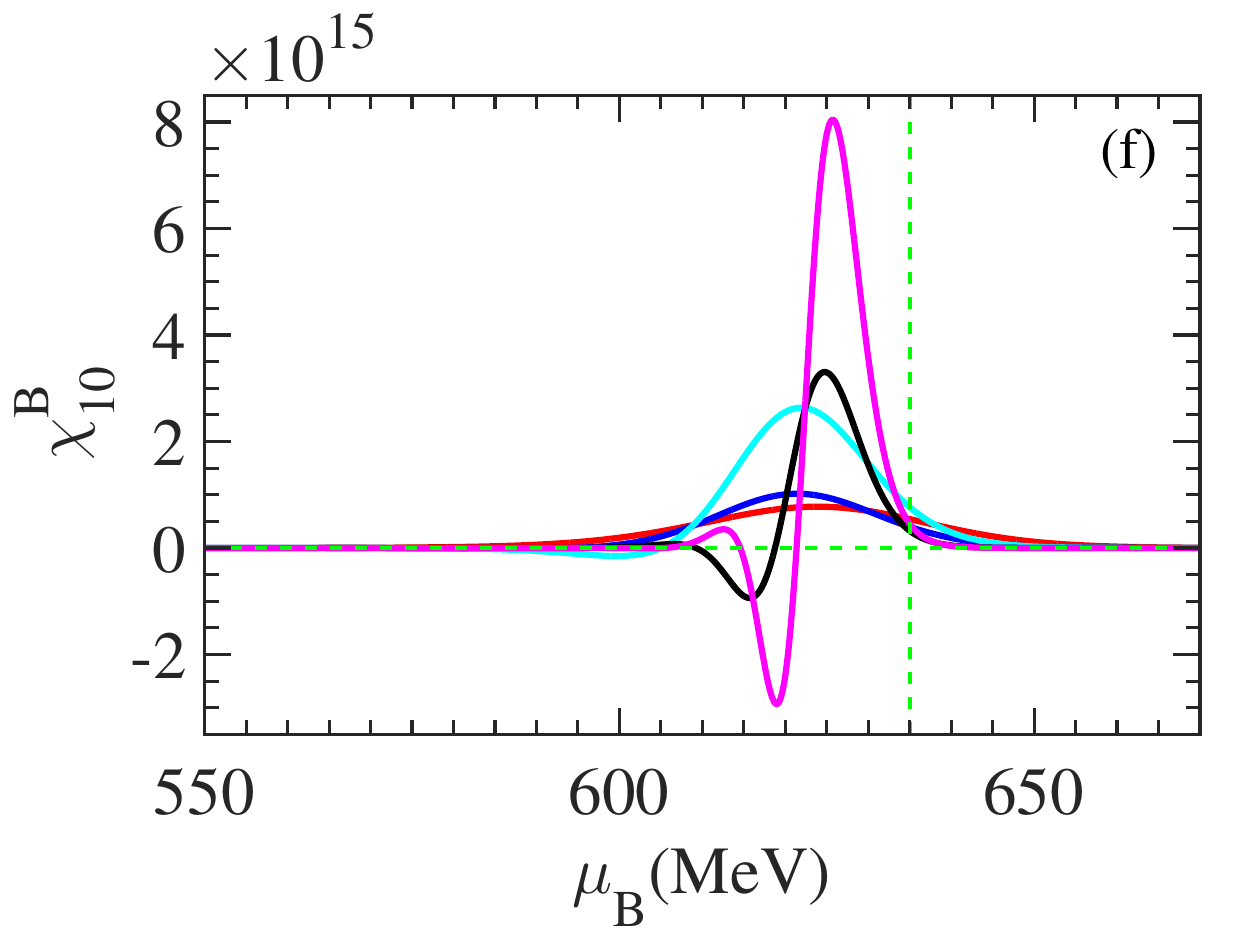}
	\includegraphics[width=0.32\textwidth]{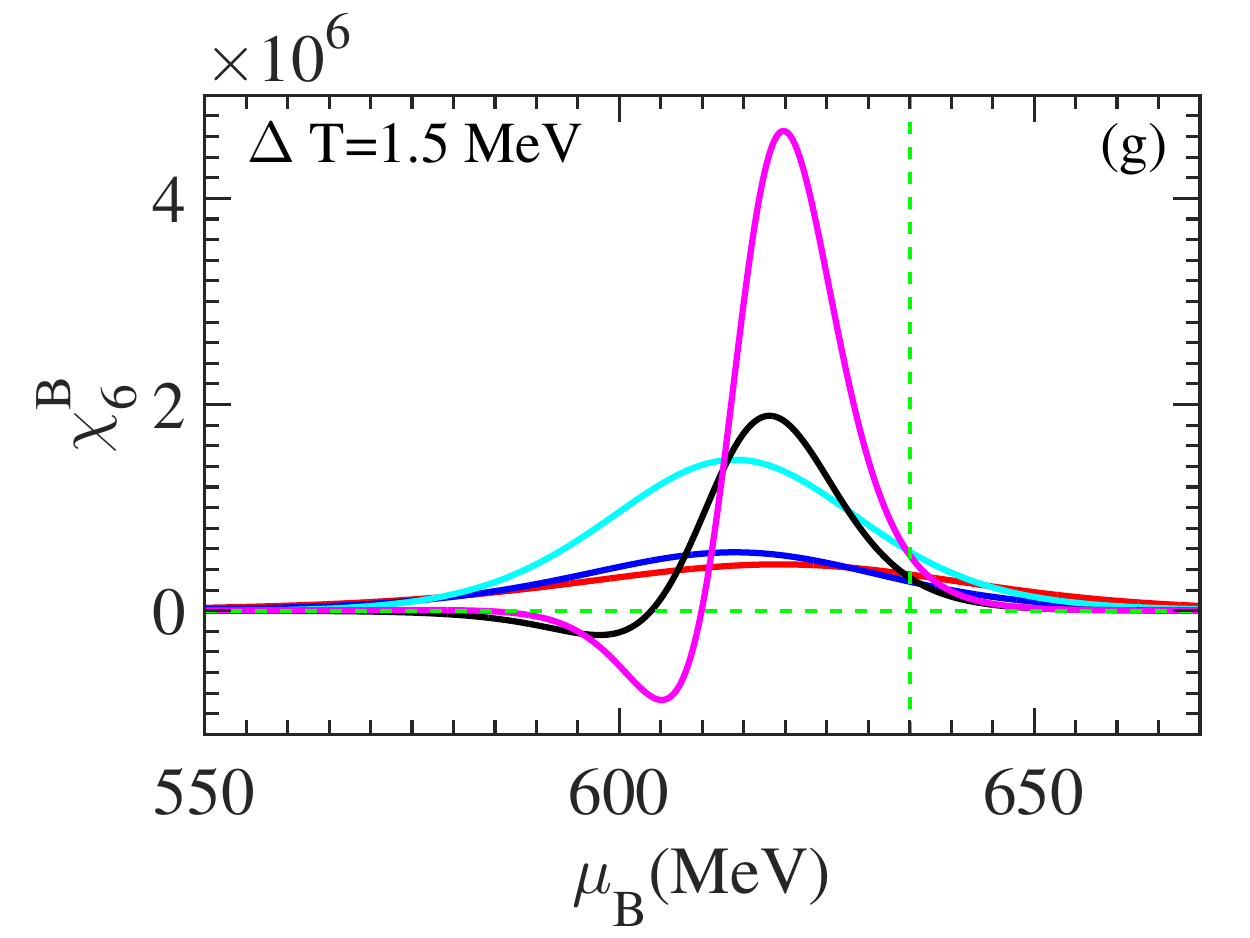}
	\includegraphics[width=0.32\textwidth]{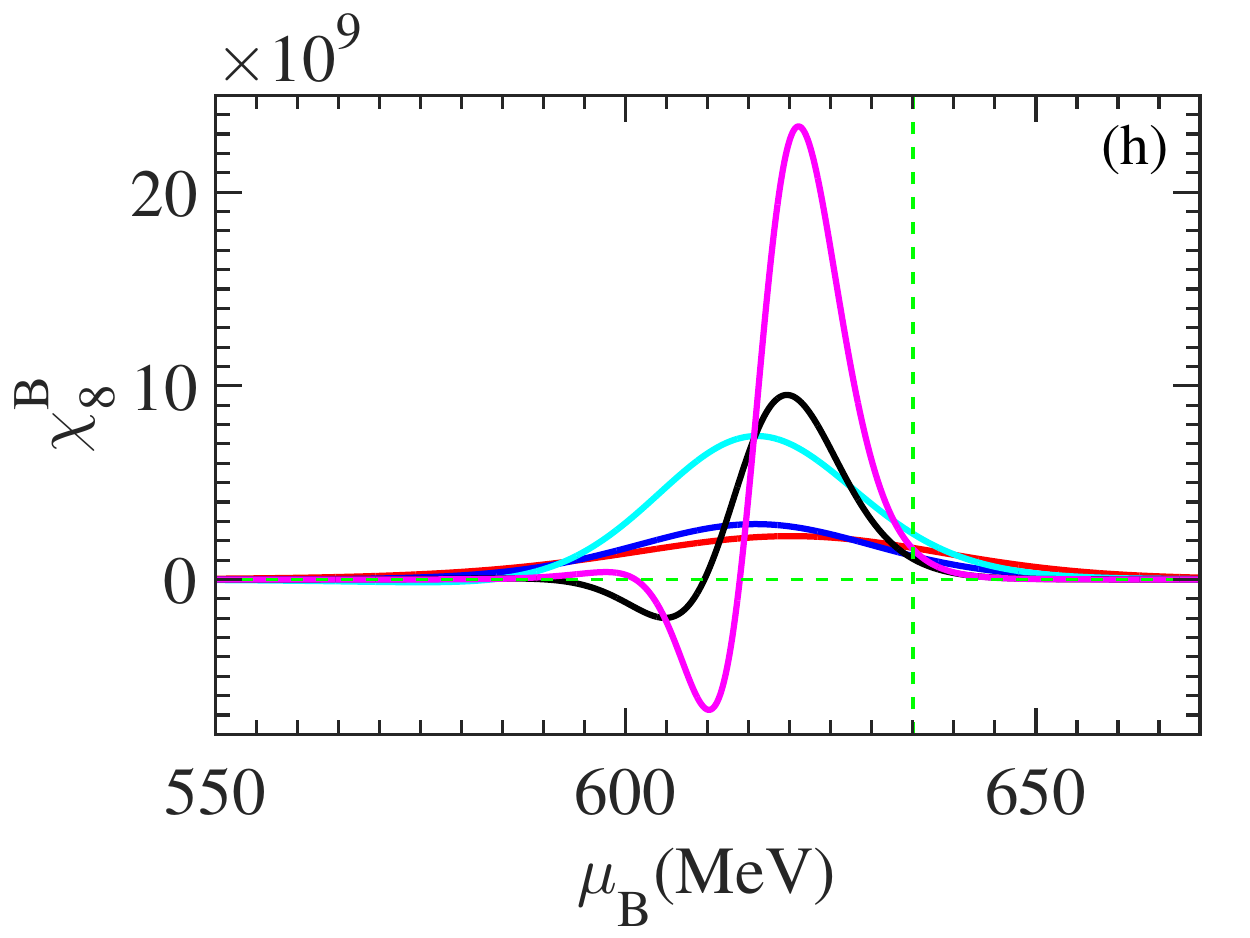}
	\includegraphics[width=0.32\textwidth]{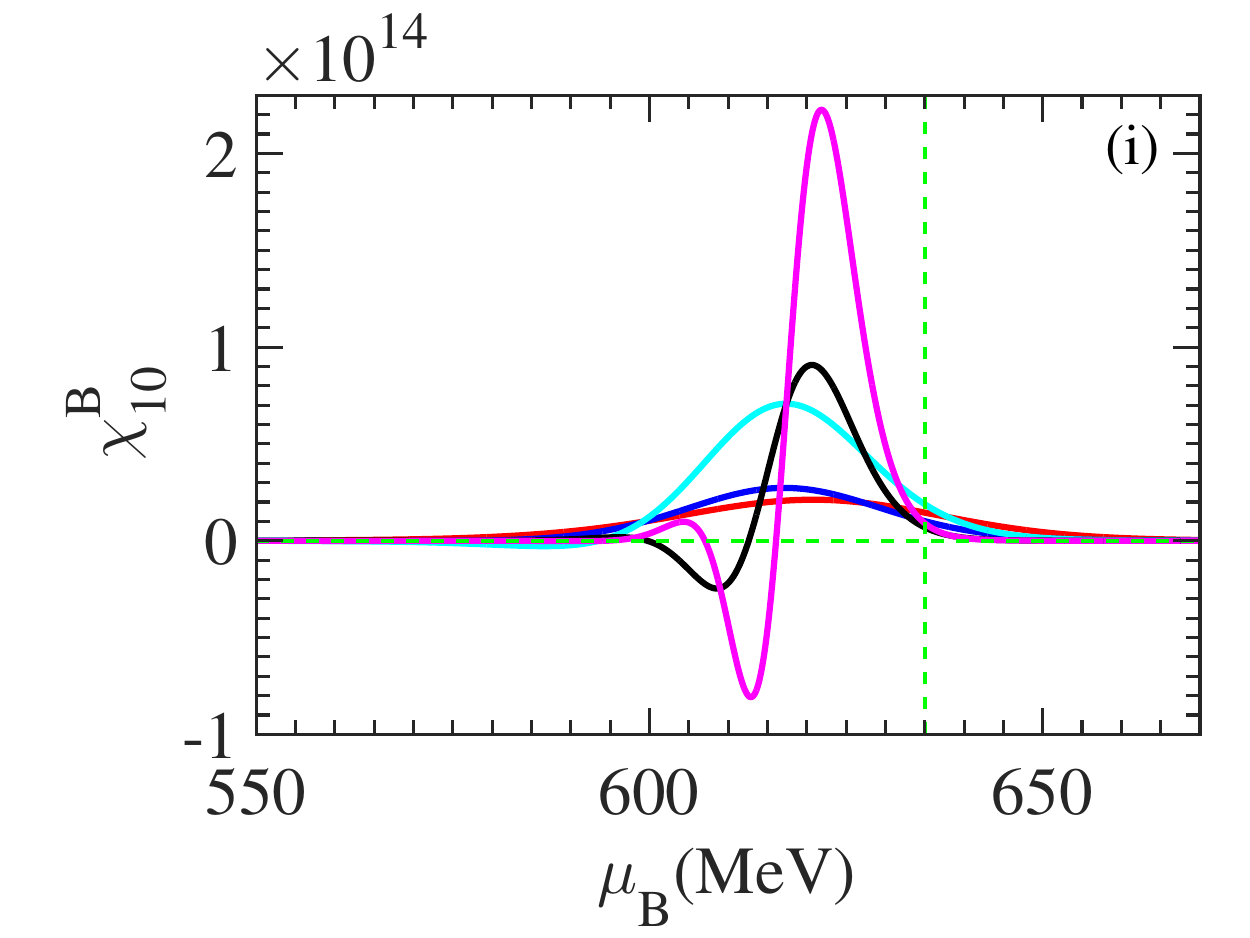}
	\caption{\label{Fig. 5}(Color online). $\mu_B$ dependence of $\chi_{6}^{B}$, $\chi_{8}^{B}$, and $\chi_{10}^{B}$ along the freeze-out curves at $\Delta T=0.2$ MeV (top row), $\Delta T=1.0$ MeV (middle row) and $\Delta T= 1.5$ MeV (bottom row) with different values of $w$ and $\rho$ where $\alpha_2=1.8^{\circ}$.}
\end{figure*}

\begin{figure*}[hbt]
	\centering
	\includegraphics[width=0.32\textwidth]{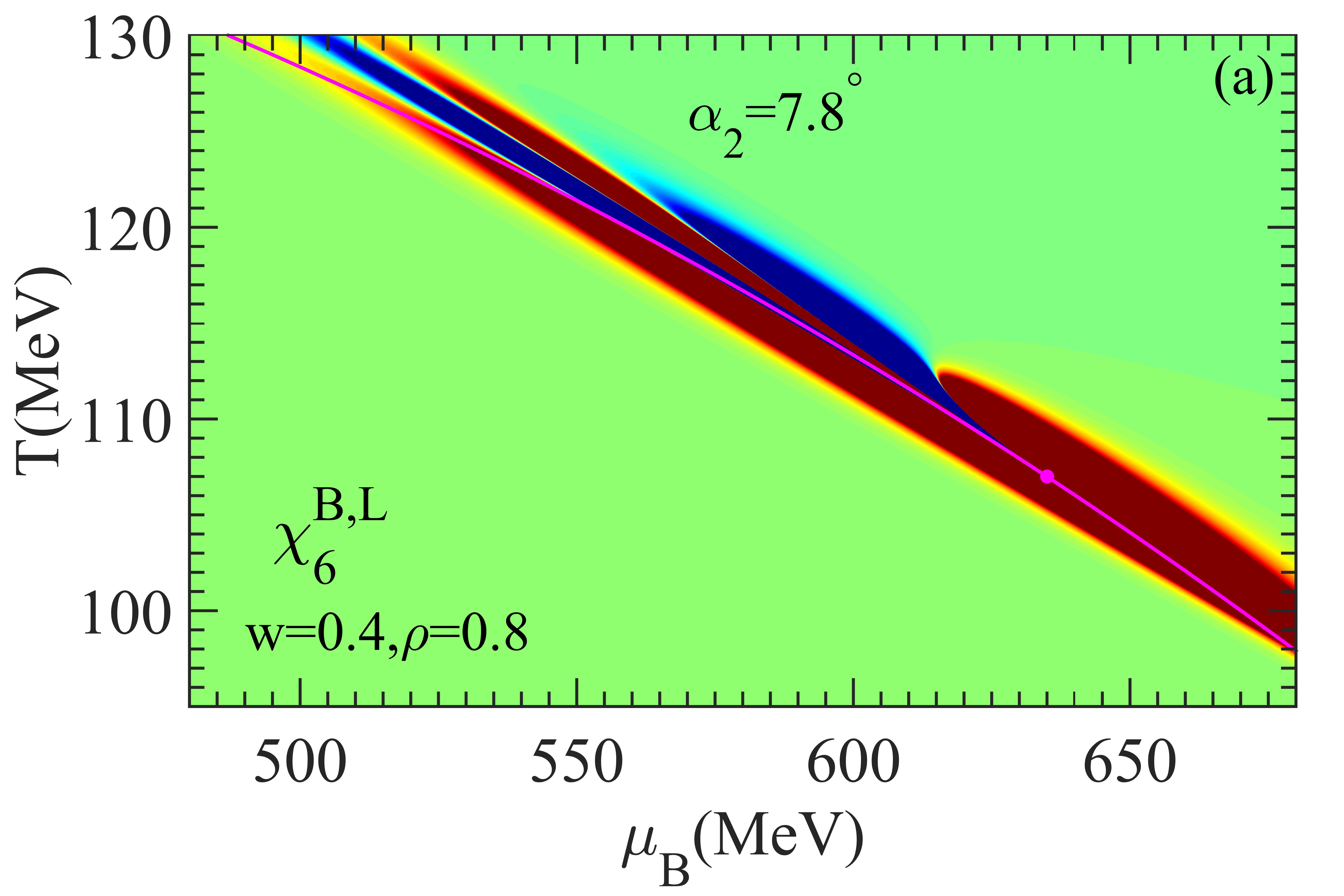}
	\includegraphics[width=0.32\textwidth]{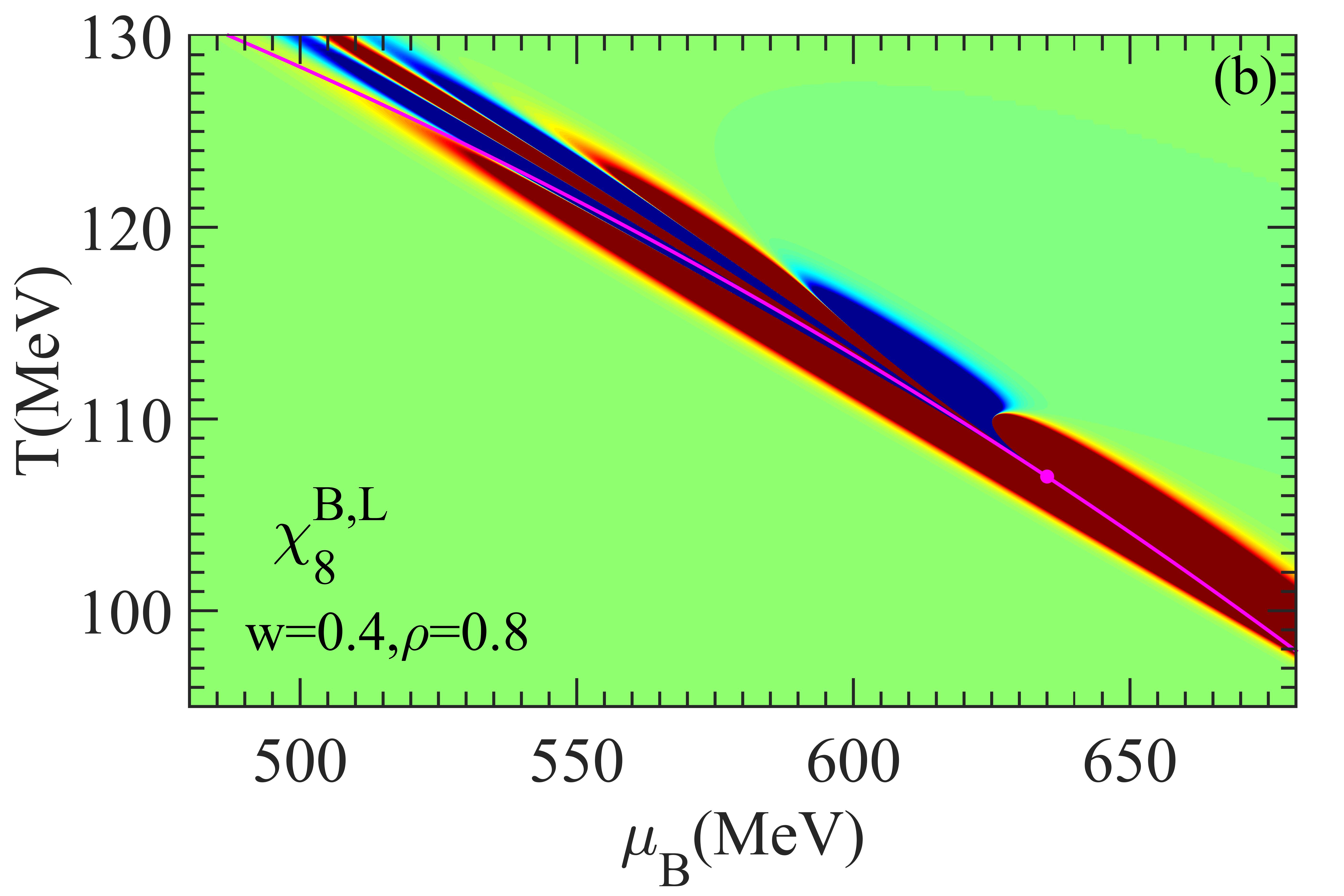}
	\includegraphics[width=0.32\textwidth]{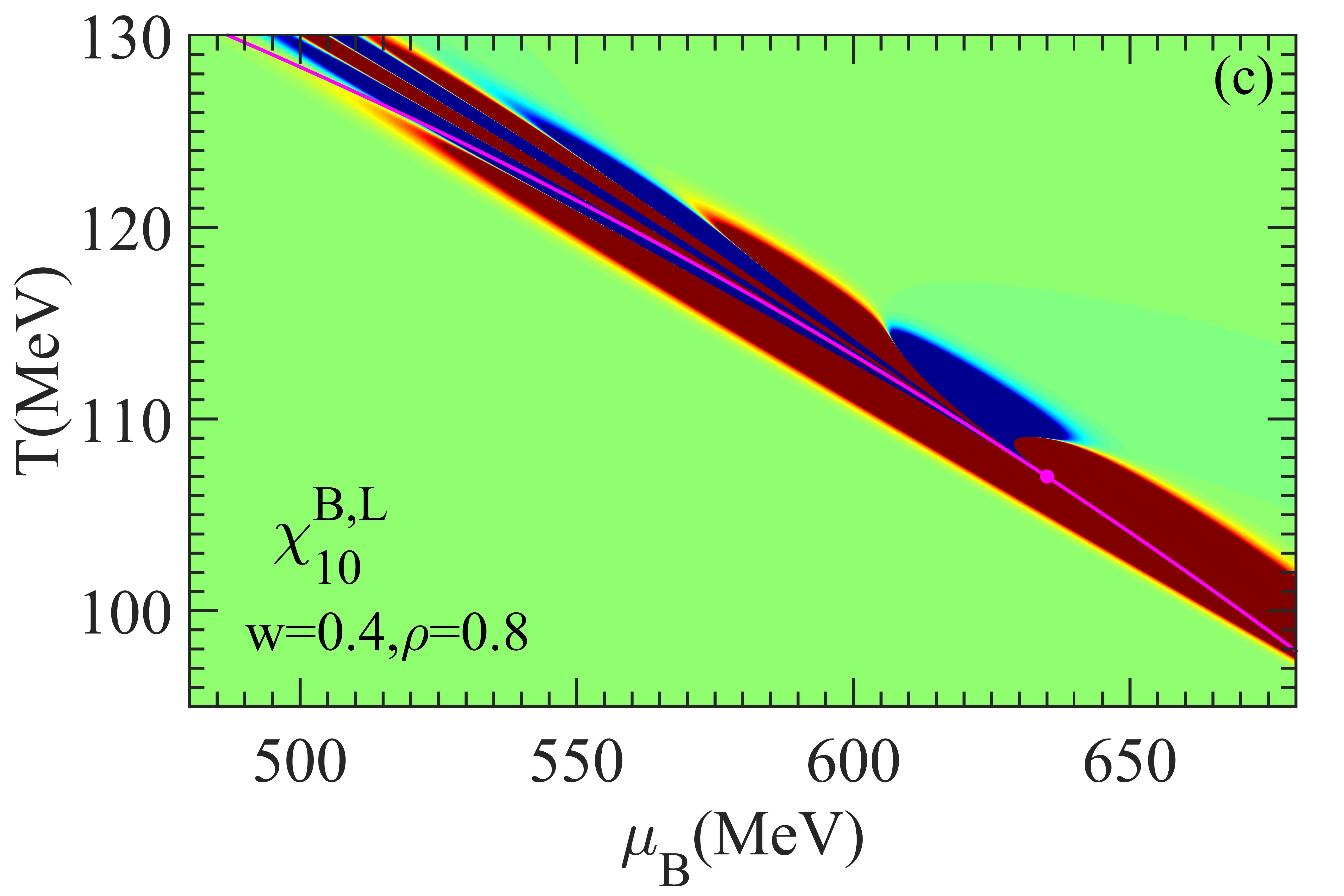}
	\includegraphics[width=0.32\textwidth]{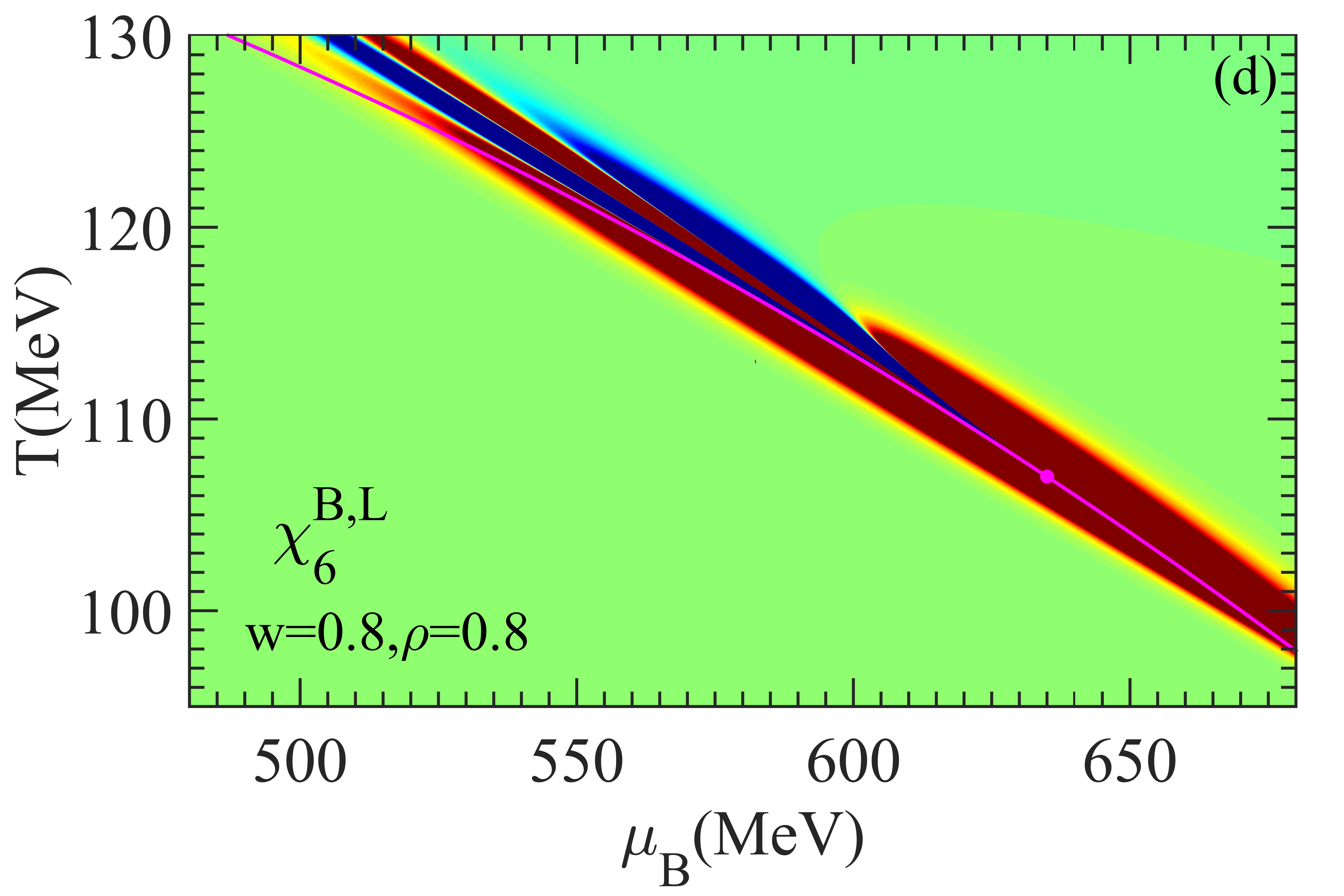}
	\includegraphics[width=0.32\textwidth]{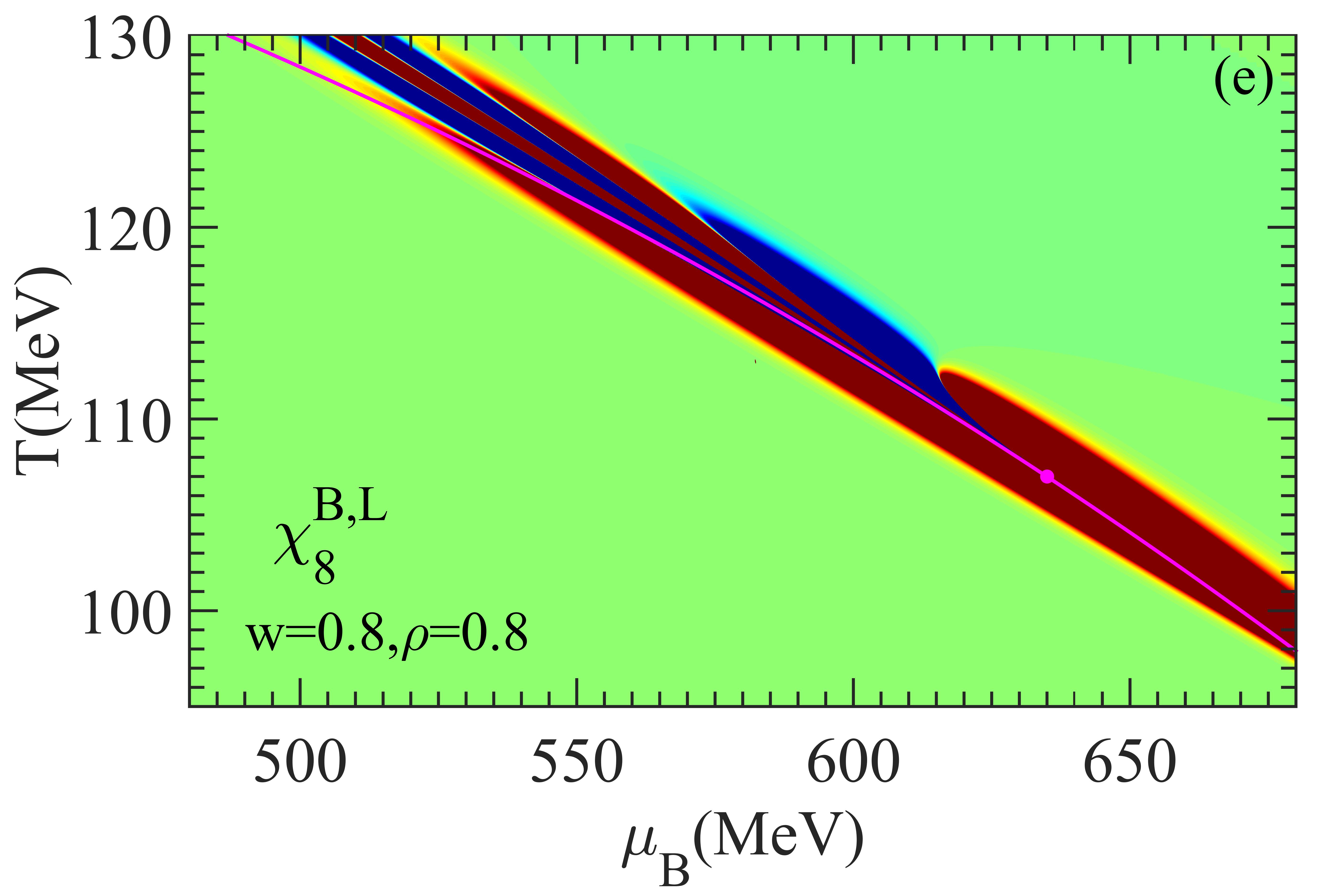}
	\includegraphics[width=0.32\textwidth]{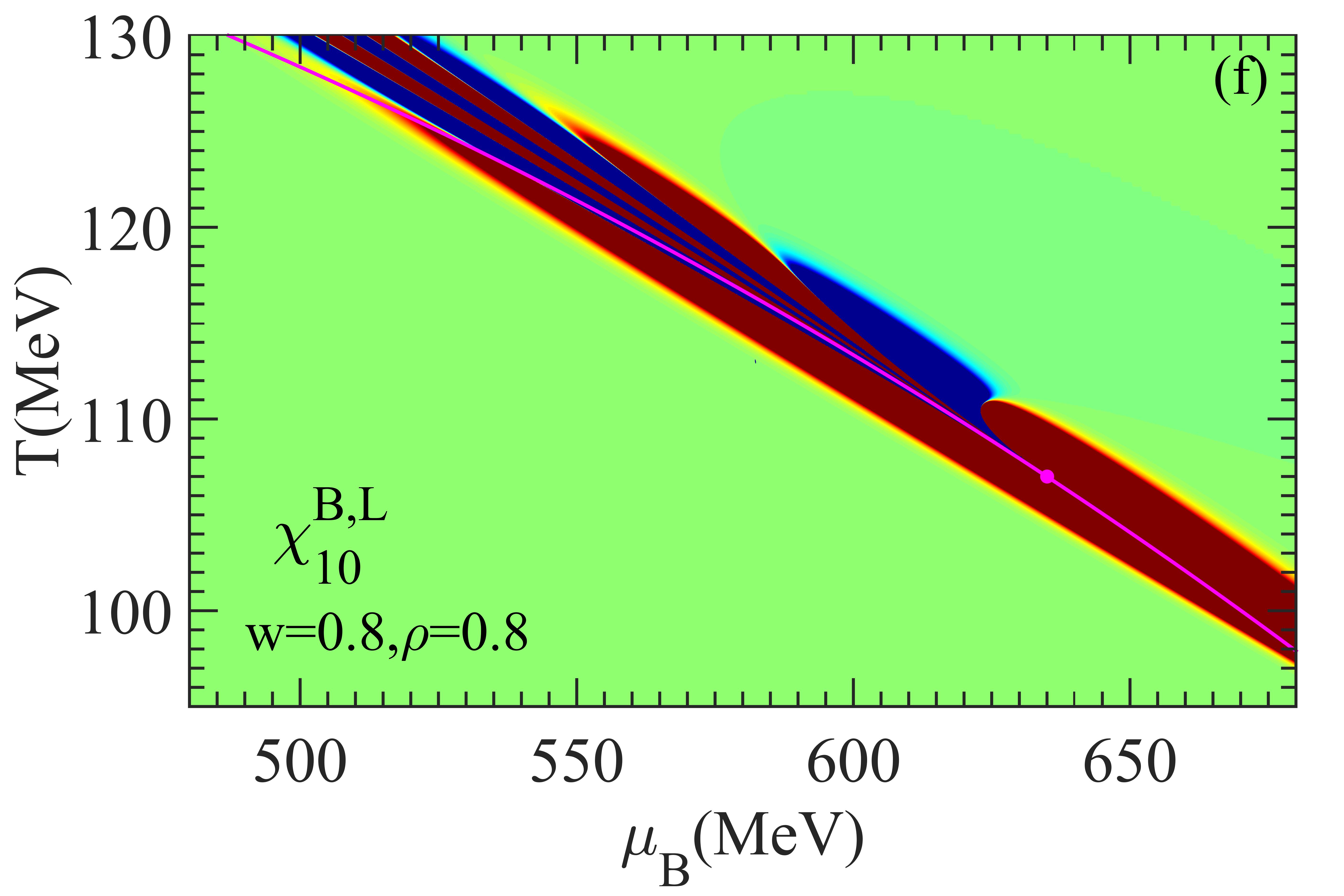}
	\includegraphics[width=0.32\textwidth]{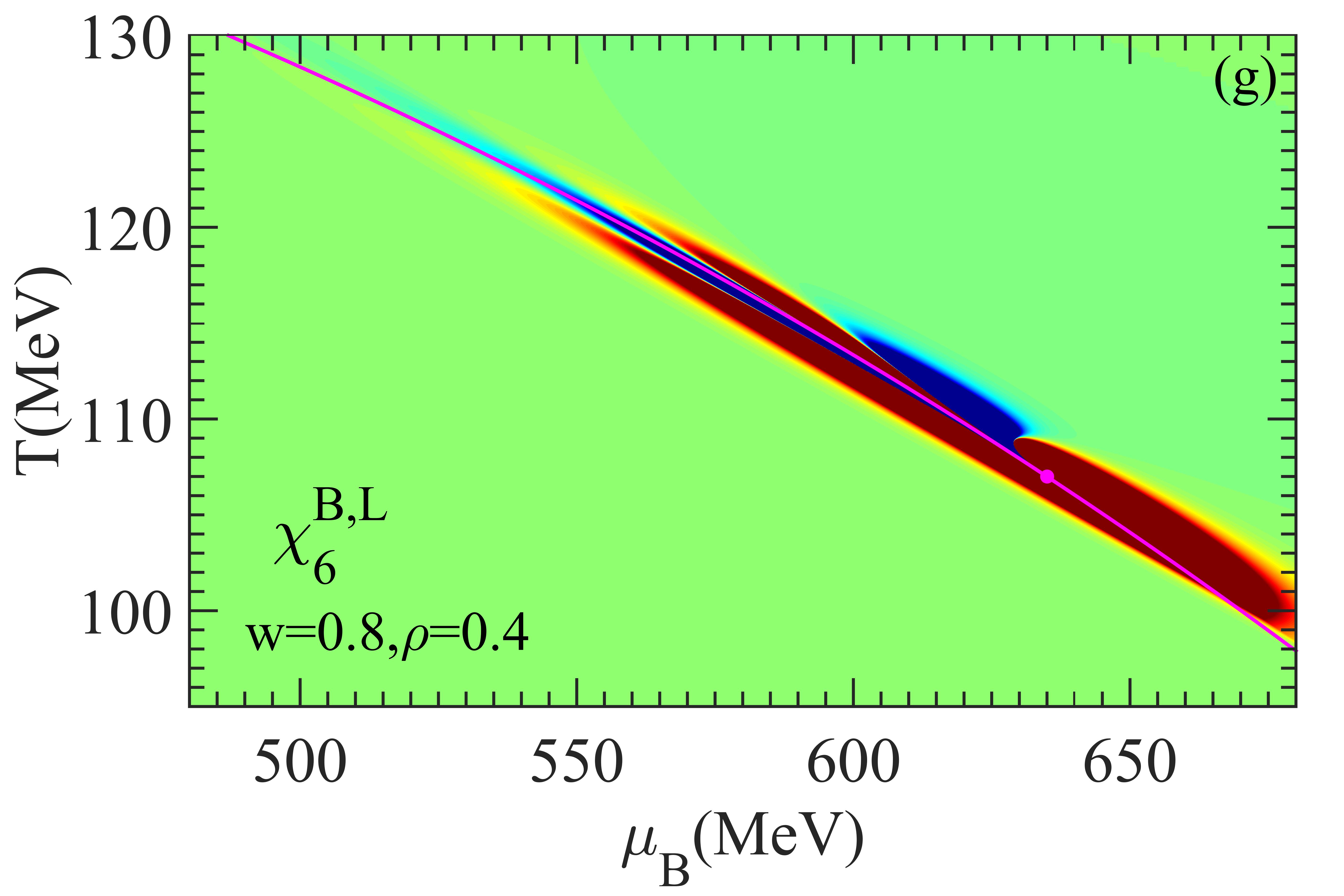}
	\includegraphics[width=0.32\textwidth]{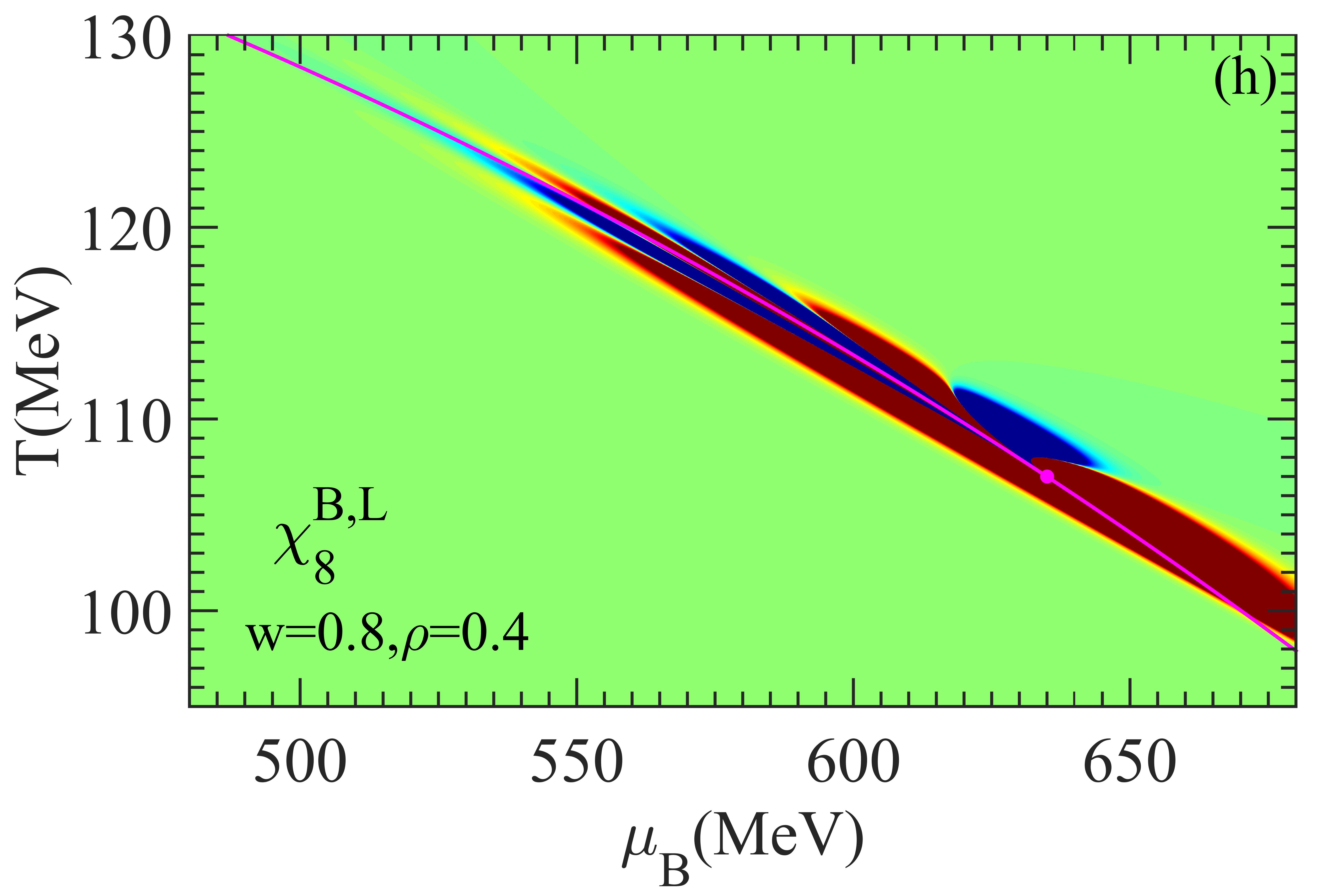}
	\includegraphics[width=0.32\textwidth]{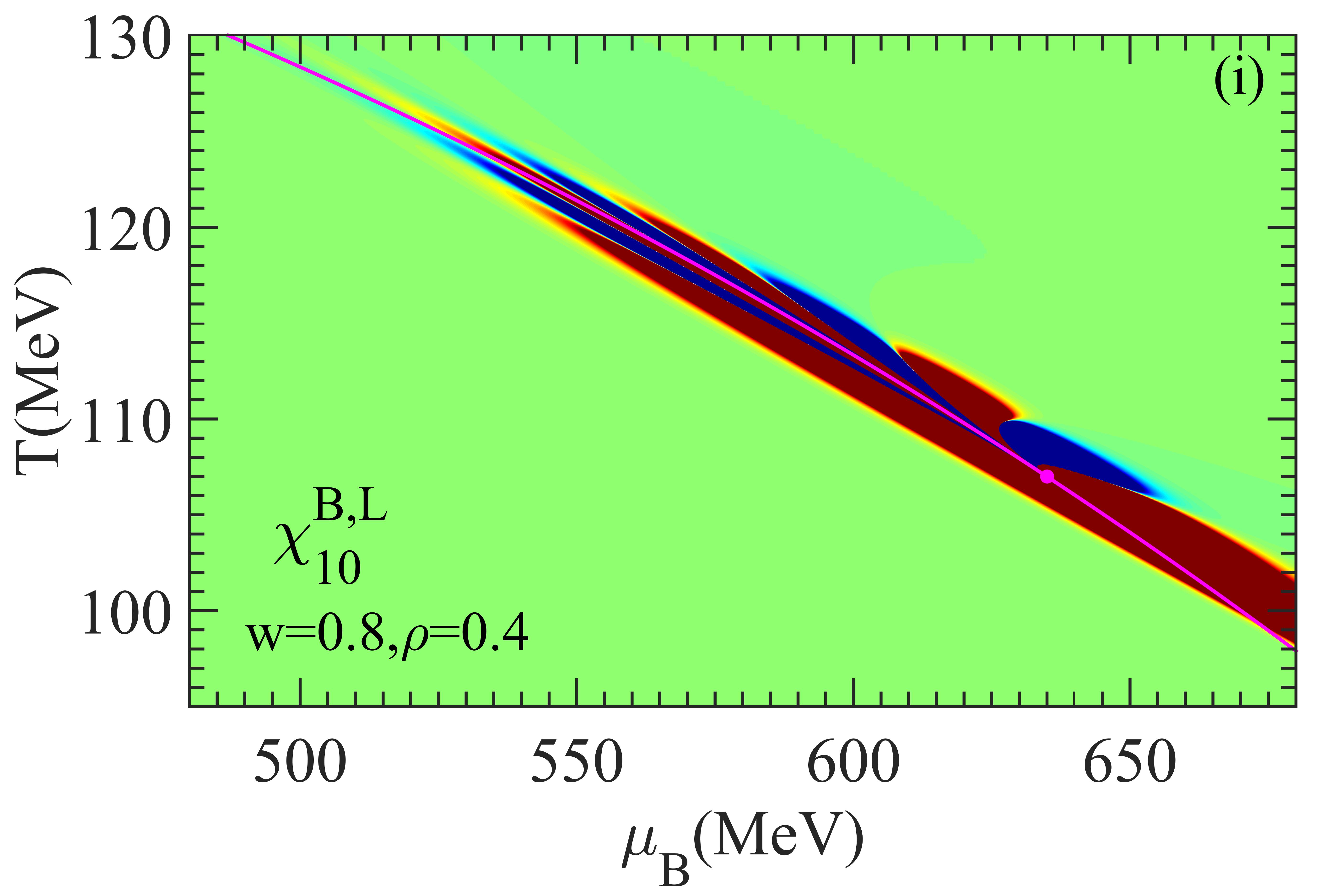}
	\caption{\label{Fig. 6}(Color online). Density plots of critical contribution to $\chi_{6}^{B,L}$, $\chi_{8}^{B,L}$, and $\chi_{10}^{B,L}$ in the QCD $T-\mu_B$ phase plane with $w=0.4, \rho=0.8$ (top row), $w=0.8, \rho=0.8$ (middle row) and $w=0.8, \rho=0.4$ (bottom row) at $\alpha_2=7.8^{\circ}$. The critical point is indicated by a purple dot, while the chiral phase transition line is represented by the solid purple line. }
\end{figure*}

\begin{figure*}[hbt]
	\centering
	\includegraphics[width=0.32\textwidth]{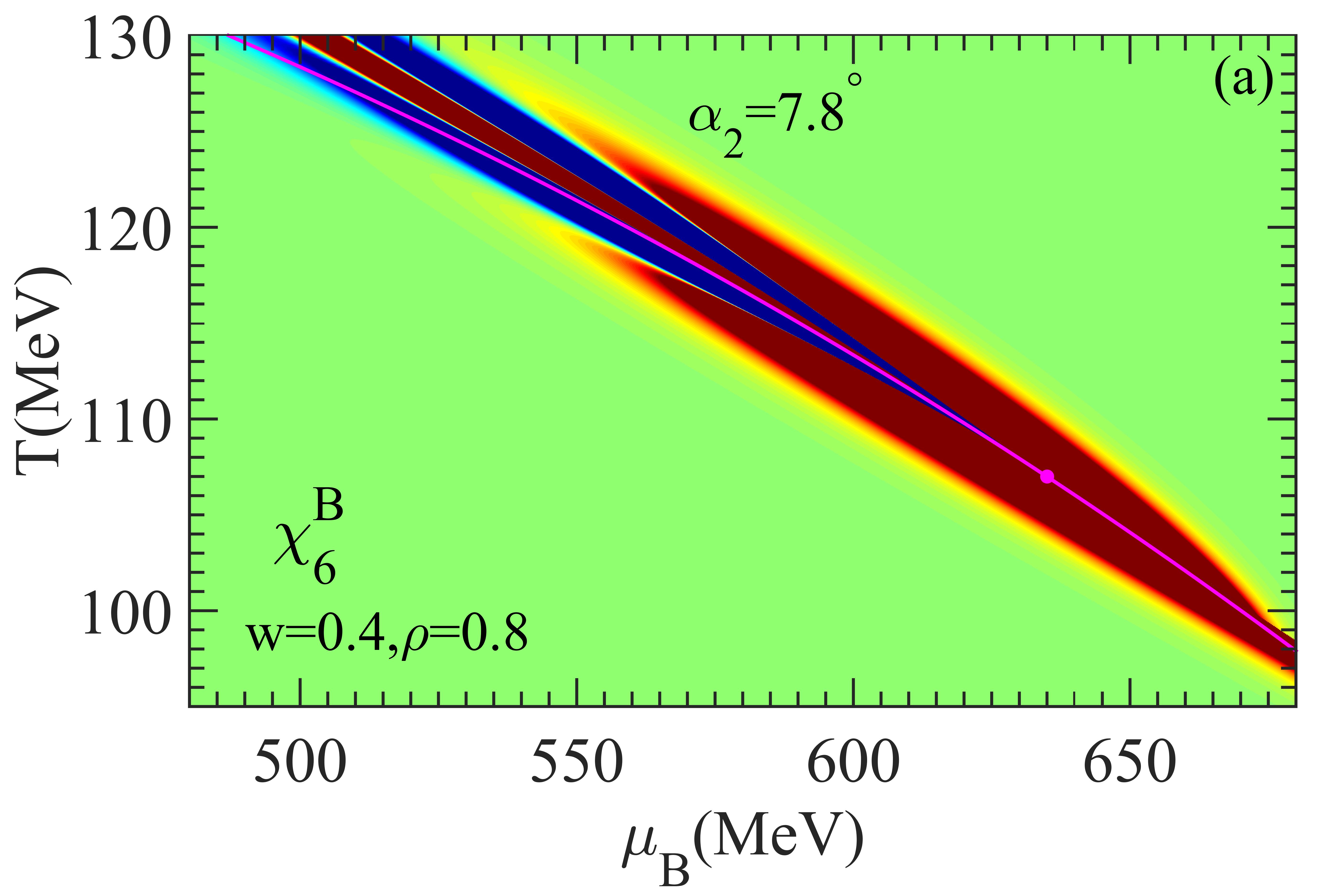}
	\includegraphics[width=0.32\textwidth]{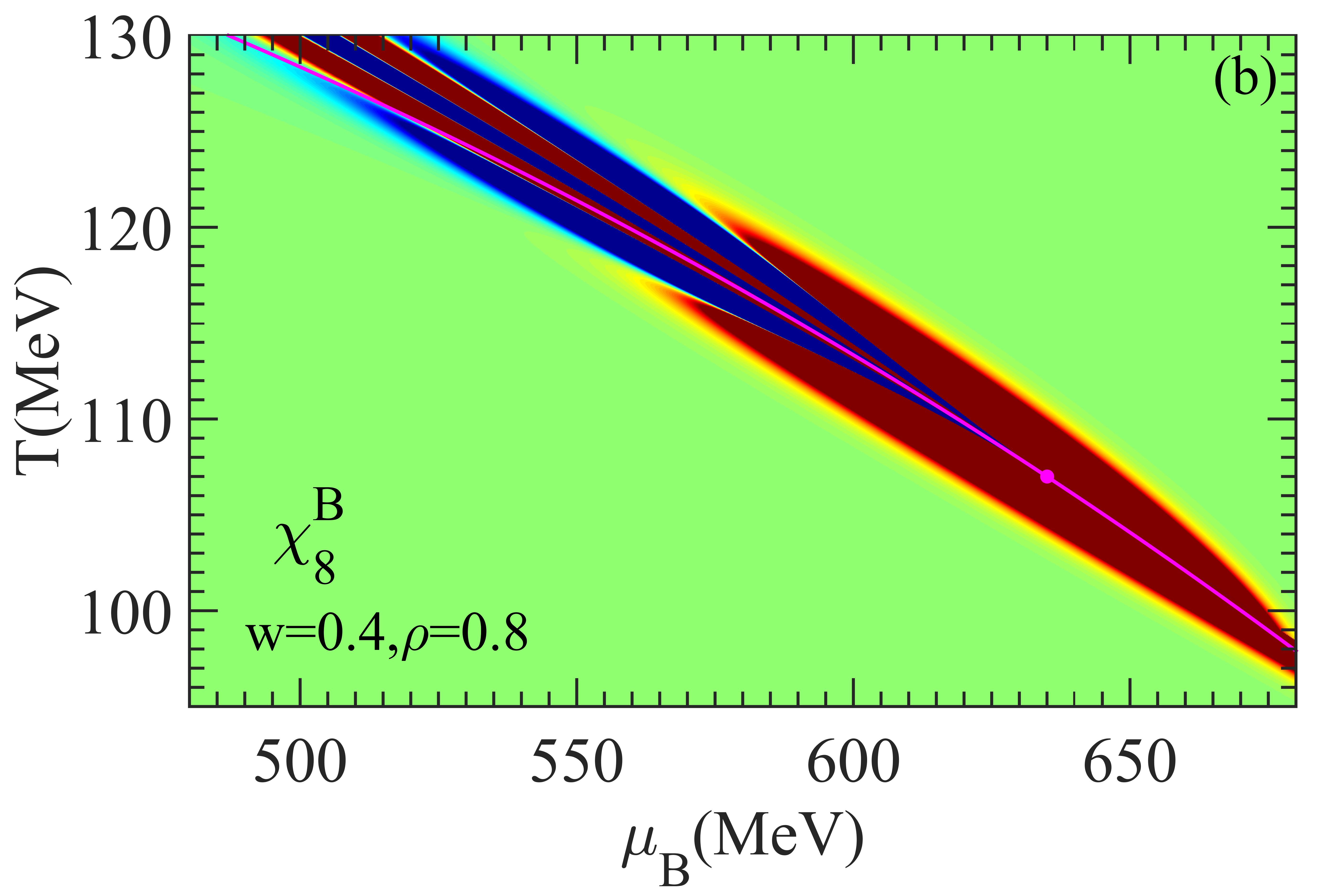}
	\includegraphics[width=0.32\textwidth]{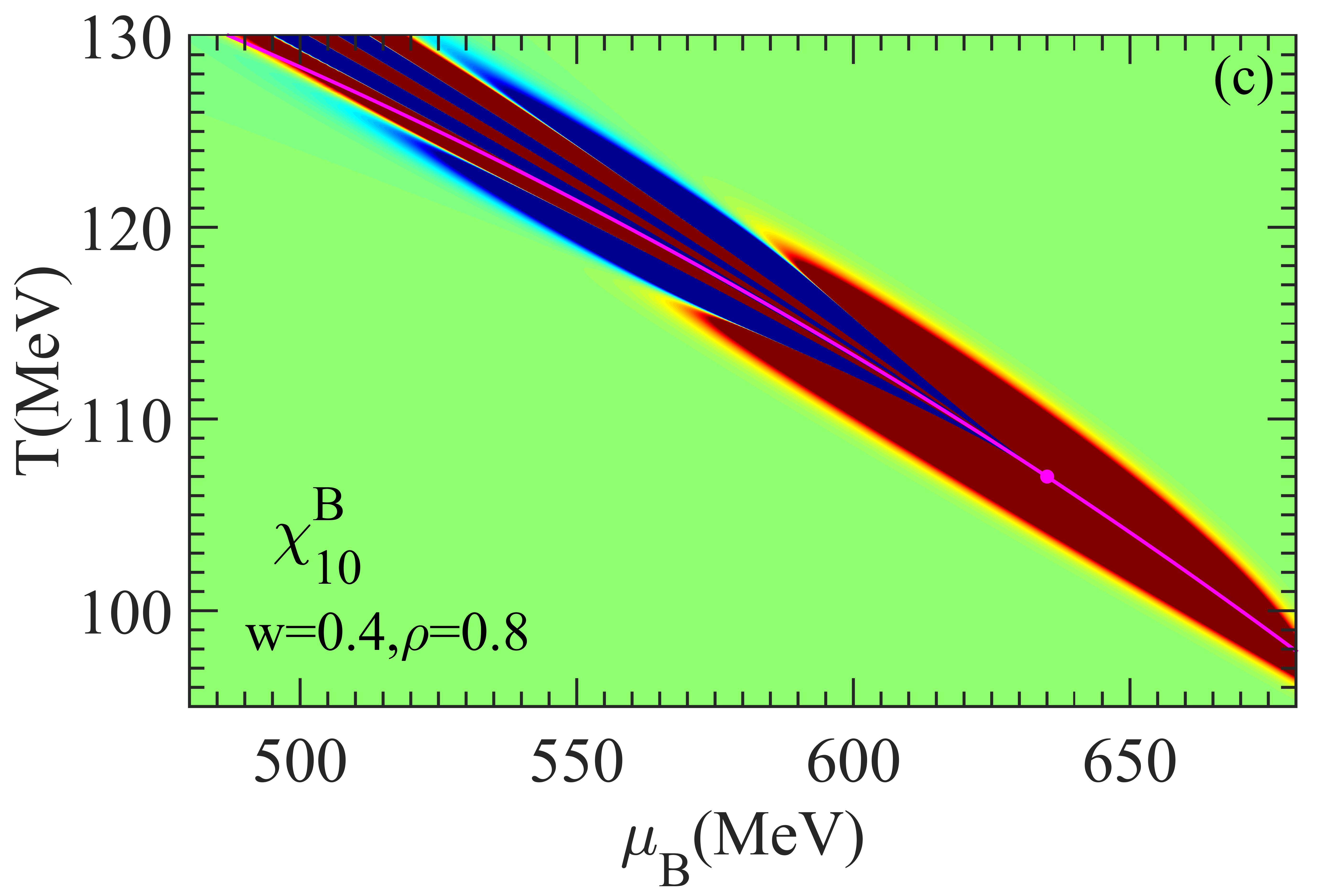}
	\includegraphics[width=0.32\textwidth]{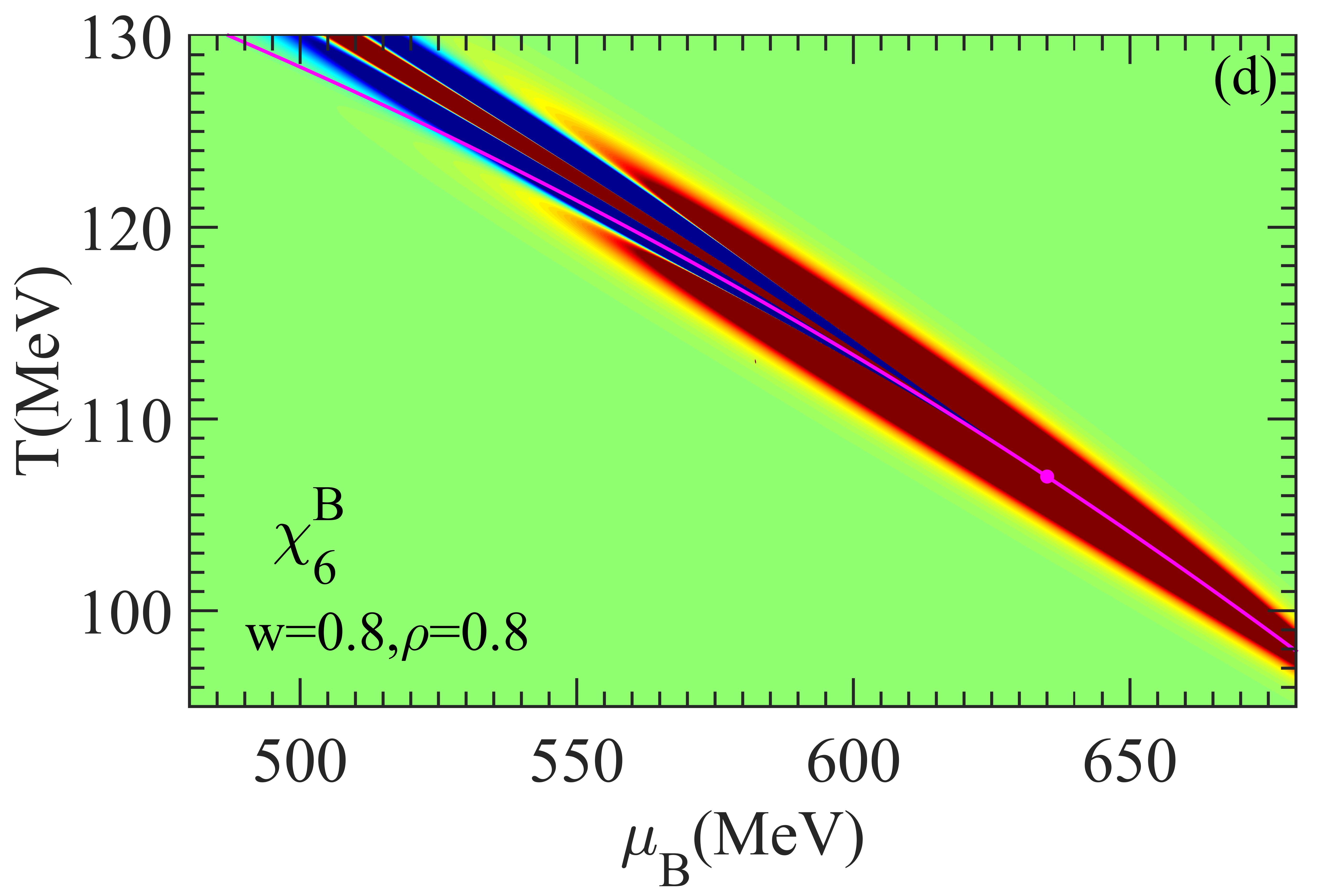}
	\includegraphics[width=0.32\textwidth]{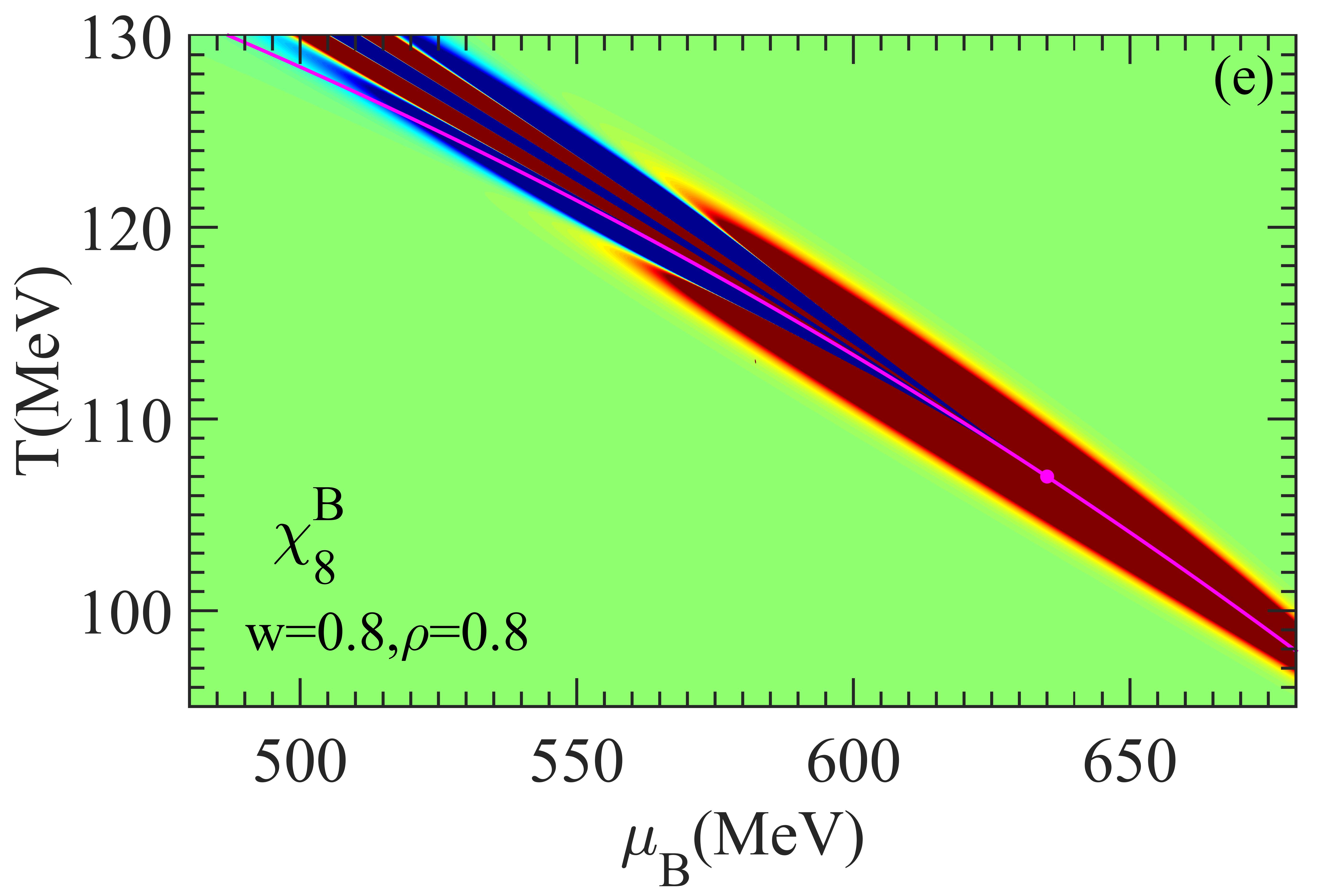}
	\includegraphics[width=0.32\textwidth]{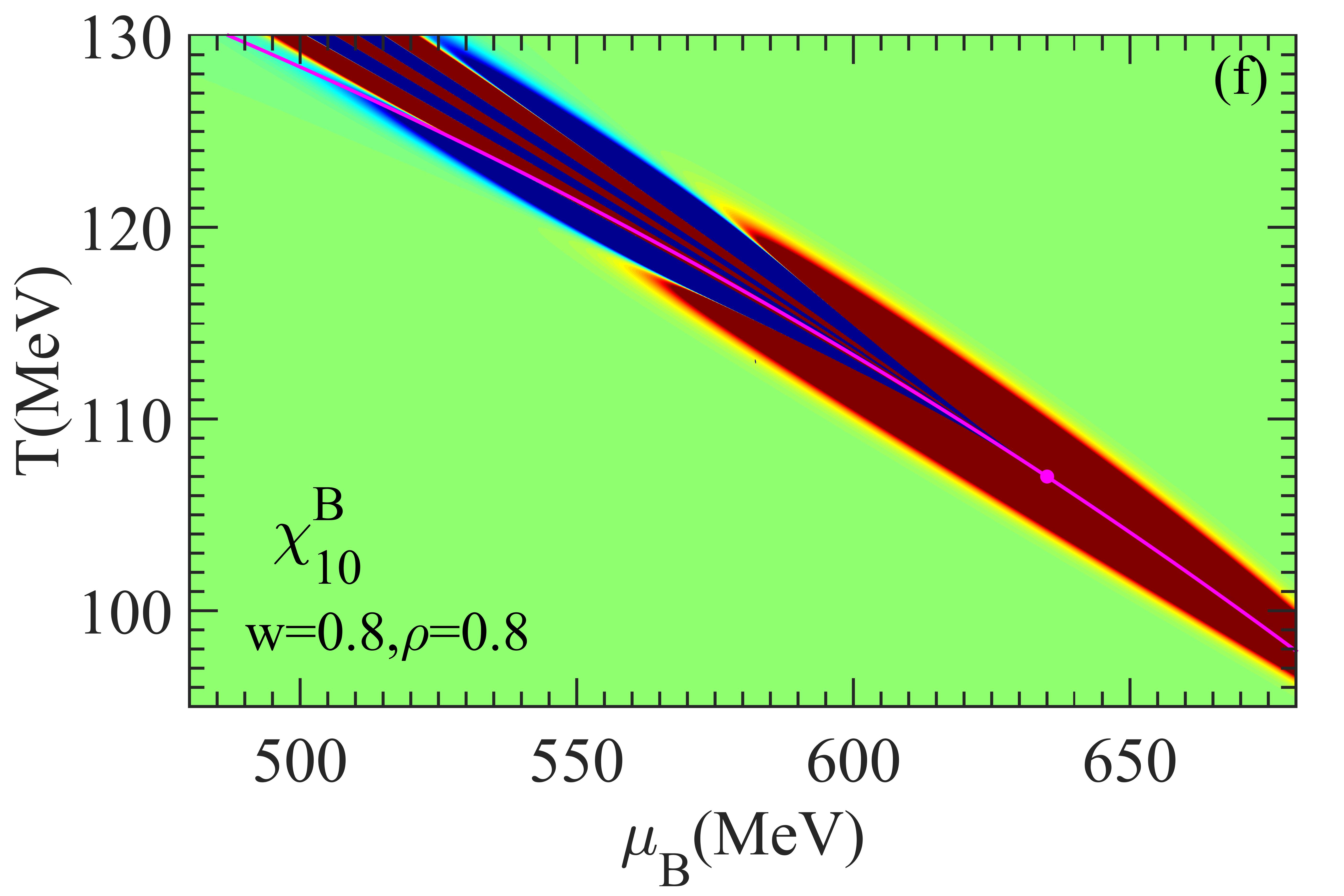}
	\includegraphics[width=0.32\textwidth]{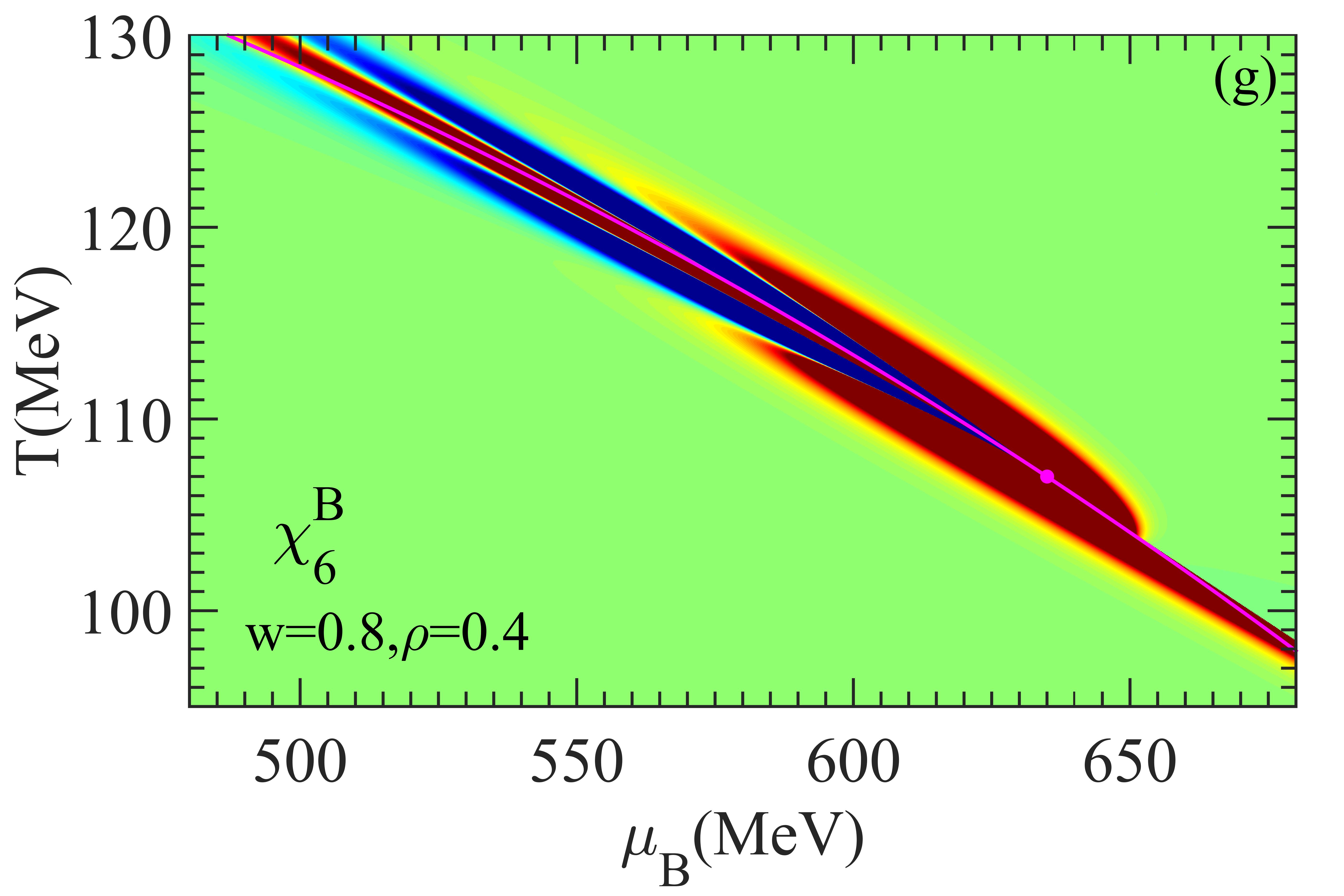}
	\includegraphics[width=0.32\textwidth]{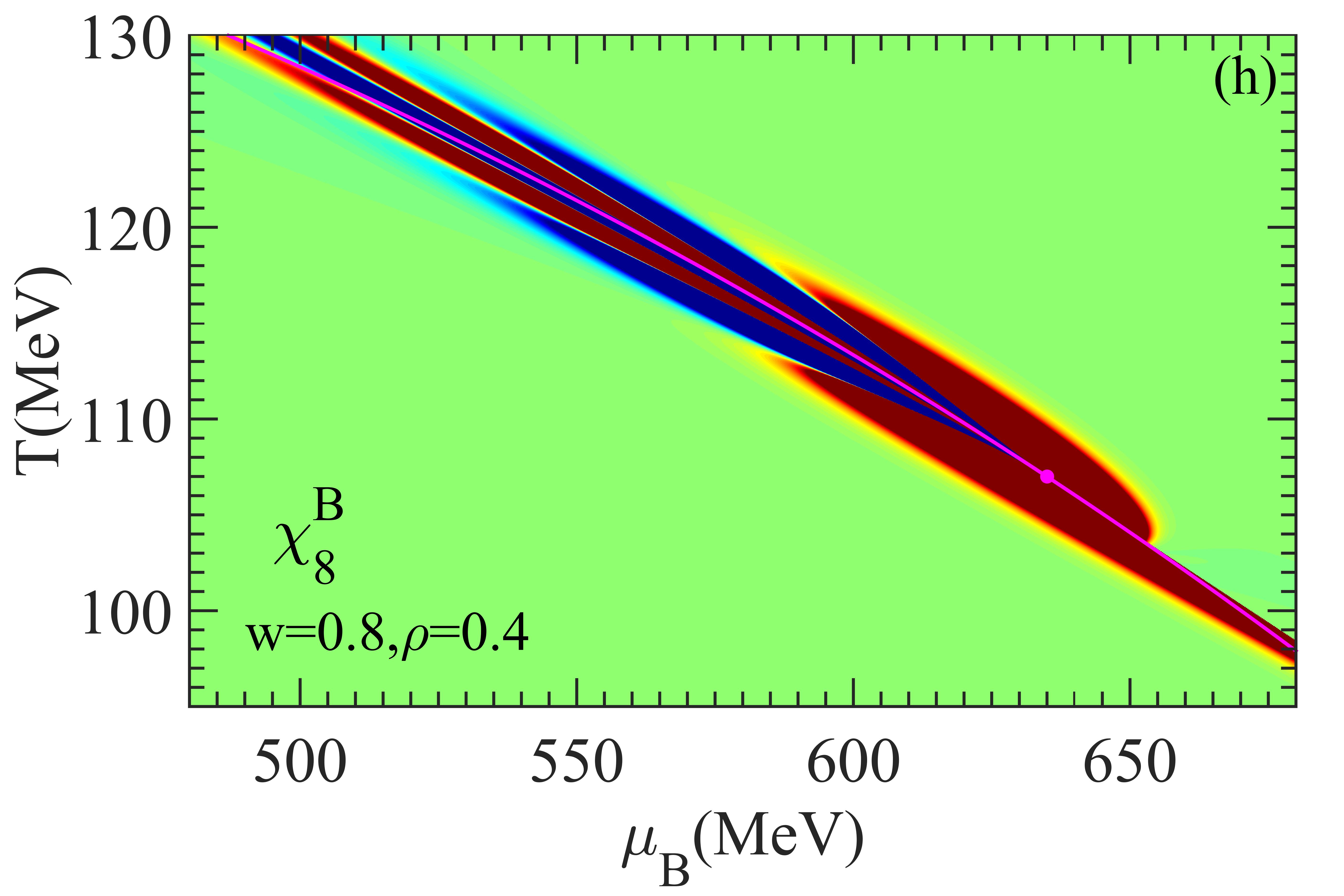}
	\includegraphics[width=0.32\textwidth]{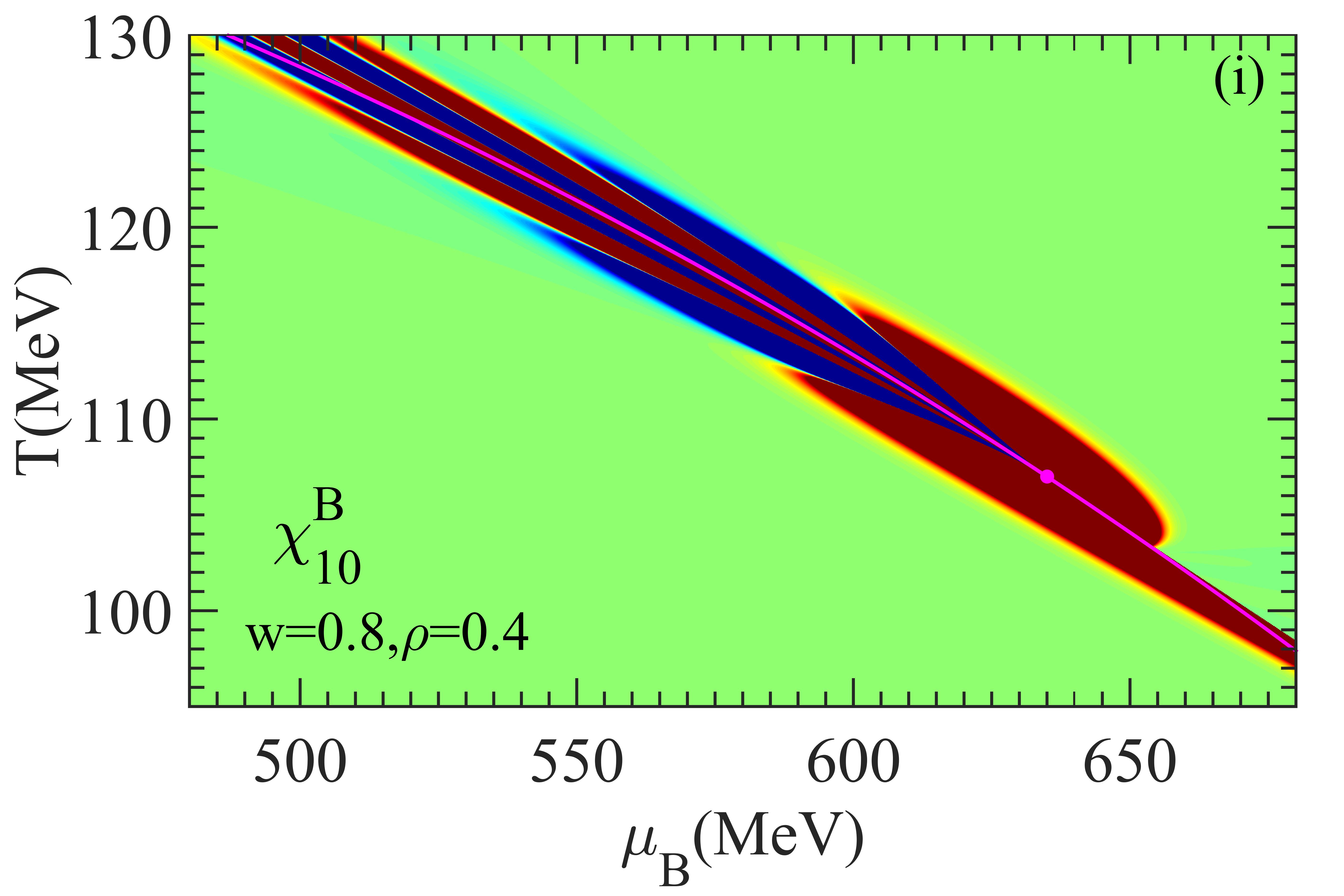}
	\caption{\label{Fig. 7}(Color online). Density plots of critical contribution to $\chi_{6}^{B}$, $\chi_{8}^{B}$, and $\chi_{10}^{B}$ in the QCD $T-\mu_B$ phase plane with $w=0.4, \rho=0.8$ (top row), $w=0.8, \rho=0.8$ (middle row) and $w=0.8, \rho=0.4$ (bottom row) at $\alpha_2=7.8^{\circ}$. The critical point is indicated by a purple dot, while the chiral phase transition line is represented by the solid purple line. }
\end{figure*}

\section{Generalized susceptibilities of the net-baryon number with $\alpha_2=1.8^{\circ}$}

For the density plots of generalized susceptibilities of the net-baryon number, the color function remains consistent for the same order. And considering the big difference of the magnitude between different orders of susceptibilities, the color schemes are such that a factor $1$ thousand in values of the eighth-order and $1$ million in values of tenth-order separates the sixth-order susceptibilities, for the same color.

The green, yellow and red areas correspond to positive values. The regions where the value is largest and smallest are indicated in red and green, respectively. The blue areas correspond to negative values, and the darker, the larger in its magnitude. 

In the case $\alpha_2=1.8^{\circ}$, three sets of values for $w$ and $\rho$ are chosen. They are $(w,\rho)=(0.5, 0.5)$, $(1.0, 1.0)$ and $(2.0, 2.0)$, respectively.  

The density plots of the sixth-, eighth- and tenth-order susceptibilities considering only the leading critical contributions are shown in the left, middle, and right column of Fig.~2, respectively. Incorporating the sub-leading critical contribution, the corresponding results are shown in Fig.~3. In each sub-figure, the purple curve shows the QCD phase transition line represented by Eq.~\eqref{QCD transition line}. The purple dot marks the critical point.

In the left column of Fig.~2, one can found that the general patterns in density plots of $\chi_6^{B,L}$ do not change with varying values of $w$ and $\rho$. So do that of $\chi_8^{B,L}$ and $\chi_{10}^{B,L}$ in the middle and right columns. The positive and negative lobes appear alternately in the vicinity of the critical point. The higher the order of the susceptibility, the more frequent this alternation becomes.

The increase of values of $w$ and $\rho$ only influence the area size occupied by the main pattern (consisted of the red and dark blue lobes) around the critical point. It is clear that, from the up to the bottom row, the main pattern becomes wider in the $\mu_B$ direction and narrower in the $T$ direction. This is because larger $w$ leads to narrower critical region in $T$ direction, while larger $\rho$ leads to wider critical region in $\mu_B$ direction~\cite{stephanov-prc103}.

One should notice that as the increase of $w$ and $\rho$, the lobes under the phase transition line in the density plot of $\chi_6^{B,L}$, $\chi_8^{B,L}$ and $\chi_{10}^{B,L}$ shift to the upside gradually. Only one positive lobe and no negative lobe left under the phase transition line in the bottom row of Fig.~2.

Comparing each sub-figure in Fig.~3 with the corresponding one in Fig.~2, it is clear that the pattern of the density plot for each order of susceptibilities is nearly the same. The sub-leading critical contribution has little effects on the density plots. It is a natural result that $\alpha_2$ is close to zero, the contribution of sub-leading singular terms is significantly suppressed by $\sin \alpha_2/\sin \alpha_1$. The leading singular term, susceptibilities of the magnetization, dominate the behavior of $\chi_{6}^{B}$, $\chi_{8}^{B}$, and $\chi_{10}^{B}$.

To observe the $\mu_B$-dependence of the susceptibilities, a freeze-out curve which parallels the QCD phase transition line as described by Eq.~\eqref{QCD transition line} is assumed, but is shifted towards lower temperatures by $\Delta T$, that is
\begin{equation}\label{freeze-out curve}
	T_f(\mu_B)=T_0[1-\kappa(\frac{\mu_B}{T_0})^2-\lambda (\frac{\mu_B}{T_0})^4]-\Delta T.
\end{equation}

The $\mu_B$ dependence of $\chi_{6}^{B,L}$, $\chi_{8}^{B,L}$, and $\chi_{10}^{B,L}$ along three different freeze-out curves described by Eq.~\eqref{freeze-out curve} with $\Delta T = 0.2$, $1.0$ and $1.5$ MeV are shown in the top, middle and bottom row of Fig.~4, respectively. After considering the sub-leading contribution, the results are shown in Fig.~5. The purple, black, cyan, blue and red curves are for five different combinations of values $0.5$, $1.0$ and $2.0$ for $w$ and $\rho$, respectively. 

It is clear that in Fig.~4, for small values of $w$ and $\rho$, i.e., the purple and black curves in each sub-figure, as the increase of $\mu_B$, a negative dip followed by a positive peak can be observed. While for big values of $w$ and $\rho$, i.e., the blue and red curves, only a positive peak can be observed and no negative dip exists anymore. For the cyan curve, the negative dip followed by a positive peak can be observed only when the freeze-out curve is very close to the phase transition line, such as in the first row of Fig.~4. The negative dip fades away when it is far from the phase transition line. It is not a robust feature any more when considering only the leading critical contribution when $\alpha_2=1.8^{\circ}$.

Comparing each sub-figure in Fig.~5 with the corresponding one in Fig.~4, the qualitative behavior of $\mu_B$-dependence of the susceptibilities remains unchanged. However, quantitatively, the negative dip in the purple and black curves is more pronounced compared to the corresponding curves in Fig.~4. For instance, the ratio of the depth of the dip to the height of the peak in the purple curves for $\chi_{6}^{B,L}$, $\chi_{8}^{B,L}$, and $\chi_{10}^{B,L}$ in the first row of Fig.~4 is 0.1294, 0.222, and 0.2922, respectively. In contrast, these ratios are 0.1742, 0.2779, and 0.3547 in the first row of Fig.~5, respectively. 

Although the negative dip is slightly amplified upon incorporating the sub-leading critical contribution, it is not a robust characteristic when $\alpha_2=1.8^{\circ}$. The presence of the negative dip depends on the scaling parameters $w$ and $\rho$, as well as the distance to the phase transition line. Nevertheless, it is important to note that in each curve of both Fig.~4 and Fig.~5, the positive peak persists. This feature is independent of the scaling parameters and the distance to the phase transition line.

\begin{figure*}[hbt]
	\centering
	\includegraphics[width=0.32\textwidth]{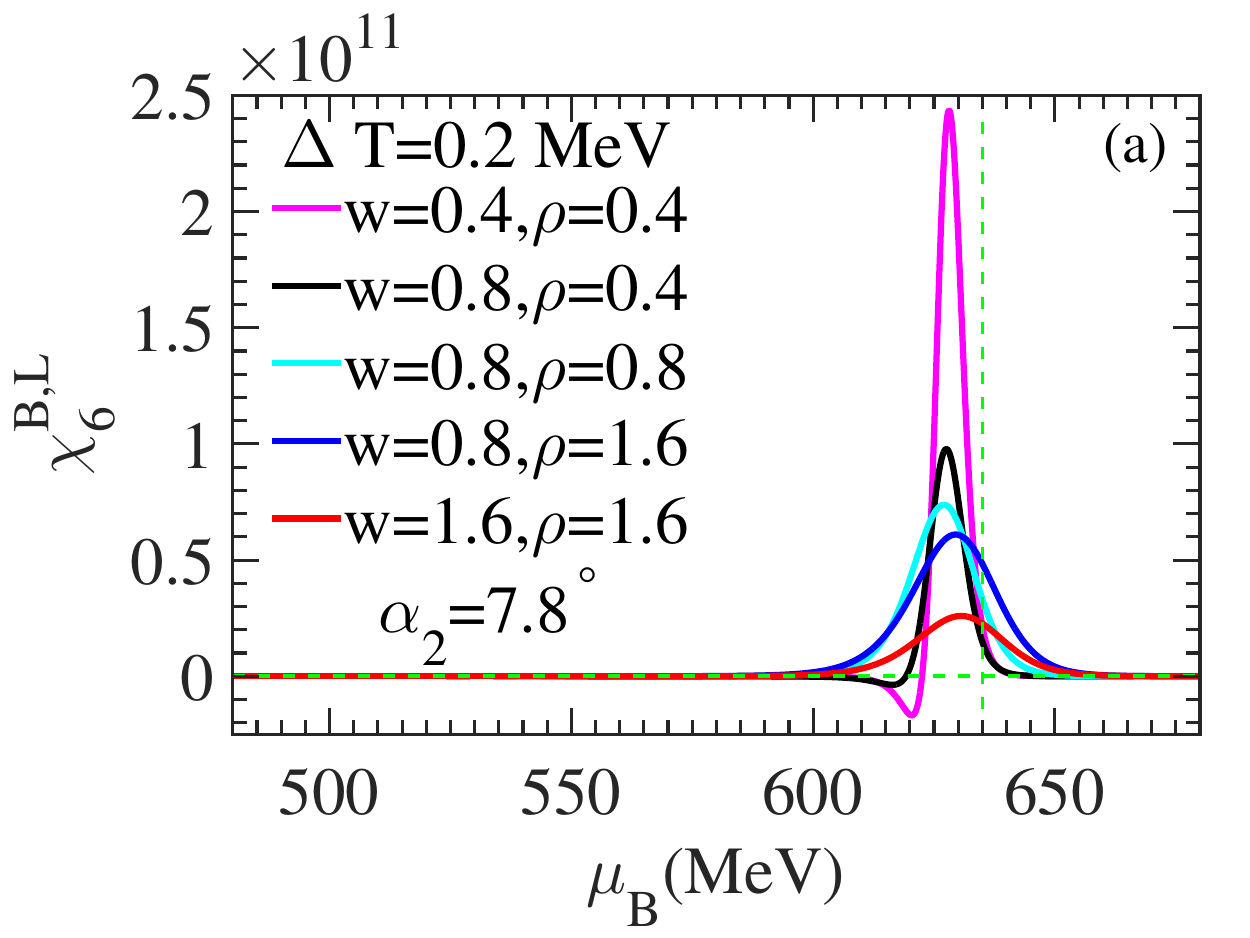}
	\includegraphics[width=0.32\textwidth]{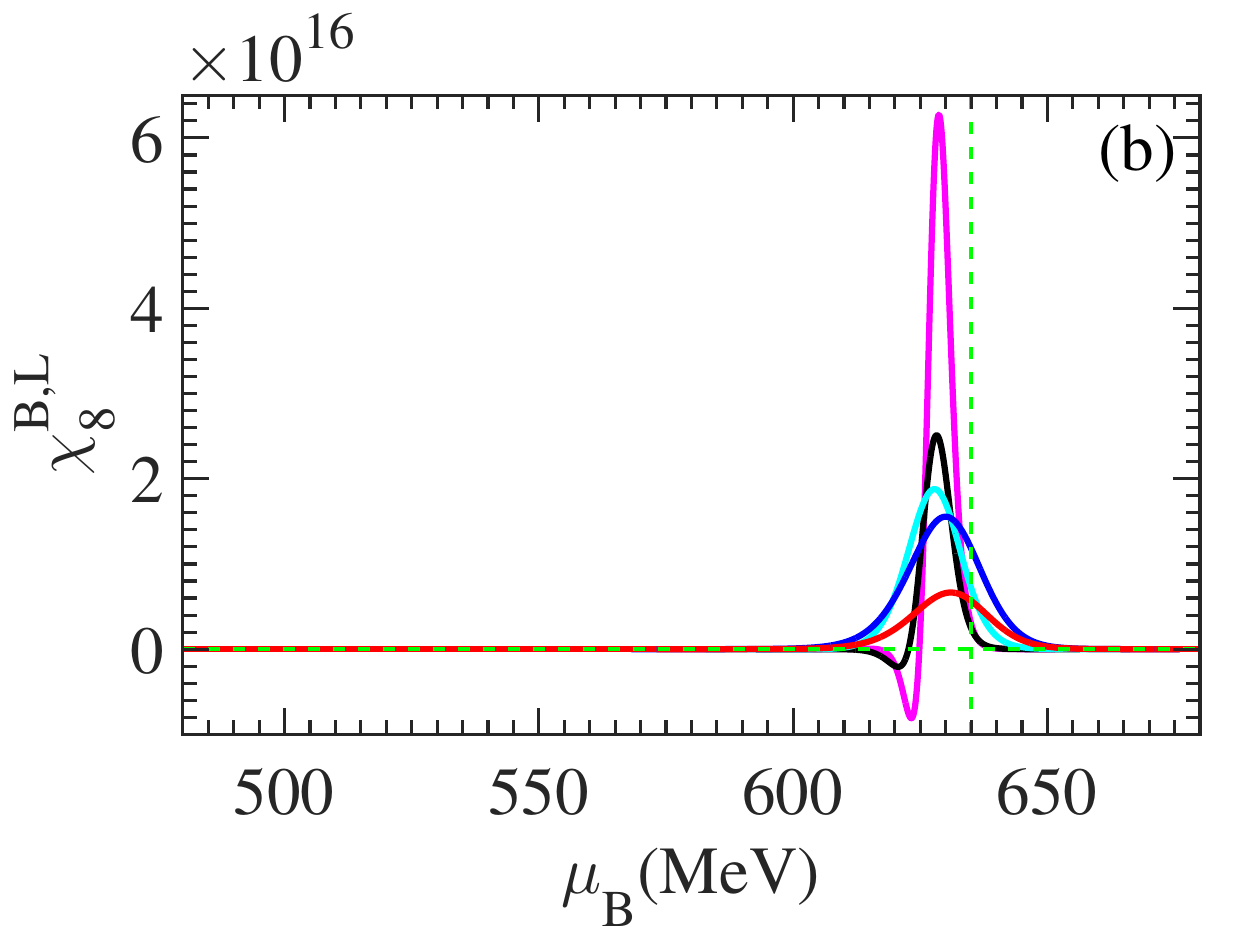}
	\includegraphics[width=0.32\textwidth]{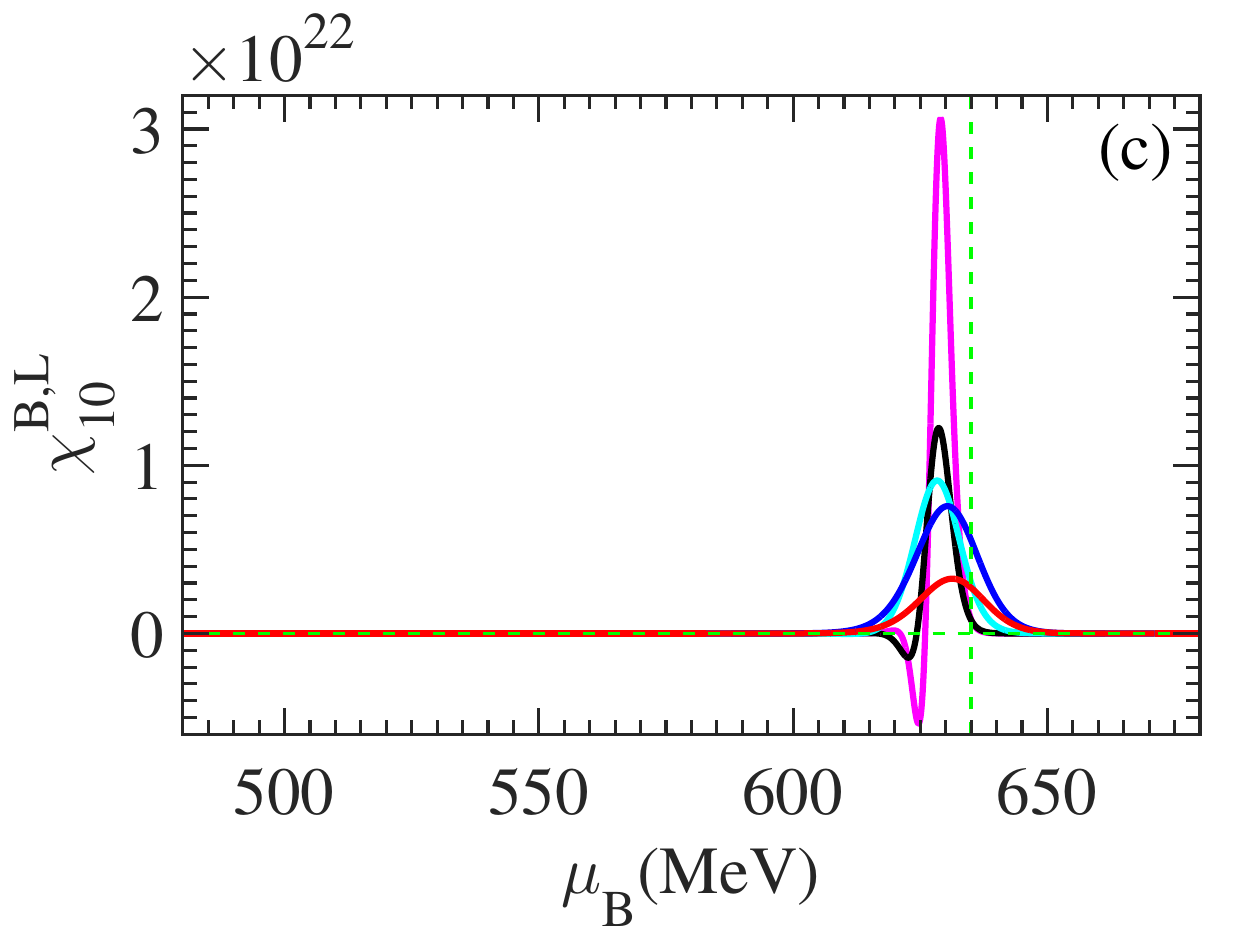}
	\includegraphics[width=0.32\textwidth]{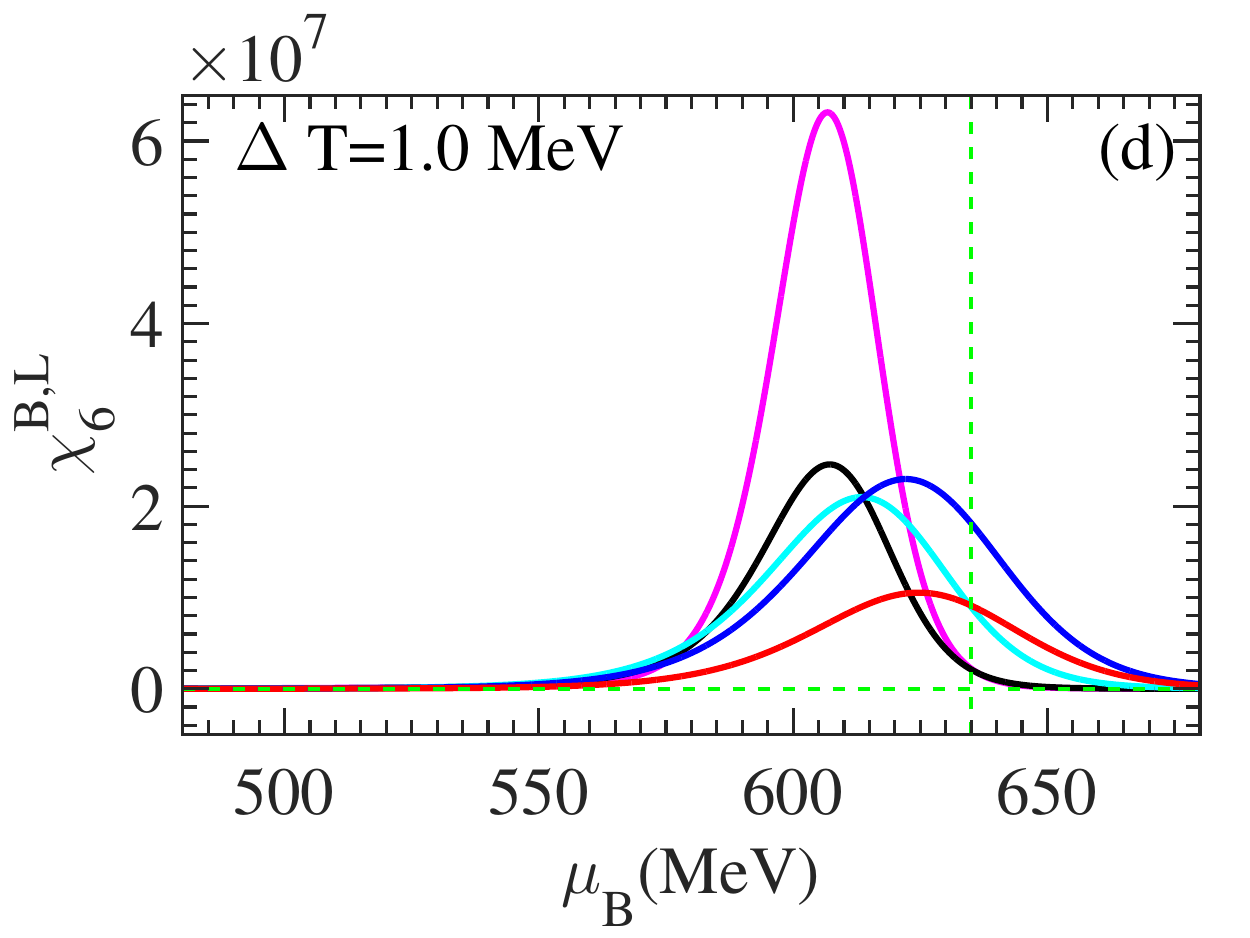}
	\includegraphics[width=0.32\textwidth]{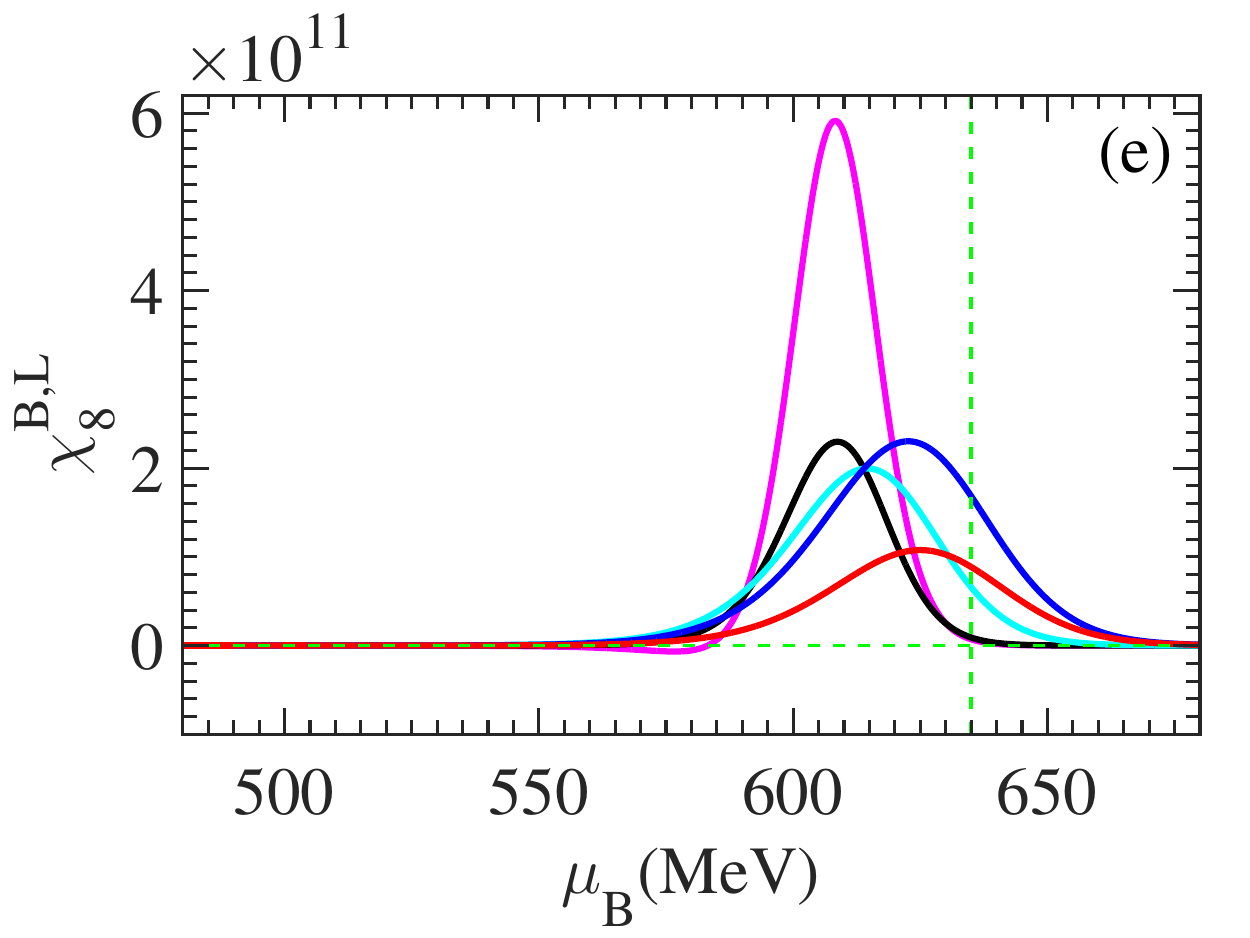}
	\includegraphics[width=0.32\textwidth]{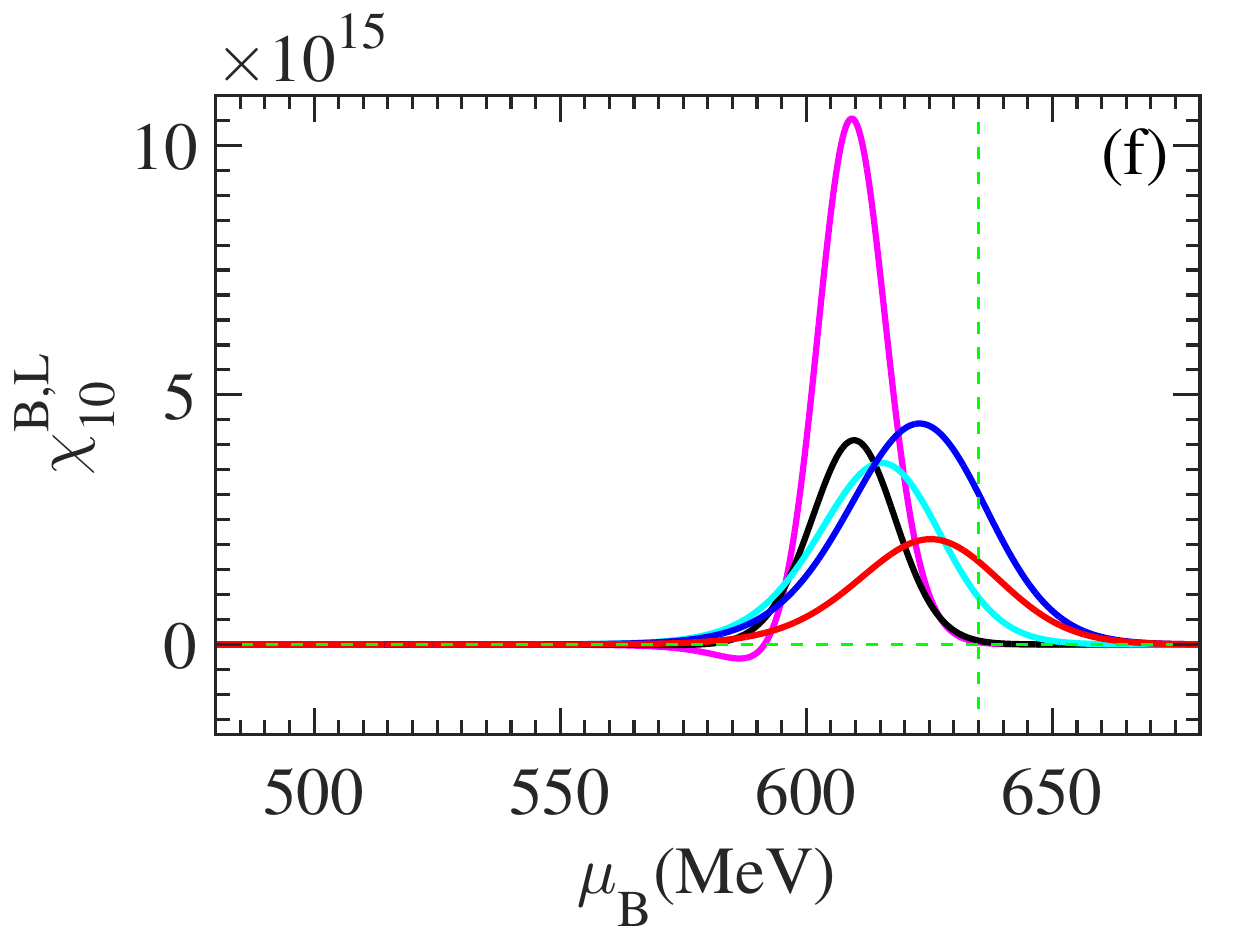}
	\includegraphics[width=0.32\textwidth]{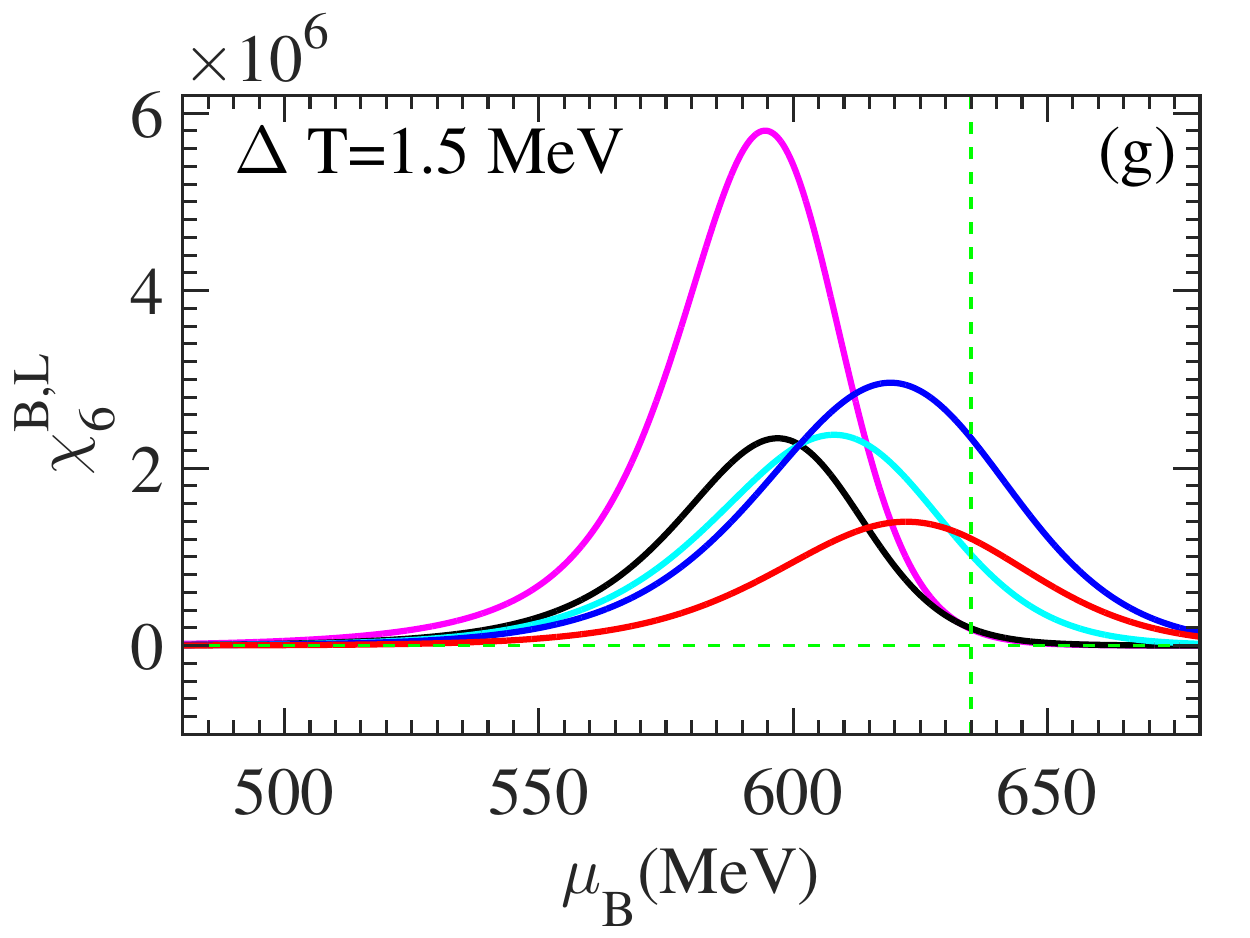}
	\includegraphics[width=0.32\textwidth]{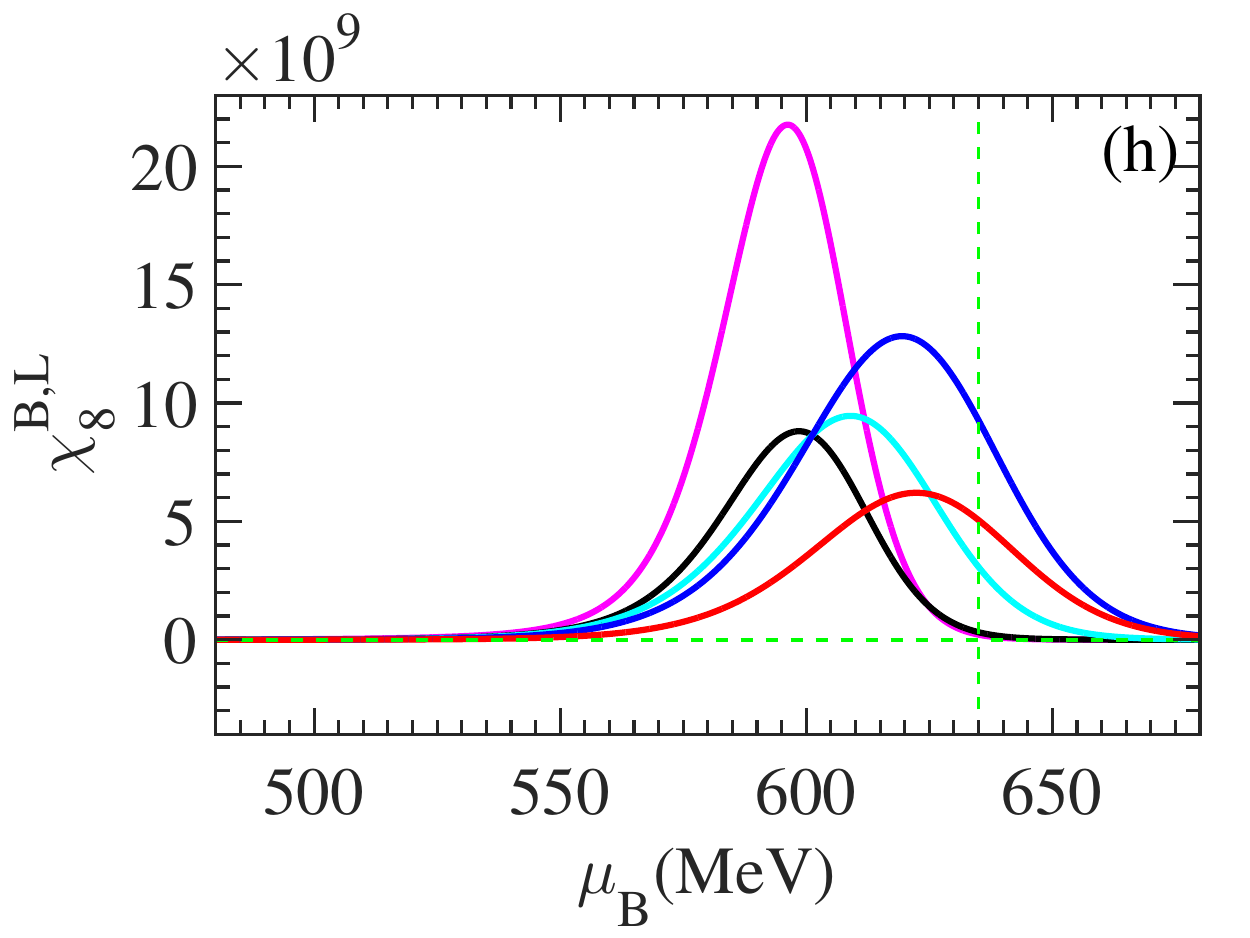}
	\includegraphics[width=0.32\textwidth]{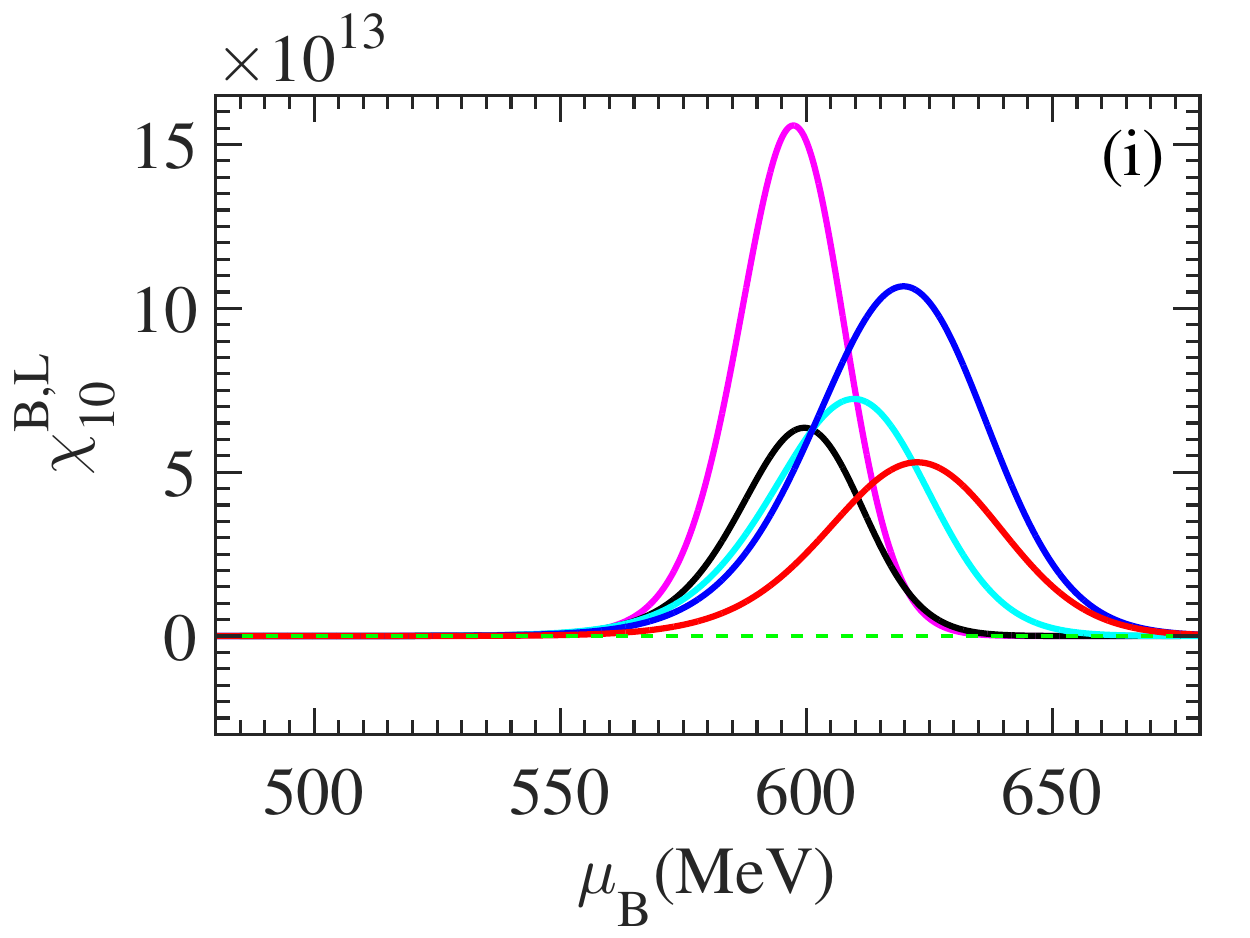}
	\caption{\label{Fig. 8}(Color online). $\mu_B$ dependence of $\chi_{6}^{B,L}$, $\chi_{8}^{B,L}$, and $\chi_{10}^{B,L}$ along the freeze-out curves at $\Delta T=0.2$ MeV (top row), $\Delta T=1.0$ MeV (middle row) and $\Delta T= 1.5$ MeV (bottom row) with different values of $w$ and $\rho$ where $\alpha_2=7.8^{\circ}$.}
\end{figure*}

\begin{figure*}[hbt]
	\centering
	\includegraphics[width=0.32\textwidth]{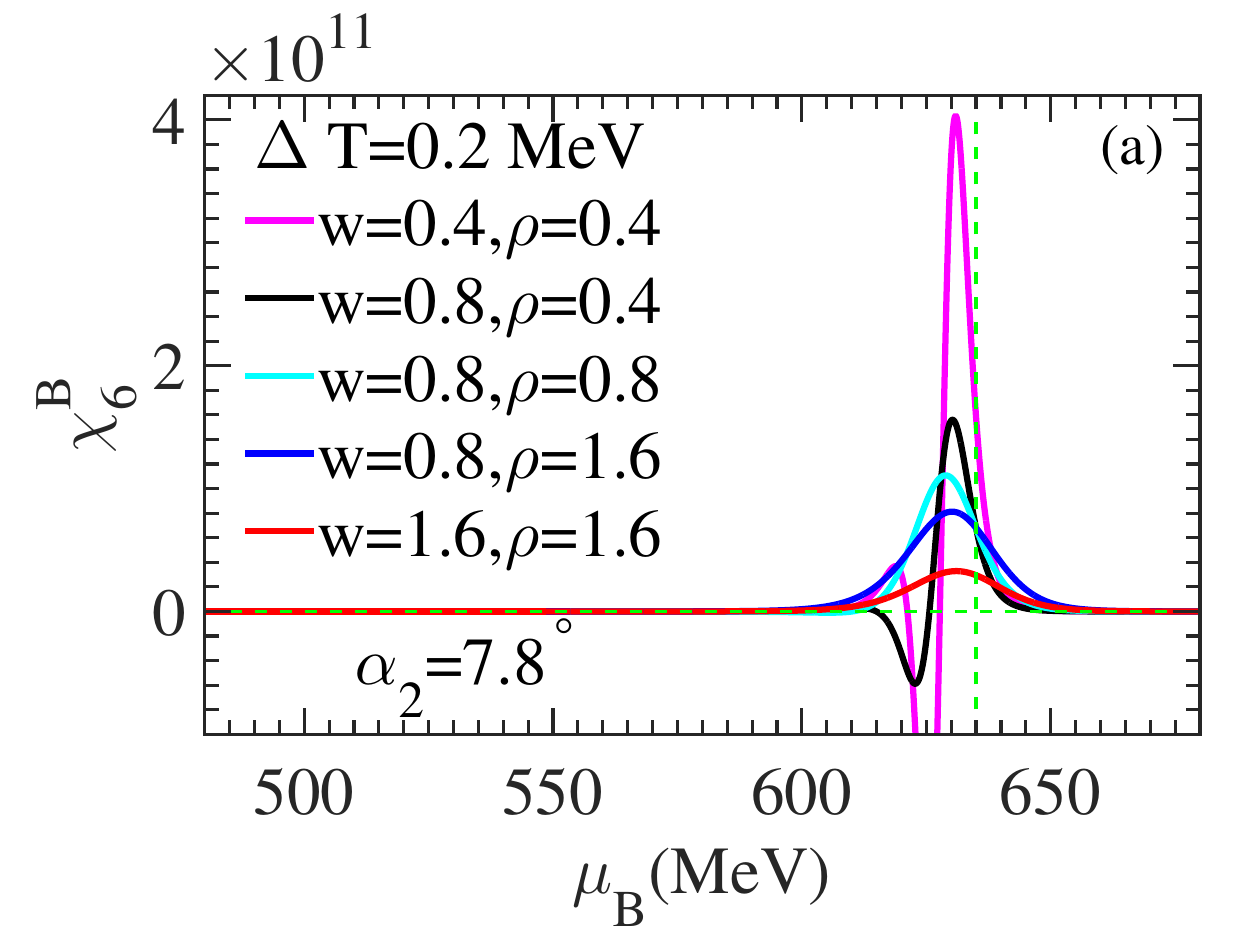}
	\includegraphics[width=0.32\textwidth]{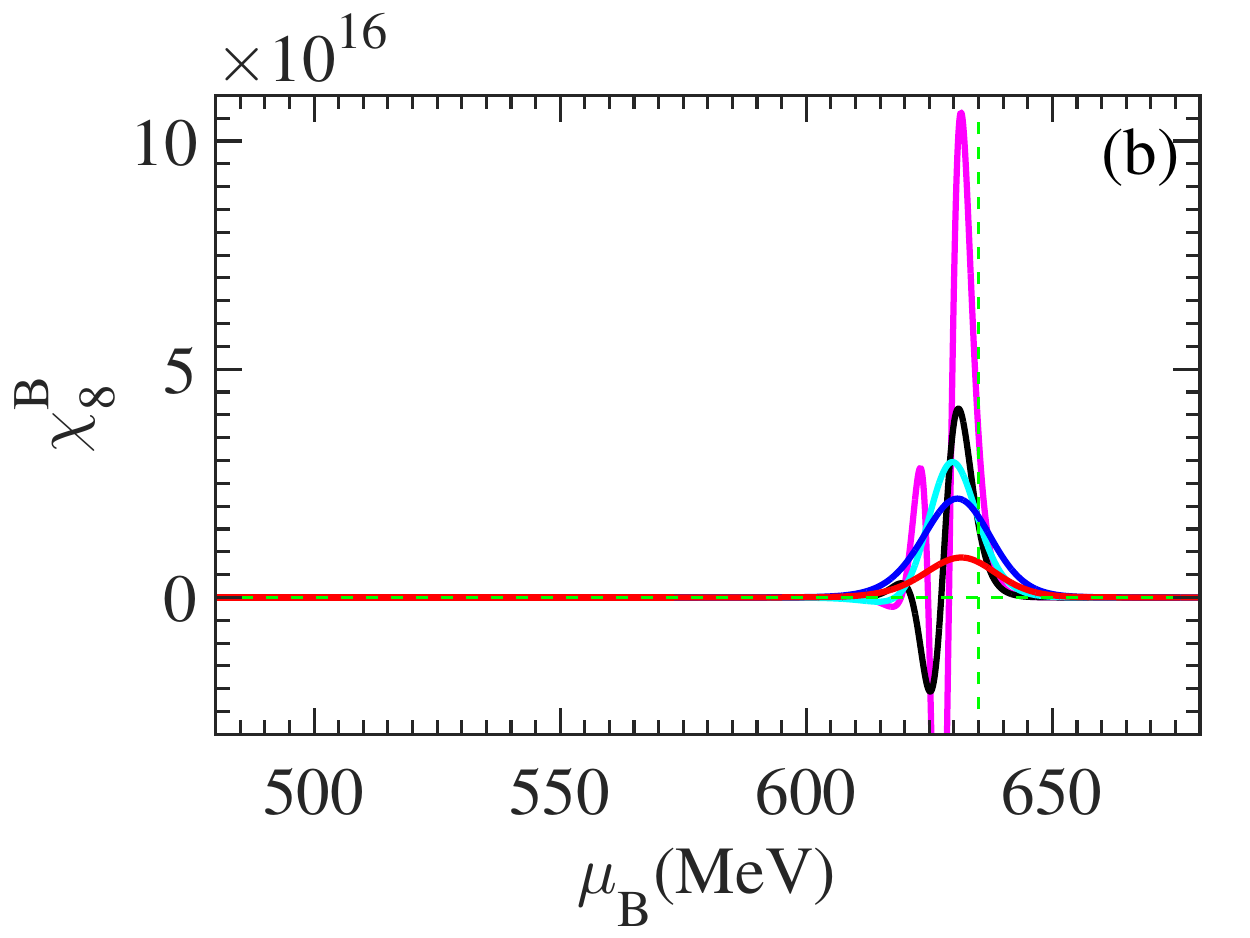}
	\includegraphics[width=0.32\textwidth]{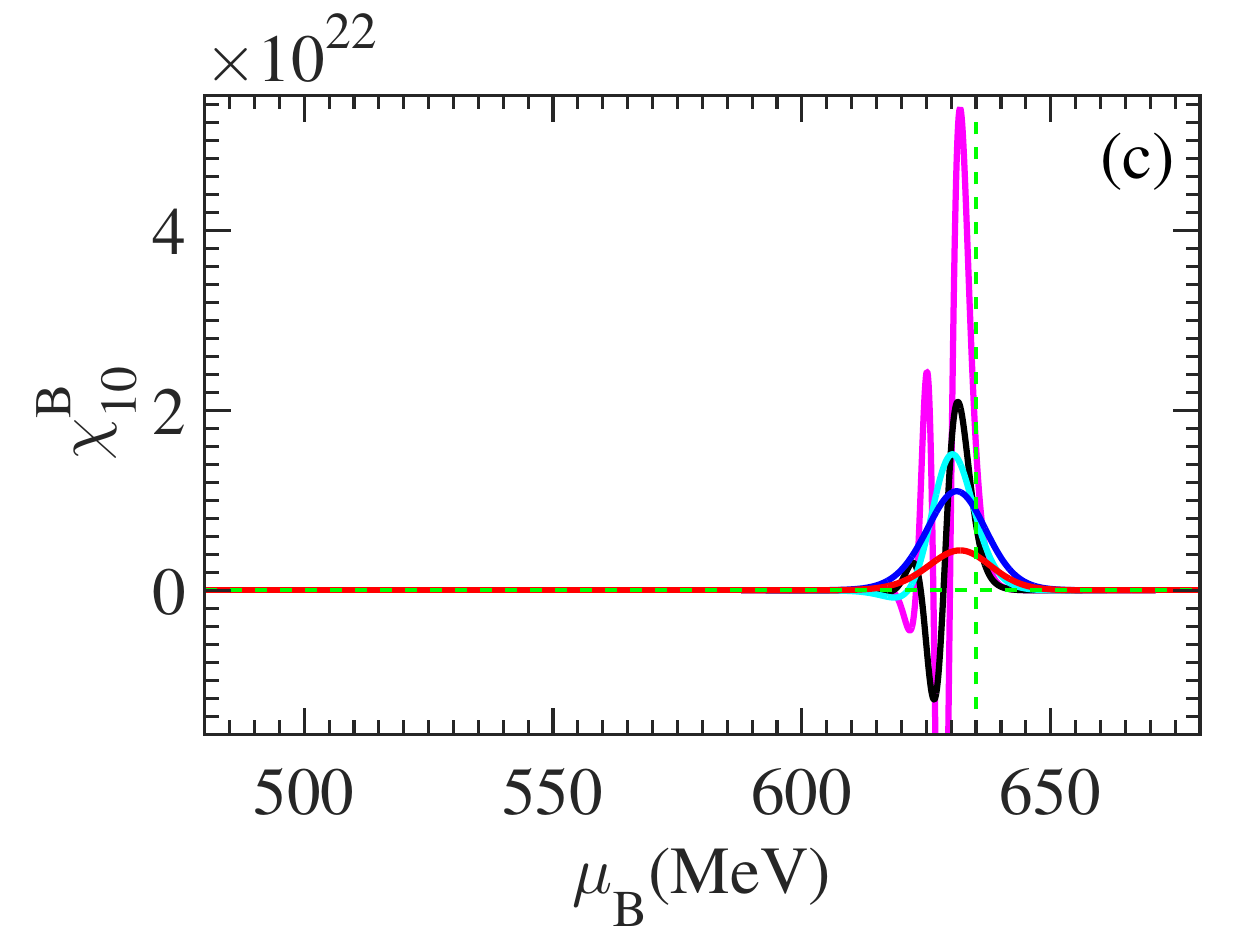}
	\includegraphics[width=0.32\textwidth]{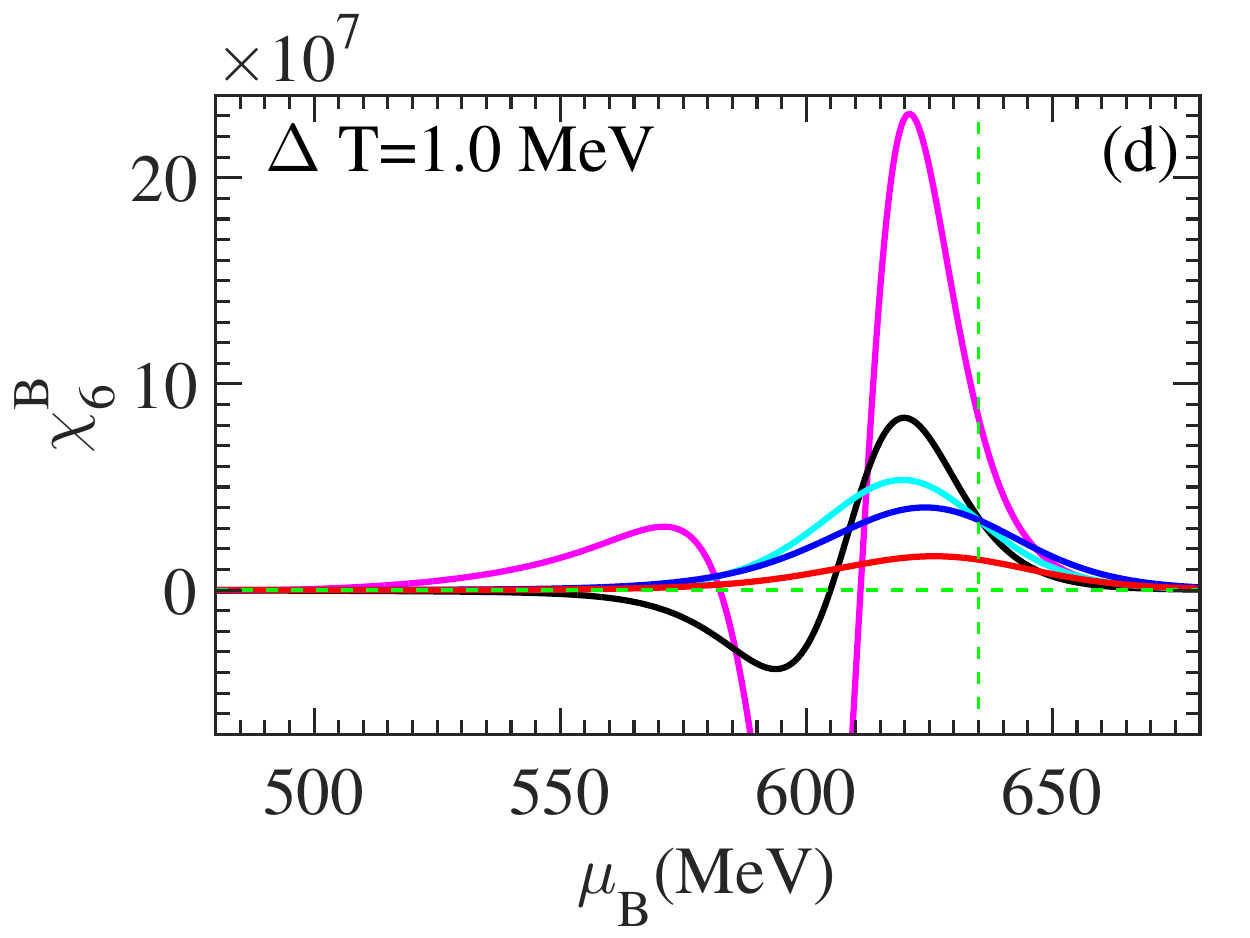}
	\includegraphics[width=0.32\textwidth]{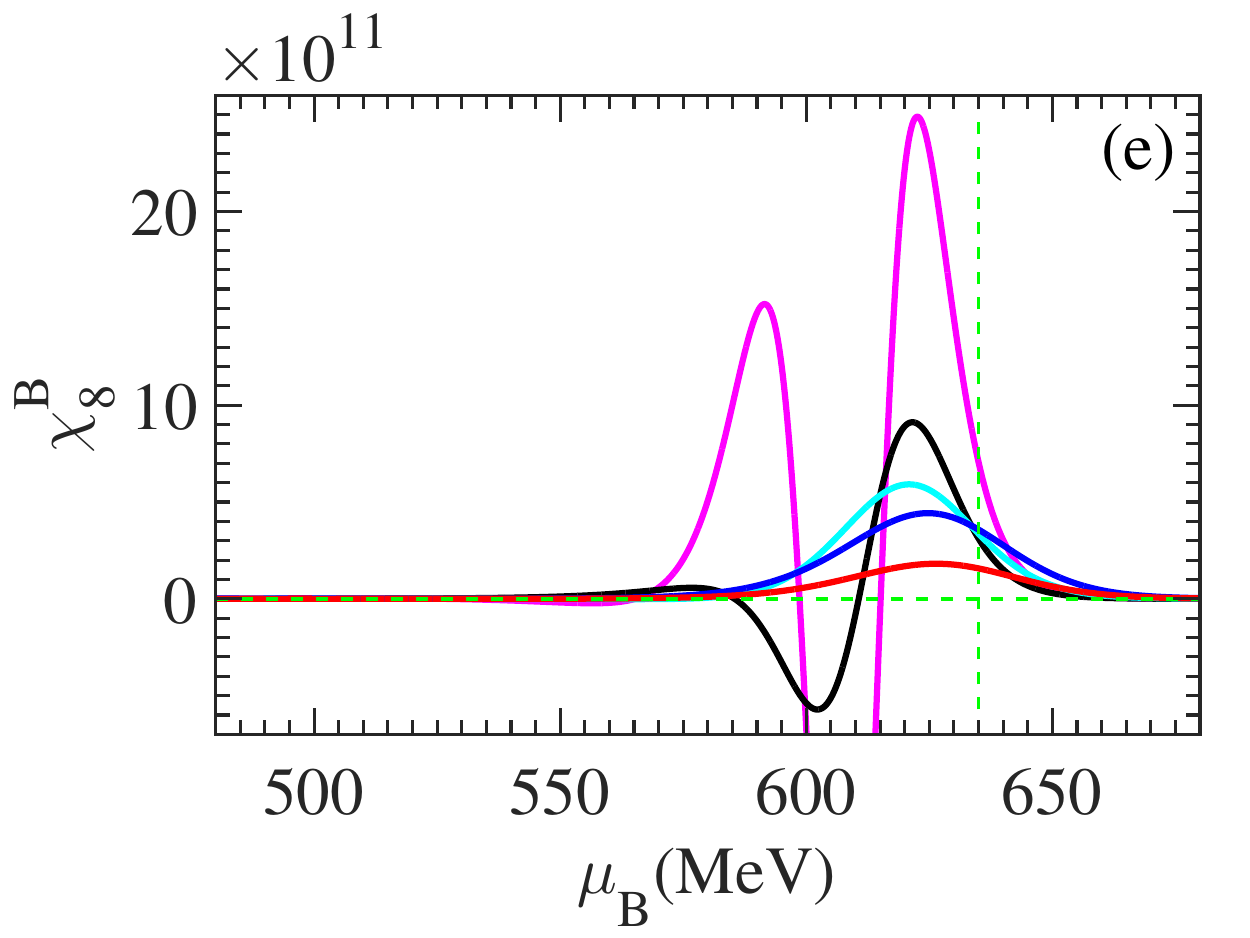}
	\includegraphics[width=0.32\textwidth]{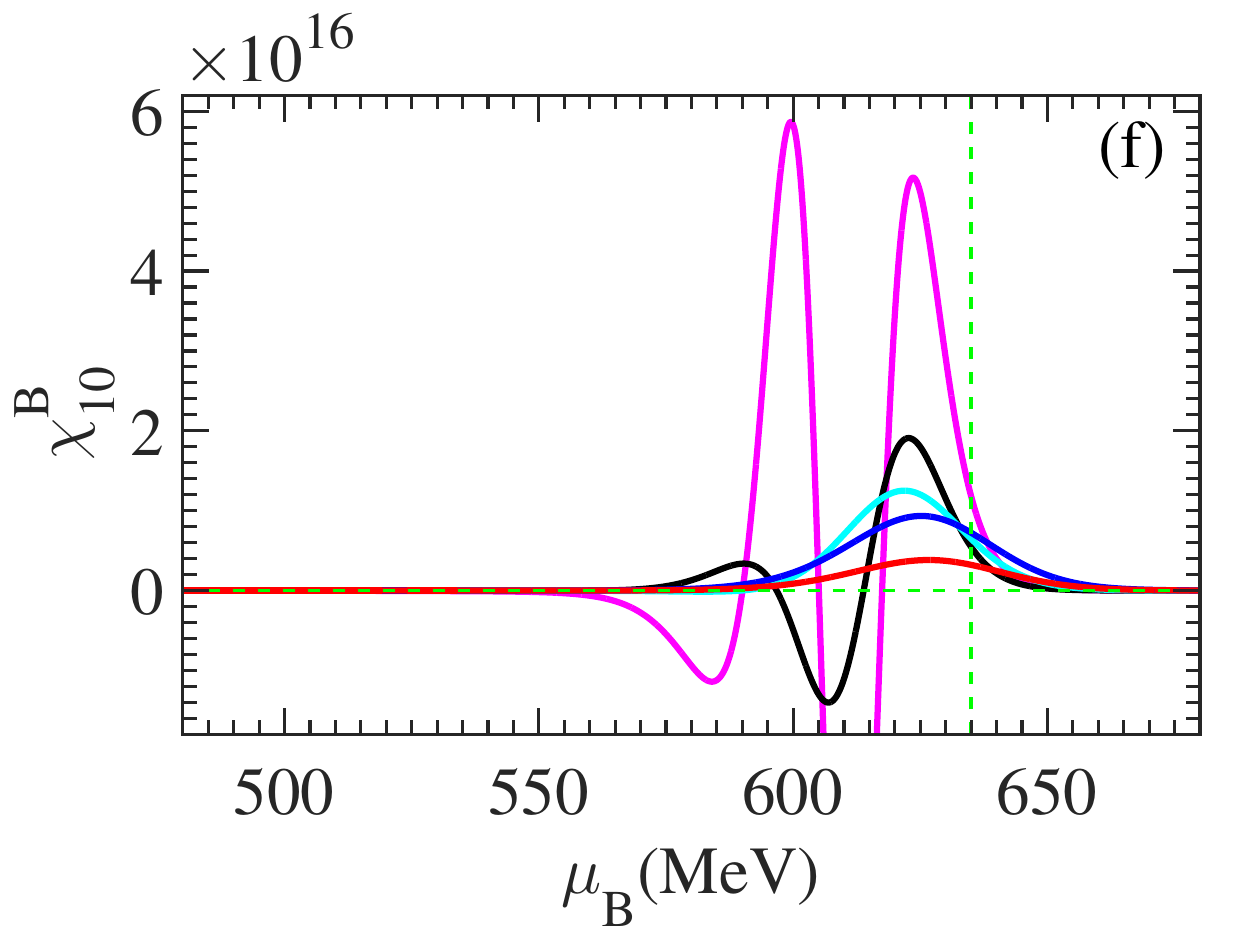}
	\includegraphics[width=0.32\textwidth]{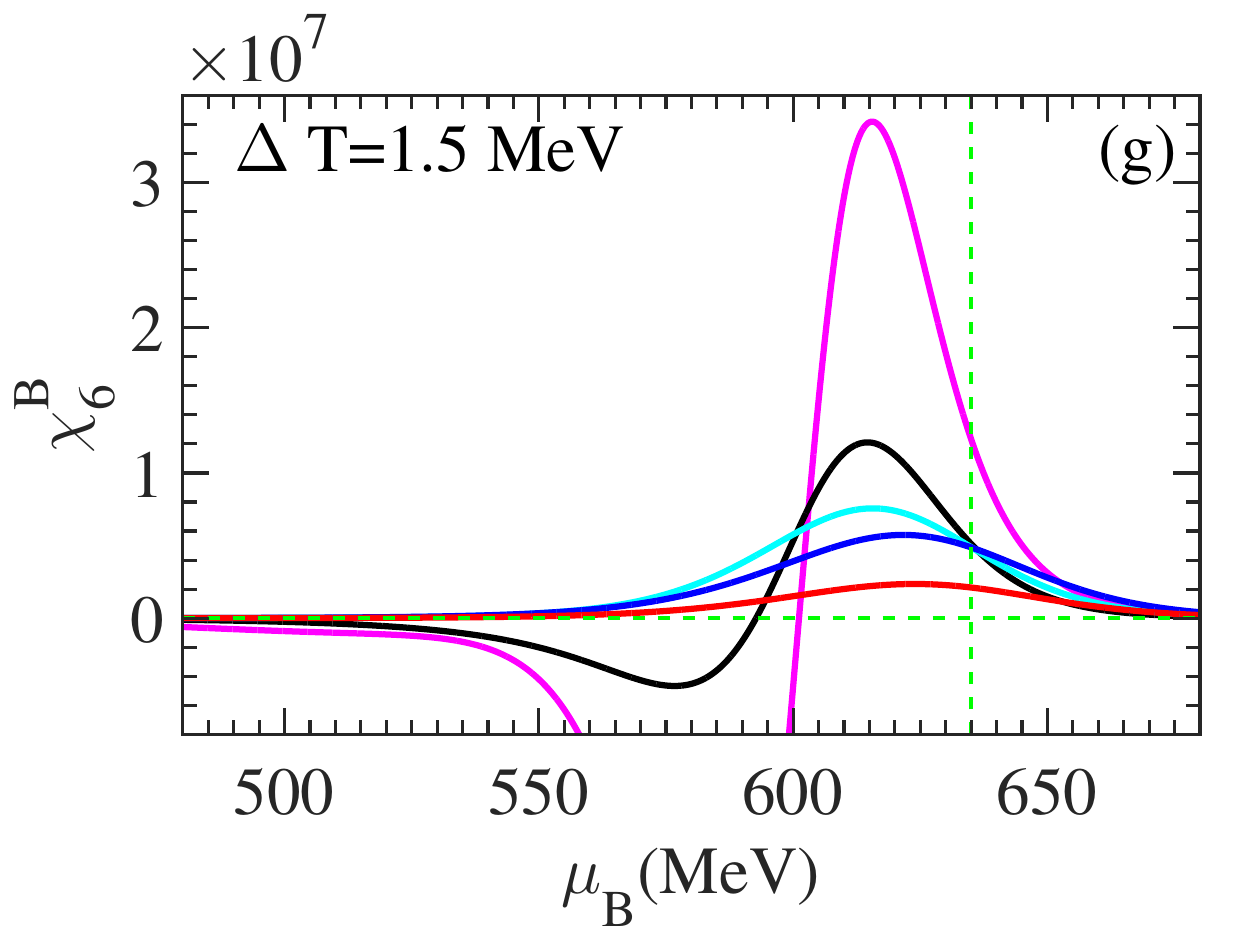}
	\includegraphics[width=0.32\textwidth]{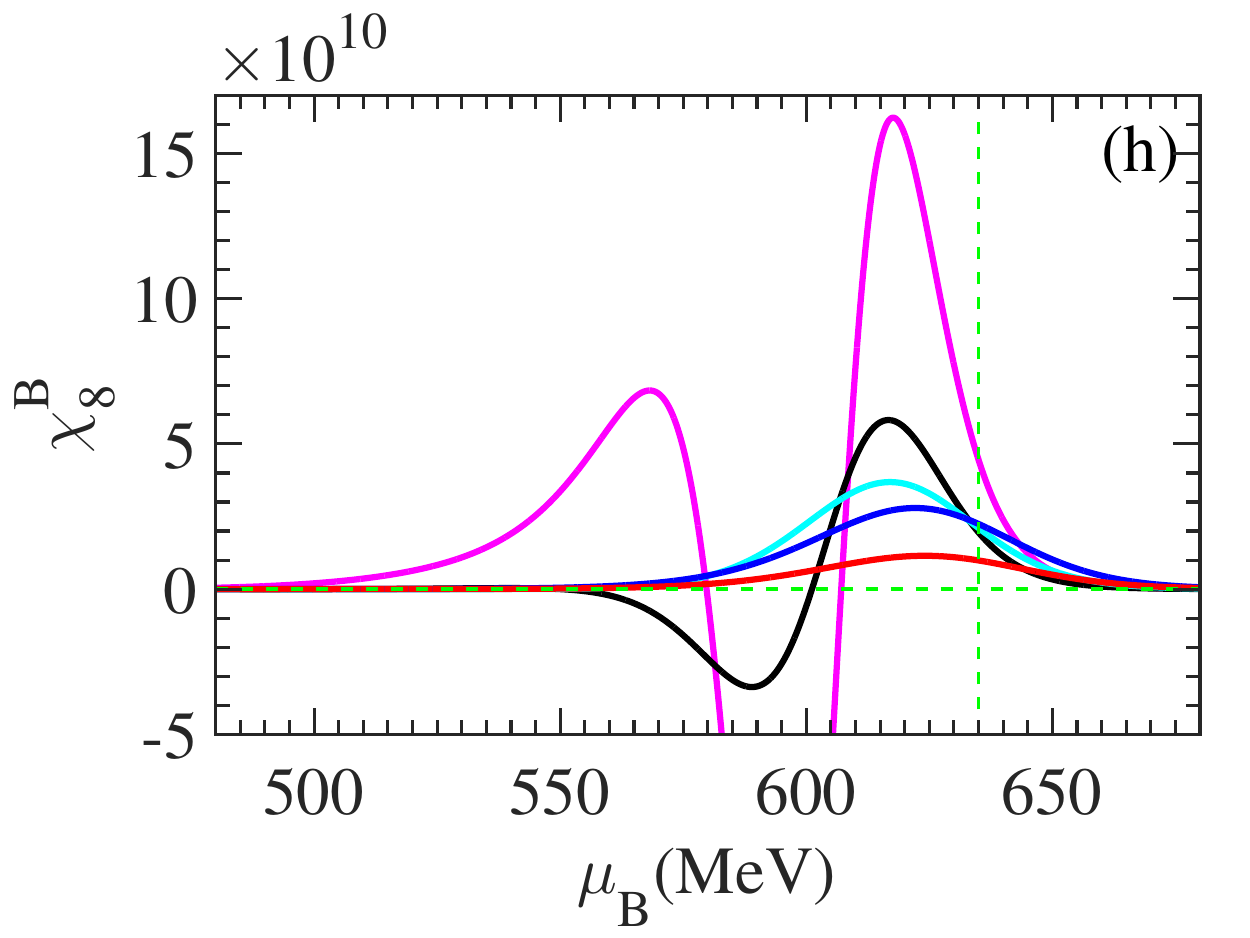}
	\includegraphics[width=0.32\textwidth]{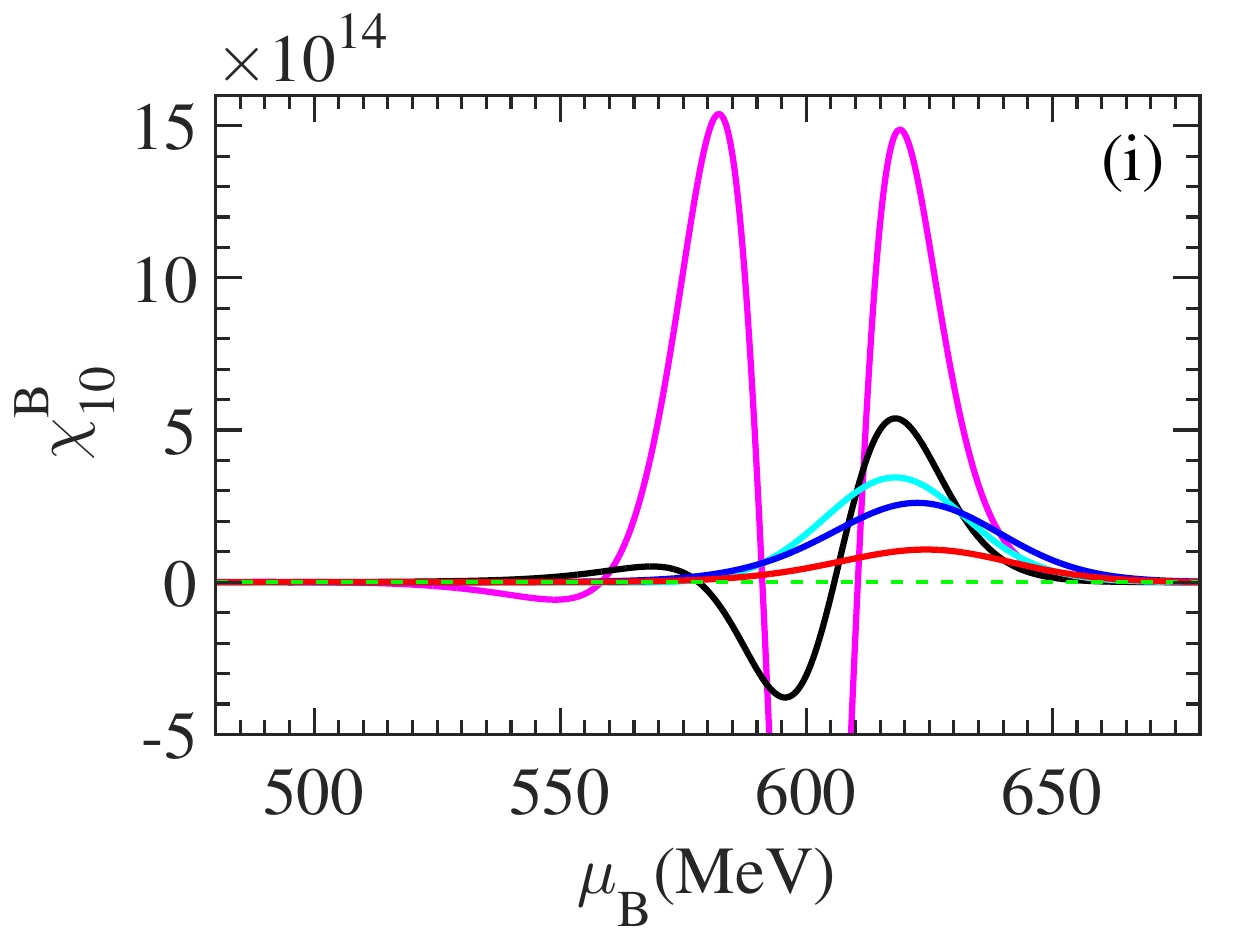}
	\caption{\label{Fig. 9}(Color online). $\mu_B$ dependence of $\chi_{6}^{B}$, $\chi_{8}^{B}$, and $\chi_{10}^{B}$ along the freeze-out curves at $\Delta T=0.2$ MeV (top row), $\Delta T=1.0$ MeV (middle row) and $\Delta T= 1.5$ MeV (bottom row) with different values of $w$ and $\rho$ where $\alpha_2=7.8^{\circ}$.}
\end{figure*}

\section{Generalized susceptibilities of the net-baryon number with $\alpha_2=7.8^{\circ}$}

For the density plots with $\alpha_2=7.8^{\circ}$, the other three sets of values for $w$ and $\rho$ are chosen. They are $(w,\rho)=(0.4, 0.8)$, $(0.8, 0.8)$ and $(0.8, 0.4)$, respectively. The color function for each order of susceptibilities remains consistent with the cases when $\alpha_2=1.8^{\circ}$ in Section 3. 

The density plots of $\chi_6^{B,L}$, $\chi_8^{B,L}$ and $\chi_{10}^{B,L}$ are shown in Fig.~6. In each sub-figure, the purple curve shows the QCD phase transition line represented by Eq.\eqref{QCD transition line}. The purple dot marks the critical point.

It is clear that in the vicinity of the critical point, the positive and negative lobes also occur alternatively in each sub-figure of Fig.~6. The higher the order of the susceptibilities, the more frequent of the alternation of the sign. It is important to note that a greater number of lobes are positioned above the phase transition line. Furthermore, negative blue lobes below the phase transition line are scarcely observable, except in regions extremely close to the phase transition line.

After taking the sub-leading critical contribution into account, the density plots of $\chi_{6}^{B}$, $\chi_{8}^{B}$, and $\chi_{10}^{B}$ are shown in Fig.~7. With small values of the parameters $w$ or $\rho$, obvious negative lobes occur under the phase transition line such as in each sub-figure of the up and bottom row. With the increase of scaling parameters, area of the negative lobes under the phase transition line becomes smaller such as in the middle row of Fig.~7. 

Comparing each sub-figure of Fig.~7 with the corresponding one in Fig.~6, it is clear that the influence of the sub-leading contribution on the density plot is significant than the cases where $\alpha_2=1.8^{\circ}$. First, the area occupied by the main pattern (consisted of the red and dark blue lobes) is bigger in Fig.~7. Second, for each sub-figure with the same values of $w$ and $\rho$, the area of negative blue lobe becomes bigger under the phase transition line.

To observe the $\mu_B$-dependence of the susceptibilities, $\chi_6^{B,L}$, $\chi_8^{B,L}$ and $\chi_{10}^{B,L}$ along the three different freeze-out curves are shown in Fig.~8. The purple, black, cyan, blue and red curve is for five different combinations of values $0.4$, $0.8$ and $1.6$ for $w$ and $\rho$, respectively. 
For $\chi_6^{B}$, $\chi_8^{B}$ and $\chi_{10}^{B}$, the results are shown in Fig.~9.

In Fig.~8, the negative dip structure exists just when the parameters $w$ and $\rho$ are small and it is very close to the phase transition line. For example, in the purple and black curves of the first row of Fig.~8, the negative dips can be observed, but they fade away gradually as it is far away from the phase transition line as shown in the middle and bottom row of Fig.~8. For bigger parameters of $w$ and $\rho$, i.e., in the cyan, blue and red curves, negative dip can not be observed even for $\Delta T = 0.2$ MeV. 

In Fig.~9, when $w$ and $\rho$ are small, such as in the purple and black curves of each row, the negative dips are all obvious, little influenced by the distance to the phase boundary. With the increase of $w$ and $\rho$, such as in the cyan, blue and red curves, it is difficult to observe negative dips even when $\Delta T=0.2$ MeV.

Comparing Fig.~8 and Fig.~9, it is clear that, the sub-leading critical contribution amplifies the negative dip. However, this dip remains non-robust as a signature of the critical point, whereas the positive peak structure persists consistently across all curves in each sub-figure of Fig.~8 and Fig.~9.

So when considering only the leading critical contribution, in the case $0<\alpha_2<\alpha_1$, the negative dip is not a robust feature in the $\mu_B$-dependence of generalized susceptibilities of the net-baryon number. This behavior markedly contrasts with the case in the conventional assumption of $\alpha_1-\alpha_2=90^{\circ}$, where the negative dip typically appears. The manifestation of this feature is now parameter-dependent, being influenced by the scaling parameters $w$ and $\rho$, as well as the proximity to the phase transition line. When incorporating the sub-leading critical contribution, although the negative dip is intensified to some degree, it remains insufficient to be a robust feature of the critical point. 

All in all, in the case $0<\alpha_2<\alpha_1$, the negative dip in the $\mu_B$-dependence of the sixth-, eighth-, and tenth-order susceptibilities is not a robust feature of the critical point. This conclusion holds true irrespective of whether one considers only the leading critical contribution or includes both leading and sub-leading terms. The existence of this dip depends on the scaling parameters $w$ and $\rho$, as well as the proximity of the freeze-out curve to the phase transition line. In contrast, the peak structure maintains its robustness as a characteristic feature of the critical point.

\section{Summary}

In this paper, we have studied the sixth-, eighth-, tenth-order susceptibilities of the net-baryon number based on the universal critical behavior in the limit of small quark masses. Where the angle between the variables, reduced temperature $t$ and magnetic field $h$ in the three-dimensional Ising model, vanishes as $m_q^{2/5}$ after mapping to the QCD $T-\mu_B$ phase plane. Keeping the condition $0<\alpha_2<\alpha_1$, we choose two different values of $\alpha_2$. One is close to zero and the other one is close to $\alpha_1$.

In both cases, when considering only the leading critical contribution, the existence of negative dip in the $\mu_B$-dependence of the susceptibilities depends on the scaling parameters $w$ and $\rho$, as well as the distance to the phase transition line. This behavior contrasts with the case in the conventional assumption of $\alpha_1-\alpha_2=90^{\circ}$, where the negative dip typically appears.

After incorporating the sub-leading critical contribution, the negative dip is amplified to some degree. This behavior markedly contrasts with the case in the conventional assumption of $\alpha_1-\alpha_2=90^{\circ}$, where the original universal negative dip becomes parameter-dependent due to the sub-leading critical contribution. Nevertheless, the negative is still not a robust feature of the critical point.

On the other hand, the positive peak persists consistently, independent of parameter choices and the proximity to the phase transition line. It is a robust signal of the critical point.

\vskip 0.5cm
This work is supported by Research Start-up Fund Project of Chengdu University (X2110) and the National Supercomputing Center in Chengdu-Chengdu University Branch.

\ed